\documentclass[acmtog,nonacm]{acmart}

\definecolor{darkgreen}{rgb}{0.0,0.65,0}

\newcommand{\add}[1]{{\color{blue}#1}}

\DeclareUnicodeCharacter{2713}{\unicheck}

\usepackage{layouts}
\usepackage{tabularx}
\usepackage{float}
\usepackage{enumitem}
\usepackage{multirow}
\usepackage{booktabs}
\usepackage[normalem]{ulem}
\usepackage{xcolor}    

\AtBeginDocument{%
  }

\setcopyright{acmlicensed}
\copyrightyear{2018}
\acmYear{2018}
\acmDOI{XXXXXXX.XXXXXXX}
\acmConference[Conference acronym 'XX]{Make sure to enter the correct
  conference title from your rights confirmation email}{June 03--05,
  2018}{Woodstock, NY}
\acmISBN{978-1-4503-XXXX-X/2018/06}

\newif\ifshownotes
\shownotestrue 

\ifdefined\MakeWithNotes
  \shownotestrue
\fi
\ifdefined\MakeWithoutNotes
  \shownotesfalse
\fi

\ifshownotes
  \newcommand{\colornote}[3]{{\color{#1}\bf{#2: #3}\normalfont}}
  \newcommand{\colornoteTwo}[3]{{\color{#1}\bf{#3}\normalfont}}
  \newcommand{\colornoteThree}[2]{{\color{#1}\bf{#2}\normalfont}}      
\else
  \newcommand{\colornote}[3]{}
  \newcommand{\colornoteTwo}[3]{}
  \newcommand{\colornoteThree}[2]{}      
\fi

\begin{document}

\title{Aggregating LLM-Based Weak Verifiers for Spatial Layout Generation}

\author{Sharon Zhang}
\email{szhang25@stanford.edu}
\orcid{0000-0002-6738-8906}
\affiliation{%
  \institution{Stanford University}
  \city{Stanford}
  \state{California}
  \country{USA}
}

\author{R. Kenny Jones}
\email{rukjones4@gmail.com}
\orcid{0009-0005-1169-0507}
\affiliation{%
  \institution{Stanford University}
  \city{Stanford}
  \state{California}
  \country{USA}
}

\author{Jiajun Wu}
\email{jiajunwu@cs.stanford.edu}
\orcid{0000-0002-4176-343X}
\affiliation{%
  \institution{Stanford University}
  \city{Stanford}
  \state{California}
  \country{USA}
}

\author{Maneesh Agrawala}
\email{maneesh@cs.stanford.edu}
\orcid{0000-0002-8996-7327}
\affiliation{%
  \institution{Stanford University}
  \city{Stanford}
  \state{California}
  \country{USA}
}
\affiliation{%
  \institution{Roblox}
  \city{San Mateo}
  \state{California}
  \country{USA}
}

\renewcommand{\shortauthors}{Zhang et al.}

\begin{teaserfigure}
    \includegraphics{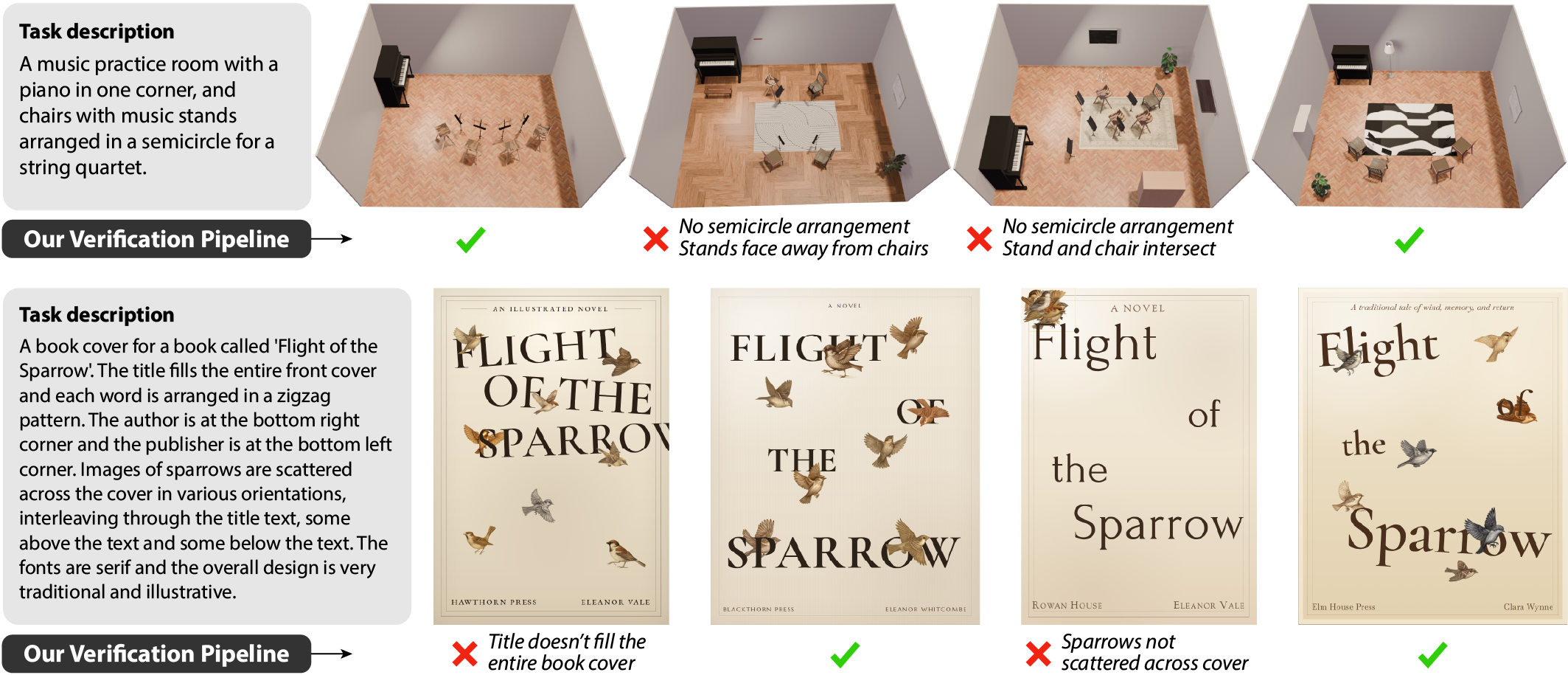}
    \caption{
      We introduce a pipeline to verify the outputs of spatial layout generators against specific task descriptions. Our approach builds a collection of LLM-based weak verifiers using a layout verification DSL, and then aggregates them into a single strong verifier. We apply our verification pipeline to 3D room layouts and 2D poster designs.
      Here, the resulting task-specific strong verifiers
      mark layouts as positive (green check) if they match the task description and negative (red X) if they
      fail to match.
      %
      For each negative example we include a brief note on what violates the task description.
      We show how these verifiers can be used in conjunction with a layout generator to automatically
      identify generated layouts that match the task descriptions.
    }
      \label{fig:teaser}
\end{teaserfigure}

\begin{abstract}
  We present a pipeline for building and aggregating task-specific, LLM-generated weak
  (imperfect) verifiers into a strong verifier for spatial layout domains.
  Given a task description, our pipeline asks an LLM to synthesize a
  collection of verifier programs using a layout verification
  DSL.
  Each individual LLM-generated verifier usually provides an imperfect check
  for a match between the layout and the corresponding task
  description.
  We show that by aggregating the responses of many such verifiers we
  can produce a stronger verifier. Moreover, by applying techniques
  from weak learning, our pipeline can learn how to aggregate the weak
  verifiers from a very sparse set of human labeled example layouts
  (about 10).
  We find that the strong verifiers produced by our pipeline
  outperform the status-quo approach of
  using a set of LLM judges to directly check whether a layout matches
  a task description, 
raising F1-scores by up to 7X across a variety of 3D room layout and 2D poster design tasks.
  We also demonstrate that verifier-guided layout generation using natural language feedback from our strong verifiers improves layout quality of
  a base layout generator by up to 66.2\% according to a human evaluator.
\end{abstract}

\maketitle

\section{Introduction}

Generative models have demonstrated tremendous
progress in producing spatial layouts in a variety of settings
including 3D room layout and 2D poster design.
But they often require repeated sampling before producing a layout
that captures all of the properties desired by a user.
%
For example, a user might ask the model to generate a music practice room for a string quartet 
(Figure\,\ref{fig:teaser} top), manually check the generated response, and iteratively repeat the generation request until the response passes
 manual verification.
%
%
Such manual verification is typically slow and requires expertise in
layout design.
One way to automate this verification process is to simply ask an LLM
to judge how well a generated layout matches the task
description\,\cite{zhu2023judgelm,zheng2023judging}. 
However, LLMs typically lack strong spatial understanding\,\cite{wang2024spatial, mayer2025ivispar} and produce noisy, inconsistent evaluations that are poorly
calibrated to human judgments of layout design\,\cite{lin2024evaluating}.
%
Another approach is to ask the user to
label a set of example layouts as positive or negative (i.e., passing
or failing their verification judgments) and then train a
verification model on this data.
However, this type of human verification labeling does not scale and
it is difficult to learn an accurate verifier from sparse training
data.



Weak learning offers an appealing approach to solve this problem.
Given a set of weak (imperfect) verifiers, the weak learning paradigm analyzes the voting patterns of these verifiers on unlabeled data to produce a strong verifier~\cite{10.5555/3157382.3157497}. 
Originally, these weak verifiers were hand-written procedural ``labeling functions'' authored by domain experts, but more recent approaches have considered whether LLMs can serve as the weak verifiers.
Weaver\,\cite{2025weaver} is a state-of-the-art method that treats different LLMs as imperfect judges, and learns a weighting of these LLM votes that has been shown to surpass the naive majority vote alternative.
However, Weaver and other LLM judge approaches have primarily been
applied to problems in natural language QA\,\cite{cook2024tickingboxesgeneratedchecklists, gundawar2024robustplanningcompoundllm}, coding\,\cite{shinn2023reflexion, pmlr-v267-zhuge25a} and math
QA\,\cite{cobbe2021trainingverifierssolvemath, zhang2024generative}, areas where the capabilities of LLMs are stronger.

In this paper we show how to adapt these ideas from weak learning into
a pipeline for building and aggregating task-specific, LLM-generated
weak verifiers into a strong aggregate verifier for spatial layout. 
%
Rather than directly 
using an LLM as a judge, our approach is to 
ask the LLM (GPT-5.4) to synthesize a set of task-specific verifier programs using a layout
verification DSL that provides access to low-level spatial queries
over layout elements.
%
%
We analyze several methods for aggregating the resulting weak verifiers, including naive majority vote, logistic regression, the top-1 weak verifier and Weaver\,\cite{2025weaver}.



We also use our strong verifiers to perform {\em verifier-guided layout generation}---a process that automates the iterative generate-then-verify loop.
We investigate two different forms of verifier feedback in this loop. {\em Binary} feedback simply uses the verifier True/False response. 
{\em Detailed} feedback provides natural language messages explaining the verifier response, 
which we obtain in two steps. 
First, we ask an LLM to modify each programmatic weak verifier by annotating each exit point with a descriptive feedback message.
Then, we combine these feedback messages using the learned aggregation weights of each weak verifier.



%


Our experiments find support for three main claims;
%
(1) Using our programmatic weak verifiers in combination with any of the aggregation techniques outperforms directly using LLMs as judges by up to 7X in F1-score; 
(2) Weaver aggregation generally outperforms the other aggregation methods with as few as 10 labeled layout examples; 
(3) Using our strong verifiers in verifier-guided layout generation with Detailed feedback improves a base layout generator by 66.2\% 
according to a human evaluator.

\section{Related Work}


{\noindent\bf\em Weak supervision for data labeling.} Weak supervision is a paradigm of
machine learning which aims to learn a model of data from both labeled
{\em and} unlabeled data\,\cite{SSL}. One primary application of weak
supervision is in data programming\,\cite{dataprog}, which aims to
create large labeled datasets by learning how to aggregate multiple
weak labeling functions into one stronger-performing label predictor. These weak labeling functions are often
designed based on domain heuristics and can be both incomplete (i.e.,
only label a subset of the data and abstain on the rest) and noisy
(i.e., not always accurate). A common approach is to learn a
generative model on top of the votes of these weak labeling functions
that can predict a single probabilistic label\,\cite{Snorkel, MeTaL, FlyingSquid}. Our work builds on ideas from weak learning to create verifiers for spatial layout tasks.


\vspace{0.3em}
{\noindent\bf\em Verification with LLMs.}
LLM judges have
been used to evaluate and select responses to generative queries\,\cite{zhu2023judgelm, zheng2023judging,wu2023gpteval3d} or provide reward signals to improve models during
training\,\cite{wang2023math}. However, LLM judges on their own are
not entirely reliable\,\cite{judgebench2024} and in particular
suffer from poor performance on direct spatial reasoning
tasks\,\cite{wang2024spatial, mayer2025ivispar, ma2026real3dqa}.
A common approach to
address this is to aggregate multiple
models\,\cite{verga2024replacingjudgesjuriesevaluating,
  2025weaver}.
  In this
paper, we take inspiration from Weaver\,\cite{2025weaver}, which uses
weak supervision to aggregate the responses of
LLM judges. Rather than directly querying an
off-the-shelf LLM to verify the outputs of our generators, we query
the LLM to write weak verification {\em programs} using a layout
verification DSL that we provide. The idea of using LLM-generated
verification programs has been explored in the context of 2D vector
graphics animation\,\cite{mover}, image synthesis\,\cite{dreamsync} and
scene generation\,\cite{scenecraft,viga}.
However, instead of relying on a
single (possibly imperfect) LLM-generated verification program we aggregate feedback from
multiple LLM-generated weak verifier programs.
SceneEval ~\cite{tam2026sceneeval} evaluates scene generators with a set of predefined task-agnostic checks and some prompt-specific verifiers, but  requires a human-in-the-loop.

\vspace{0.3em} {\noindent\bf\em Layout generation.} 
Creating layouts is an
important step in many visual creation pipelines, and layout generation methods have been proposed in both 2D graphic design\,\cite{feng2023layoutgpt, chen2024textlap, 10.1007/978-3-032-04614-7_12, Wu_2025_CVPR} and computer-aided
design\,\cite{Jones2025, du2024blenderllmtraininglargelanguage, Li2025}. Many recent generation systems focus on creating 3D scenes, as they have applications in interior design, game environments and embodied-AI training\,\cite{Paschalidou2021NEURIPS, littlefair2025flairgpt, feng2023layoutgpt, artiscene2024, zhou2024gala3d, 10.1145/3588432.3591561, 10.1145/3757377.3763930, zhang2023scenewiz3d, 10.1145/3197517.3201362,wang2020sceneformer, scenesmith2026}. 
Recently, LLMs have been incorporated into scene generation as critics for verifier-guided layout generation~\cite{viga,scenecraft}. Our work uses feedback from our aggregated strong verifiers, which we show results in stronger generators.



\begin{figure*}[t!]
  \centering
    \includegraphics[width=\linewidth]{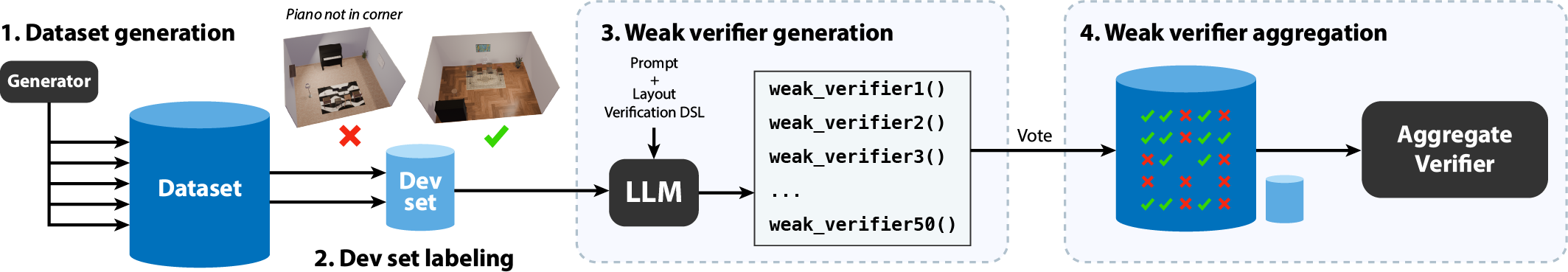}
    \caption{The four stages of our verification pipeline.
      {\bf 1. Dataset generation.} We first curate a dataset of training examples from a generator or an existing dataset. {\bf 2. Dev set labeling.} The user then labels a small random subset (about 10) of these examples as positive or negative, possibly providing notes alongside each label, e.g. ``the piano is not in the corner." 
      {\bf 2. Weak verifier generation.} We provide an LLM with the user prompt, the labeled dev set examples, and documentation of our DSL and query it to generate a set of weak verifiers. 
      {\bf 3. Weak verifier aggregation.} We run all the weak verifiers on our entire training dataset and aggregate them into a single verifier. On a new unlabeled example, the aggregate verifier determines a single {\tt True} or {\tt False} label from the labels of the weak verifiers. 
      }
    \label{fig:pipeline}
    \vspace{-1em}
\end{figure*}


\section{Method}
\label{sec:method}

Our verification pipeline (Figure\,\ref{fig:pipeline})
takes a user's task description describing the desired layout as input
and outputs a strong verifier that evaluates whether a generated
layout matches the prompt. The pipeline includes includes four main
stages.

\vspace{0.3em}
\noindent
{\bf \em Stage 1: Dataset generation.} We sample a
layout generator to produce a dataset of layouts. These
generated layouts serve as training data for the aggregation in Stage 4
of our pipeline.  This dataset is unlabeled but is assumed to contain
some layouts which meet all the properties in the task description and
some that fail to meet them. The task of the verifier is to
distinguish the positive samples from the negative samples. 


\vspace{0.3em}
\noindent
{\bf \em Stage 2: Dev set labeling.} The user assigns ground truth
labels (positive/negative) for a small number of
the layouts (about 10) to form a {\em development set} (or ``dev
set''). These labels may also be accompanied with user notes about why each label was assigned.  The dev set is a small set of ground truth data that is used for generating weak verifiers in Stage\,3 and aggregating them in Stage\,4.
%

\vspace{0.3em}
\noindent
{\bf \em Stage 3: Weak verifier generation.} 
We use an LLM to generate a collection of
weak verification programs using our layout verification DSL (Section\,\ref{sec:wvg}). This LLM has
access to the task description as well as the labeled dev set of
positive/negative example layouts.

\vspace{0.3em}
\noindent
{\bf \em Stage 4: Weak verifier aggregation.} 
We use the information from the dev set with the votes of the LLM-generated weak verifiers on the unlabeled dataset to learn how to aggregate the weak verifiers into a single strong verifier (Section\,\ref{sec:wva}).


\subsection{Weak Verifier Generation}
\label{sec:wvg}
A weak (imperfect) verifier is a function which takes in a layout and outputs a response in \{{\tt True}, {\tt False}, {\tt Abstain}\}.
In the weak learning paradigm, the outputs of many weak
verifiers are aggregated together to form a stronger
verifier. 
The approach is based on the idea that it is often far easier for people 
to manually write multiple imperfect verification functions
than to write a single high-quality verification function that can perfectly distinguish between
positive (desired) outputs and negative (undesired) outputs.
Yet, in our setting of layout design, manually writing even imperfect verifiers requires
design expertise, time and effort.

Thus, our approach is to ask an LLM to write verification functions
using a domain specific language (DSL) that we have developed to
support layout verification.  LLMs have made significant strides in
programming tasks~\cite{livebench} but are known to lack strong
spatial understanding\,\cite{wang2024spatial, mayer2025ivispar}. Our
layout verification DSL is therefore designed to let the LLM
programmatically access accurate spatial information about the
layout. 
We first present our layout verification DSL and then discuss 
how we use the LLM to generate the collection of weak verifiers. Full details on our weak verifier generation are included in Section~3 of the supplemental materials. 

\vspace{0.1in}
\noindent
{\bf \em Layout verification DSL.}
Our layout verification DSL consists of a data model for representing
spatial layouts and a set of low-level spatial queries that can
query properties of any such layout.
Our data model treats a layout $\mathbf{L}$ as a collection of geometric elements $e_i \in \mathbf{L}$, each with {\tt id}, {\tt class}, {\tt position}, {\tt size}, {\tt orientation}, {\tt bbox} and {\tt mesh} attributes.
%
This set is extensible so that each layout domain
can include domain specific attributes (e.g. elements in 2D poster
designs might also include a {\tt font-family} attribute).
In general, the data model should include attributes that the layout generator can control and whose values the verifiers might consider when labeling positive or negative layouts. 
%
%

Our DSL also includes a variety of low-level spatial queries (e.g. {\tt intersects(), facing(), distance()}) that can help
separate positive and negative layouts. At a high-level, these queries fall into five major groups: {\em layout-level queries}, which check how individual elements relate to the bounds of the overall layout; {\em position queries}, which check relationships between the {\tt position} attributes of elements; {\em size queries}, which check relationships between the {\tt size} attributes of elements; {\em orientation queries}, which check relationships between the {\tt orientation} attributes of elements; and {\em intersection queries}, which check how elements overlap.
By default, the queries operate on the {\tt bbox} geometry of elements, but they can also be invoked on the {\tt mesh} attribute for more fine grained geometric queries. 
Using these low-level queries, we can express a variety of higher-level spatial relationships used in layout descriptions. For example, we can express whether a set of chairs ``surrounds" a table by checking whether each chair is facing towards the table, ensuring that the chairs are all within a certain distance of the table, and that none of the chairs intersect with each other. A full description of our data model and DSL is included in Tables~1-4 of the supplemental materials.

\vspace{0.3em}
\noindent
{\bf \em LLM-based verifier generation.}
Given the dev set, the DSL and the task description, we ask an LLM to write a set of weak verifier functions that address the entire task (including any user notes in the dev set). 
Each weak verifier is generated in two steps.
%
First, we ask the LLM to generate a {\em plan} as a natural language docstring coarsely describing 
the set of checks the verifier should perform.
%
Next, we query the LLM to provide an {\em implementation} of the function using our DSL according to the docstring plan. 
By separating planning and implementation we split the complex task of crafting a holistic program that captures the entire task description into simpler subtasks. Moreover, the LLM can generate different plans for each weak verifier in the collection, thereby encouraging greater diversity 
which is useful for aggregation (Section\,\ref{sec:adapt-weaver}). 
Finally, for each weak verifier we ask the LLM to add natural language {\em feedback messages} 
to each exit point that results in the verifier voting {\tt False}. The messages describe specific reasons for why a layout does not match the task description and is useful for verifier-guided layout generation (Section~\ref{sec:vglg}).
Additional details for each of these steps are provided in Section~3.1 of the supplemental materials.

\vspace{-0.1em}
\subsection{Weak Verifier Aggregation}
\label{sec:wva}
The aggregation stage of our pipeline is responsible for learning how
to weight each weak verifier's responses to produce a single, more accurate verification label. 
Intuitively the weights represent how much we should trust each weak verifier to correctly label a layout.
We consider four aggregation methods that use different approaches for
learning these weights.
%

\vspace{0.3em}
\noindent
{\bf \em Naive Majority.}\label{sec:nm} This method weights each weak verifier 
equally and labels a layout as positive if the majority of the
verifiers which do not output {\tt Abstain} vote \texttt{True}.
%
This is a conservative approach that assumes all of the verifiers are
equally accurate and therefore warrant equal trust. 
However, if there is strong variance in the accuracies of the verifiers, naive majority will split the difference between them,
and so the presence of good verifiers can be counterbalanced by inaccurate verifiers.

\vspace{0.3em}
\noindent
{\bf \em Logistic Regression.}\label{sec:lr} This method trains a regression
model to find an optimal weighting for the verifiers based on the
ground truth labels in the dev set.  Thus, it assigns high weights
to verifiers that are likely to be more accurate based on their
performance on the dev set and low weights to those that are
likely to be inaccurate. It therefore has a higher upper-bound
than Naive Majority when it finds an optimal weighting. However, with a very small dev set to learn from,
this method has a higher likelihood to assign inaccurate
weights and may thus perform sub-optimally. In general, this method is susceptible to high variance in
its overall accuracy depending on the dev set.

\vspace{0.3em}
\noindent
{\bf \em Top-1.}\label{sec:top1} This method identifies the best
weak verifier as the one that is most accurate over the dev set 
(randomly choosing between ties). It assigns all the weight
to this top verifier and zero weight to all the others.
If the collection of verifiers contains some that are highly accurate
this method can identify them, while down-weighting ones that are
especially inaccurate. However, the weighting scheme over the verifiers
is especially sharp and it is upper-bounded in performance by the
accuracy of the single top verifier. Like Logistic Regression, when the dev set is
small this method may pick a sub-optimal verifier and yield worse results.

\vspace{0.3em}
\noindent
{\bf \em Weaver.}\label{sec:weaver}
Unlike the other aggregation methods,
Weaver\,\cite{2025weaver} uses the voting patterns of the weak
verifiers on both the dev set and the {\em unlabeled} dataset 
to determine how to weight the verifiers.
%
At a high level, verifiers that vote in a consistent pattern in
relation to other verifiers tend to be upweighted, while those whose
outputs appear noisy or uninformative are downweighted.
More precisely, these weights estimate the true positive rate (TPR) and true negative rate (TNR) of each weak verifier.
%
Weaver relies on two core assumptions about the collection of weak verifiers.
First, it assumes sufficient {\em diversity} between verifiers---enough differences in the verifier voting patterns. 
Second, it assumes each verifier offers {\em predictive signal} that is better than random.
\vspace{-0.1em}
\subsection{Adapting Weaver to Spatial Layout Verification}
\label{sec:adapt-weaver}

To successfully apply Weaver in spatial layout domains 
our approach addresses three key factors.

\vspace{0.3em}
\noindent
{\bf \em Programmatic weak verifiers. }
While Weaver uses LLM-judges, we find that LLMs are not well-suited for spatial layout verification (Section~\ref{sec:wvg}).
Thus, we use Weaver to aggregate the votes from LLM-generated programmatic verifiers written with our spatial layout verification DSL. 

\vspace{0.3em}
\noindent
{\bf \em Weak verifier diversity.}
Weaver assumes diversity in the voting patterns of the weak verifiers.
To encourage such diversity, our weak verifier generation uses a two-step procedure (Section~\ref{sec:wvg}): first generating a docstring plan, and then a verifier function implementing that plan.
In this way, each LLM generated weak verifier is conditioned on a different docstring plan, thereby promoting diverse solutions.
Further, for both plan and implementation generation, we sample the LLM with high temperature.

\vspace{0.3em}
\noindent
{\bf \em Pre-filtering to increase predictive signal.}
Weaver also assumes each verifier offers predictive signal that is better than random. 
Weaver addresses this filtering out verifiers with extreme marginal voting behavior (e.g. vote uniformly \texttt{True} or \texttt{False}).
We find that with this filtering alone Weaver fails in our setting.
We develop a stronger filter by checking the performance of the weak verifiers on the dev set.
Specifically we select for verifiers with at least one of the following properties:
balanced precision and recall;
very high precision alone;
or very high negative predictive value (negative precision).
In our setting there is a strong negative class imbalance, so we aim to filter out verifiers that produce false negatives, as these are especially difficult for Weaver to incorporate. 

\vspace{-0.1em}
\subsection{Verifier-Guided Layout Generation}\label{sec:vglg}
We can leverage our strong verifier within an iterative pipeline which we call {\em verifier-guided layout generation}. The layout generator synthesizes a layout, the verifier evaluates it. If the layout fails verification, the verifier provides feedback to the generator and asks it to generate another layout. 

As noted in Section\,\ref{sec:wvg}, we designed our LLM-generated weak verifiers to 
to provide {\em feedback messages} for this application. 
After generating each weak verifier, we query the LLM to identify each exit branch in the verifier body 
resulting in the verifier voting {\tt False} and 
write a natural language string that describes the reason for that exit. This message provides context about what causes a verifier to fail. 
We incorporate such feedback messages from aggregated weak verifiers 
into a verifier-guided layout generation pipeline using two possible approaches. 

\vspace{0.3em}
\noindent
{\bf \em Binary feedback.}
This feedback consists of the most recent negative layout and the corresponding verifier result ({\tt False}). There is no explicit natural language text describing the reason for the rejection. Such pass/fail feedback enables a baseline form of verifier-guided generation that is similar to rejection sampling.


\vspace{0.3em}
\noindent
{\bf \em Detailed feedback.} 
Detailed feedback consists of the most recent negative layout and the verifier result ({\tt False}), as well as a natural language feedback describing reasons for the negative result.
While this type of 
feedback can be difficult to elicit from a verifier in many domains, our strong 
strong verifier is composed of weak verifiers which output such feedback messages. 
 
We aggregate those feedback messages into a single feedback message as follows.
For each aggregation method (Section~\ref{sec:wva}), we define the {\em reliability} weight of a weak verifier as its aggregation weight. For example, the reliability weights in Logistic Regression are determined by normalizing the logits for each verifier. In Top-1 only one weak verifier is used and therefore has a reliability weight of $1.0$. In Naive Majority all the verifiers are aggregated with equal reliability weight. In Weaver, we define the reliability weight of a feedback message as the harmonic mean of the TPR and TNR of the corresponding weak verifier~(see Section~\ref{sec:wva}). 
 
We create an aggregate feedback message by identifying which weak verifiers voted {\tt False}, annotating that feedback message with the weak verifier reliability weight, and then concatenating the feedback messages in order from highest to lowest reliability (Figure~\ref{fig:itergen}). Additional details and examples are in Section~7 of the supplemental materials. 

\subsection{Implementation Details}
The implementation for Stage 1 is domain-dependent, but we have experimented with datasets of size 100-1000 per task. 
In Stage 2, we randomly sample from the unlabeled dataset until we reach 10 examples or until the user has labeled at least one positive and one negative example, whichever is later. 
In Stage 3, we independently sample the LLM (GPT-5.4) for 50 different weak verifiers. 
%
In Stage 4, 
our filtering step for Weaver (Section~\ref{sec:adapt-weaver}) only keeps weak verifiers which have an F1-score of at least 0.5, precision of at least 0.75 or predictive negative value of at least 0.75 on the dev set.
For complete details about our implementation, see Section~3 of the supplemental materials.

\section{Results and Evaluation}
\setlength{\fboxsep}{0.8pt}
\newcommand{\phl}[1]{#1}

\begin{table}[t]
\centering
\small
\caption{Comparison against aggregating LLM judges. Our programmatic weak verifiers outperform using LLM Judges as weak verifiers by a significant margin, regardless of the aggregation method being used. 
} \label{tab:claim1}
\vspace{-\baselineskip}
\begin{tabularx}{0.48\textwidth}{@{}>{\raggedright\arraybackslash}p{2.3cm}|>{\raggedleft\arraybackslash}X>{\raggedleft\arraybackslash}X>{\raggedleft\arraybackslash}X>{\raggedleft\arraybackslash}X@{}}
\multicolumn{5}{@{}l@{}}{\textbf{3D Rooms}} \\
\noalign{\global\arrayrulewidth=1.5pt}
\hline
\noalign{\global\arrayrulewidth=0.4pt}
\multirow{2}{*}{\textbf{Verifier Type}} & \multicolumn{4}{c}{\textbf{F1-Scores ($\uparrow$)}} \\
\cline{2-5}
 & Weaver & NM & LR & Top-1 \\
\hline
LLM Vision Judges & 0.26 & 0.35 & 0.12 & 0.34 \\
LLM Judges & 0.25 & 0.49 & 0.11 & 0.43 \\
Ours & \textbf{0.84} & \textbf{0.79} & \textbf{0.77} & \textbf{0.79} \\
\hline

\noalign{\vskip 0.55em}
\multicolumn{5}{@{}l@{}}{\textbf{2D Posters}} \\
\noalign{\global\arrayrulewidth=1.5pt}
\hline
\noalign{\global\arrayrulewidth=0.4pt}
\multirow{2}{*}{\textbf{Verifier Type}} & \multicolumn{4}{c}{\textbf{F1-Scores ($\uparrow$)}} \\
\cline{2-5}
 & Weaver & NM & LR & Top-1 \\
\hline
LLM Vision Judges & 0.46 & 0.52 & 0.35 & 0.59 \\
LLM Judges & 0.21 & 0.21 & 0.21 & 0.23 \\
Ours & \textbf{0.79} & \textbf{0.62} & \textbf{0.69} & \textbf{0.74} \\
\hline

\end{tabularx}
\vspace{-1em}
\end{table}

\begin{table*}[t]
    \small
    \caption{
    F1-scores of our verifiers with four aggregation methods. The top performing verifier per-task is bolded in each column, though in the case of ties we bold Weaver. 
    Overall, our Weaver verifier performs strongest on average and is the top-performing verifier on the most number of tasks (18 out of 26 tasks). 
    } 
    \label{tab:claim2}
    \vspace{-0.5em}
\begin{tabularx}{\textwidth}{@{}>{\raggedright\arraybackslash}p{2.6cm}|*{13}{>{\centering\arraybackslash}X}|>{\centering\arraybackslash}X@{}}
\multicolumn{15}{@{}l@{}}{\textbf{3D Rooms (F1-Scores $\uparrow$)}} \\
\noalign{\global\arrayrulewidth=1.2pt}\hline\noalign{\global\arrayrulewidth=0.4pt}
\textbf{Method} & T1 & T2 & T3 & T4 & T5 & T6 & T7 & T8 & T9 & T10 & T11 & T12 & T13 & Avg \\
\hline
Naive Majority & 0.82 & \textbf{0.92} & 0.94 & 0.66 & 0.86 & 0.75 & 0.83 & 0.80 & 0.56 & 0.80 & 0.82 & \textbf{0.85} & 0.70 & 0.79 \\
Logistic Regression & 0.82 & 0.91 & \textbf{0.95} & 0.59 & 0.84 & 0.46 & 0.80 & 0.80 & 0.73 & 0.80 & 0.81 & 0.81 & 0.68 & 0.77 \\
Top-1 & 0.80 & 0.91 & \textbf{0.95} & 0.67 & 0.85 & 0.68 & 0.77 & 0.79 & \textbf{0.75} & 0.83 & 0.79 & 0.78 & 0.73 & 0.79 \\
Weaver & \textbf{0.84} & 0.91 & 0.94 & \textbf{0.74} & \textbf{0.86} & \textbf{0.76} & \textbf{0.86} & \textbf{0.82} & 0.74 & \textbf{0.90} & \textbf{0.84} & 0.82 & \textbf{0.91} & \textbf{0.84} \\
\hline
\end{tabularx}
\vskip 0.5em
\begin{tabularx}{\textwidth}{@{}>{\raggedright\arraybackslash}p{2.6cm}|*{13}{>{\centering\arraybackslash}X}|>{\centering\arraybackslash}X@{}}
\multicolumn{15}{@{}l@{}}{\textbf{2D Posters (F1-Scores $\uparrow$)}} \\
\noalign{\global\arrayrulewidth=1.2pt}\hline\noalign{\global\arrayrulewidth=0.4pt}
\textbf{Method} & T14 & T15 & T16 & T17 & T18 & T19 & T20 & T21 & T22 & T23 & T24 & T25 & T26 & Avg \\
\hline
Naive Majority & 0.94 & 0.87 & 0.91 & 0.06 & 0.92 & 0.91 & 0.20 & \textbf{0.83} & 0.29 & \textbf{0.85} & 0.22 & 0.72 & 0.29 & 0.62 \\
Logistic Regression & 0.94 & 0.84 & 0.91 & 0.93 & 0.59 & 0.90 & 0.95 & 0.77 & 0.21 & \textbf{0.85} & 0.13 & 0.66 & 0.28 & 0.69 \\
Top-1 & 0.93 & 0.81 & 0.89 & \textbf{0.95} & 0.90 & 0.87 & \textbf{0.96} & 0.67 & 0.38 & 0.81 & 0.30 & 0.69 & 0.46 & 0.74 \\
Weaver & \textbf{0.94} & \textbf{0.90} & \textbf{0.91} & 0.93 & \textbf{0.94} & \textbf{0.91} & 0.94 & 0.77 & \textbf{0.38} & 0.84 & \textbf{0.46} & \textbf{0.74} & \textbf{0.59} & \textbf{0.79} \\
\hline
\end{tabularx}
\end{table*}

Our verification pipeline includes two main components: weak
verifier generation and weak verifier aggregation.
Our evaluation considers two main questions about these components.
First, how does our DSL-based weak verifier generation compare to
using off-the-shelf LLM as black-box judges
(Section~\ref{subsec:llm-judge-comp})?  Second, when is a particular
aggregation method most effective and why
(Section~\ref{subsec:analysis-wva})? Finally, we compare verifier-guided layout generators using our verifiers against the standard approach of using an LLM judge as a critic (Section~\ref{subsec:vglg}).

\subsection{Experiment Design}\label{sec:exp-design}

\vspace{0.3em}
\noindent
{\bf \em Domains.}
%
%
We apply our verification pipeline to two layout domains, 3D rooms and 2D posters. Our 3D rooms focus on enclosed and furnished spaces and our 2D posters focus on static graphic designs with typography. 
For both layout domains, we find that existing generation systems are unable to handle complex spatial arrangements when specified (e.g., shapes, alignment, symmetry, etc.), so we implement two LLM-based generators.
The 3D Rooms generator outputs layouts directly in our data model, readily populated with 3D assets and textures from BlenderKit~\cite{BlenderKit2018}. 
The 2D Posters generator outputs layouts in HTML/CSS, which we then parse into our data model. 
For each dataset in Stage 1 of the verification pipeline (Section~\ref{sec:wvg}) we sample 100 examples using these generators. In general, we sample a dataset for each task, but we also experiment with amortizing dataset generation by sharing datasets across tasks (e.g., three of our 3D Rooms tasks use the same dataset). Section~6.3 of the supplemental materials also includes multiple 3D layout tasks which all use the pre-existing 3D-FRONT dataset~\cite{3dfront}.
Additional details on both generators as well as layouts from existing generators are included in Section~4 of the supplemental materials. 

\vspace{0.3em}
\noindent
{\bf \em Layout verification tasks.}  We define a layout verification
task as {\em user-centric}; given some domain and decision criteria,
the user can sort layout instances into positive or negative buckets.
We ask users to convert this decision criteria into text, and call
this a \textit{task description}. The job of our weak verification
pipeline is to learn a verifier per task from limited labeled data.
We have developed a representative set of such
layout tasks (13 tasks for 3D Rooms, 13 tasks for 2D Posters) that range
in complexity with respect to layout requirements. Figures\,\ref{fig:scene3d-task8}-\ref{fig:poster2d-task9} show examples from our
poster and room layout domains along with possible task descriptions a
user might specify.
Examples include:
\begin{itemize}[leftmargin=1em]
\item {\bf T1: Symmetric Nightstands (3D Rooms).} ``A bedroom with symmetric identical nightstands." (supplemental materials, Figure~7)
\item {\bf T10: College Dorm (3D Rooms).} ``A dorm room for two college students. One of them is very messy (leaving books and clothes on the floor) and the other is very neat." (Figure~\ref{fig:scene3d-task10})
\item {\bf T16: Sparrows (2D Posters).} This task describes a book cover with specific requirements on the title text and sparrow images scattered throughout. Figure~\ref{fig:teaser} (bottom) gives the full task description. 
\end{itemize}
A comprehensive list of all 26 tasks and their descriptions is included in the supplemental materials (Section~5). 

Note that in many cases such descriptions are under-specified, as
people often have implicit notions of positive/negative layouts that
can be difficult to fully describe in text. 
For example,
{\bf T1: Symmetric Nightstands (3D Rooms)} does not explicitly specify that the nightstands should be {\em placed at the head of the bed},
but that is commonly assumed to be the right placement. Ultimately, only the user can determine whether a generated layout fully matches their implicit notions for an acceptable layout (i.e., provide ground truth labeling). In our pipeline, users may encounter examples which violate these implicit notions in the dev set, and make those notions explicit via the dev set notes. 
%

\vspace{0.3em}
\noindent
{\bf \em Evaluating verification performance.}
To evaluate verifier performance we require a test set of generated
layouts for each of our task prompts with corresponding ground truth
positive/negative labels. 
To this end we ask a user to manually provide the 100 examples in each task dataset with positive/negative ground truth labels. In this process, we found that both our layout generators are largely biased towards producing negative samples (on 23 out of 26 tasks we observe a positive label rate of less than 50\%).
Because of this ground truth imbalance we primarily evaluate performance of our weak verification pipeline using F1-scores rather than accuracy. For each task, we report the average F1-score across three different randomly sampled dev sets.
We also report accuracy, precision and recall in the supplemental materials (Section~6). 

\vspace{-0.1em}
\subsection{Comparison with LLM Judges}\label{subsec:llm-judge-comp}
\vspace{0.3em}
\noindent
{\bf \em Our DSL-based verifiers outperform LLM judges by up to 7X.}
The status quo approach for producing weak verifiers is to directly
ask multiple LLM judges 
to check that a generated output (a generated layout in our
setting) matches the task prompt. Here we compare verification performance when 
aggregating our programmatic weak verifiers against aggregating LLM judges.
Current frontier models have both text and vision capabilities, so we consider two approaches to using LLM judges: one where each LLM judge receives our data model representation of the layout along with documentation of the data model, and one where each LLM receives a top-down rendering of the layout instead (denoted LLM Vision Judges). 
In our experiments, we use an ensemble of frontier vision-capable LLMs: GPT-5.4, LLaVA, Gemma4 and Qwen3.

%
%
We report the results of this experiment in Table~\ref{tab:claim1}.
We find that across all four aggregation methods (Weaver, Naive Majority, Logistic Regression and Top-1) our programmatic weak verifiers
strongly outperform LLM judges, by at least 1.2X and up to 7X. 
%
These results show that directly using aggregated LLM judges is insufficient for verifying spatial layout, even when the LLMs have vision capabilities.
For example, in 
{\bf T9: Dining Room (3D Rooms)} the LLM judges often incorrectly accept rooms where a seating arrangement for 8 people explicitly violates the requirement that all chairs face the long side of the table (supplemental materials, Figure~15).  
In 
{\bf T15: Art Show (2D Posters)}, the LLM judges incorrectly accept posters where explicit text-alignment requirements are not satisfied (supplemental materials, Figure~21). 
In both instances, our strong verifier is able to correctly label these examples.
These results suggest that LLM-generated weak verification programs are better at analyzing spatial layout than directly using LLMs.
%
\add{

}



\vspace{-0.1em}
\subsection{Analysis of Aggregation Methods}
\label{subsec:analysis-wva}
We compare the performance of each of the four aggregation methods discussed in Section\,\ref{sec:wva} across the full set of tasks 26 in Table\,\ref{tab:claim2}. 


\vspace{0.4em}
\noindent

\noindent
{\bf \em Weaver is the top-performing aggregation method.}
Across tasks, aggregating verifiers with Weaver is 
consistently the strongest, achieving the best average F1-score overall and the best F1-score on 18 out of 26 tasks (see Table~\ref{tab:claim2}). On 4 of the remaining tasks, Weaver performs second-best and is within 0.01 of the best F1-score. 
This trend is consistent with the intuition that our programmatic weak verifiers tend to overlap in what they check, producing redundant but noisy votes with agreement structure that Weaver can exploit to downweight unreliable voters and amplify more consistent verifiers.

\vspace{0.4em}
\noindent
{\bf\em Logistic Regression and Top-1 are unreliable on small dev set.}
%
Both Logistic Regression and Top-1 greatly vary in their relative performance among the four aggregation methods. 
Logistic Regression is the best aggregation  method on 4 tasks but also the worst aggregator on 9 tasks. Similarly, Top-1 is the best aggregation method on 5 tasks and the worst on 8 tasks. 
Since the dev set is the only source of information that these methods use, this result suggests that the signal from a small dev set of 10 examples is often too noisy or misaligned for Logistic Regression. We also find that increasing the dev set size up to 50 examples does not necessarily improve the performance of Logistic Regression or Top-1 relative to Weaver, and even with larger dev sets both Logistic Regression and Top-1 still exhibit high variance in performance (supplemental materials, Figure~33). 


\vspace{0.4em}
\noindent
{\bf\em Naive Majority performs poorly with low recall weak verifiers.} 
Naive Majority does not require any labeled data and is easy to understand, hence it is a commonly used method for aggregating votes from independent voters. When many of the voters have comparably strong predictive signal, Naive Majority performs well.
Indeed, we find that it is more consistent than Logistic Regression and Top-1 (best on 8 out of 26 tasks and the second-best on 9 other tasks). 
However, we find that it underperforms when many weak verifiers have low recall. This is relatively common in our tasks. As we have seen, the base layout generators are negatively-skewed and thus the dev sets have very few positive examples.
Since the weak verifiers are generated using these biased dev sets, they tend to reject true positives and thus have low recall.
%
Because Naive Majority weights all weak verifiers equally, the low recall verifiers hurt performance.
In four layout tasks where the average recall of the weak verifiers is less than 0.33, Naive Majority performs the worst or second-worst, whereas the other aggregation methods find more optimal ways to weight the weak verifier (supplemental materials, Figure~34).

\subsection{Verifier-Guided Layout Generation}
\label{subsec:vglg}

We evaluate three different verifier-guided layout generators (Section~\ref{sec:vglg}) on our 3D Rooms layout tasks. 
The first uses a single vision-enabled LLM judge (GPT-5.4) to verify that the generated layout matches the task description and to provide natural language feedback on the failing layouts (denoted LLM+Vision). The second uses our strong Weaver-aggregated verifier with {\em Binary} feedback. The third uses our strong Weaver-aggregated verifier with {\em Detailed} feedback (Figure~\ref{fig:itergen}). Each 
verifier-guided generator is allowed a maximum of 10 iterations per task.
As a baseline for comparison we include the non-iterative base 3D layout generator operating without verifier feedback (Section~\ref{sec:exp-design}).

Table~\ref{tab:iterative-results} reports results for 5 different 3D Room layout tasks. 
We use each generator to sample 25 layouts per task. We then ask a human evaluator to 
determine whether each layout sample matches the corresponding task description. Table~\ref{tab:iterative-results} (top)
gives the percentage of samples that pass this human evaluation.
Overall we find that our Binary and Detailed feedback improve the human pass rate by 42.2\% and 66.2\% respectively on average over the base generator (no verifier).
In contrast, using the LLM+Vision feedback performs {\em worse} on average than using our Binary and Detailed feedback. For task {\bf T4} it even underperforms the base generator, because many generated layouts pass the LLM+Vision verifier but not the human evaluator.
%

Table~\ref{tab:iterative-results} (bottom) gives the average number of iterations each generator required to produce a single layout.
We find that Detailed feedback
requires an average of just over 2 iterations to produce a layout that passes verification. In contrast Binary feedback requires over 6 iterations.
%
%
While the LLM+Vision feedback requires about 3 iterations to generate a layout that passes its verifier, the resulting layouts are significantly lower in quality than those produced by using Binary or Detailed feedback from our strong verifier. This is evident in the lower human pass rates in the top part of the Table.
%
%
Together these results suggest 
that the detailed natural language feedback of our Detailed feedback approach provides useful guidance to the base generator on how to repair layouts that do not pass our strong verification. 

\begin{table}[t]
    \centering
    \small
    \caption{
    We compare three different verifier-guided 3D Rooms layout generators (LLM+Vision, Ours (Binary), Ours (Detailed)). We include a base 3D Rooms layout generator with no verifier as a baseline. 
    We report the human evaluator pass rates over 25 samples per task (top), as well as the average number of iterations required to pass each generator's verifier (bottom).
    On average, using Detailed feedback produces the highest human pass rate while requiring the fewest iterations.
    } \label{tab:iterative-results}
    \vspace{-\baselineskip}
    \begin{minipage}[t]{0.48\textwidth}
        \vspace{0.25em}
        \begin{tabularx}{\linewidth}{@{}>{\raggedright\arraybackslash}p{2.8cm}|>{\raggedleft\arraybackslash}X>{\raggedleft\arraybackslash}X>{\raggedleft\arraybackslash}X>{\raggedleft\arraybackslash}X>{\raggedleft\arraybackslash}X|>{\raggedleft\arraybackslash}X@{}}
        \multicolumn{7}{@{}l@{}}{\textbf{Human Evaluator Pass Rate ($\uparrow$)}} \\
        \noalign{\global\arrayrulewidth=1.5pt}
        \hline
        \noalign{\global\arrayrulewidth=0.4pt}
        {\bf Feedback Type} & T3 & T4 & T9 & T10 & T11 & Avg. \\
        \cline{1-7}
        Base (No verifier) & 18\% & 14\% & 30\% & 21\% & 34\% & 23.4\% \\
        \hline
        LLM+Vision & 28\% & 12\% & 60\% & 56\% & 48\% & 40.8\% \\
        Ours (Binary) & 40\% & 64\% & 84\% & 40\% & \textbf{100}\% & 65.6\% \\
        Ours (Detailed) & \textbf{96\% }& \textbf{100\%} & \textbf{84\% }& \textbf{84\%} & 84\% & \textbf{89.6\%}  \\
        \hline
        \end{tabularx}
    \end{minipage}\hfill
    \vspace{0.05em}
    \begin{minipage}[t]{0.48\textwidth}
        \vspace{0.25em}

        \begin{tabularx}{\linewidth}{@{}>{\raggedright\arraybackslash}p{2.8cm}|>{\raggedleft\arraybackslash}X>{\raggedleft\arraybackslash}X>{\raggedleft\arraybackslash}X>{\raggedleft\arraybackslash}X>{\raggedleft\arraybackslash}X|>{\raggedleft\arraybackslash}X@{}}
        \multicolumn{7}{@{}l@{}}{\textbf{Number of Iterations ($\downarrow$)}} \\
        \noalign{\global\arrayrulewidth=1.5pt}
        \hline
        \noalign{\global\arrayrulewidth=0.4pt}
        {\bf Feedback Type} & T3 & T4 & T9 & T10 & T11 & Avg. \\
        \cline{1-7}
        LLM+Vision & \textbf{2.12} & 6.88 & \textbf{1.72} & 2.20 & \textbf{1.44} & 2.87 \\
        Ours (Binary) & 6.28 & 5.52 & 2.28 & 7.76 & 10.0 & 6.37 \\
        Ours (Detailed) & 2.16 & \textbf{1.88} & 1.92 & \textbf{1.44} & 3.64 & \textbf{2.21} \\
        \hline
        \end{tabularx}
    \end{minipage}
    \vspace{-1em}
\end{table}

\subsection{Limitations}\label{subsec:limitations}
A limitation of our verification pipeline is that we require a dataset for each task, specifically when aggregating with Weaver. If the generator is expensive to sample from, this may limit its applicability. 
However, we can amortize the cost of dataset generation by reusing a single dataset across multiple similar layout tasks.
For example, in our experiments we use the same dataset for three 3D Rooms tasks ({\bf T1: Symmetric Nightstands}, {\bf T12: Reading Nook} and {\bf T13: Study Area}) as all three tasks are about bedrooms.
In supplemental materials, we report on a second experiment in which we use the 3D-FRONT dataset~\cite{3dfront} for five different 3D Room layout tasks.
The key requirement for reusing a dataset is that it contains both positive and negative layouts for every task description.
We also note that once a strong verifier has been constructed, it can be repeatedly used for verifier-guided layout generation---for example, to generate varying bedroom layouts in a video game environment.
\bibliographystyle{ACM-Reference-Format}
\bibliography{main}

\appendix
\begin{figure*}[t]
    \centering
    \includegraphics[width=\linewidth]{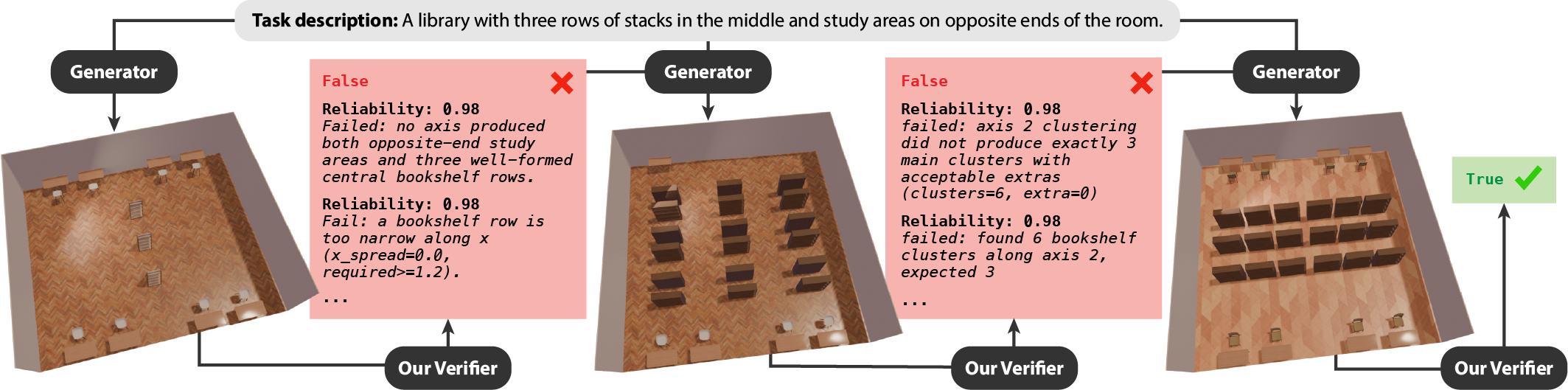}
    \caption{{\bf Verifier-guided layout generation with Detailed feedback.} A single layout example is generated by iteratively sampling a layout, verifying the output with our strong Weaver verifier, and re-generating using the verifier feedback (red boxes) if the verifier response is {\tt False}. We repeat this until the layout passes or until we reach a maximum number of iterations. 
    }
    \label{fig:itergen}
\end{figure*}

\begin{figure*}
    \centering
    \includegraphics[width=\linewidth]{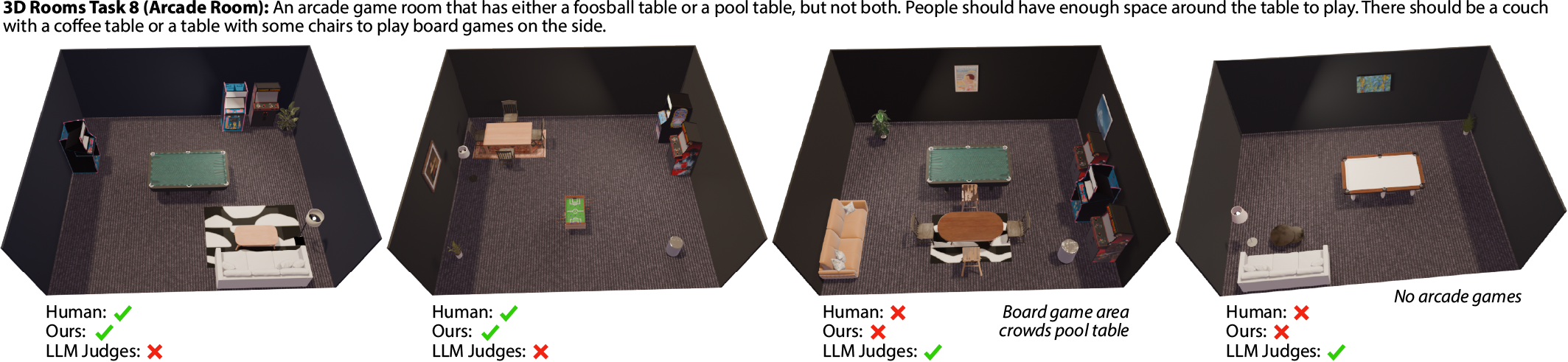}
    \caption{The LLM judges incorrectly reject the two positive layouts on the left and accept the two negative layouts on the right. In the first negative layout, the dining table area crowd the pool table and do not leave enough space to play. In the second negative layout, there is no indication that the room is an arcade game room.}
    \label{fig:scene3d-task8}
\end{figure*}

\begin{figure*}
    \centering
    \includegraphics[width=\linewidth]{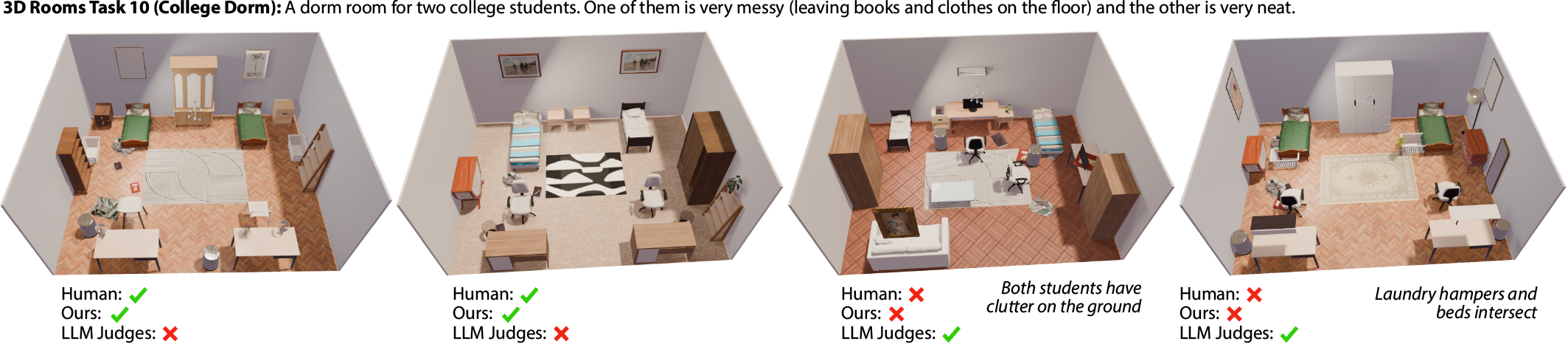}
    \caption{The LLM judges incorrectly reject the two positive layouts on the left and accept the two negative layouts on the right. In both negative layouts there are many major object intersections, such as between the wardrobes and chairs or the desk and nightstands. In the first negative layout, the paintings are also oriented perpendicularly to the walls, which is not physically plausible. In the second negative layout, it is not entirely clear that one student is very neat, as the clutter on the floor is close to both student desks. These requirements are expressed through the task description and the user's dev set notes (see supplemental).}
    \label{fig:scene3d-task10}
\end{figure*}

\begin{figure*}
    \centering
    \includegraphics[width=\linewidth]{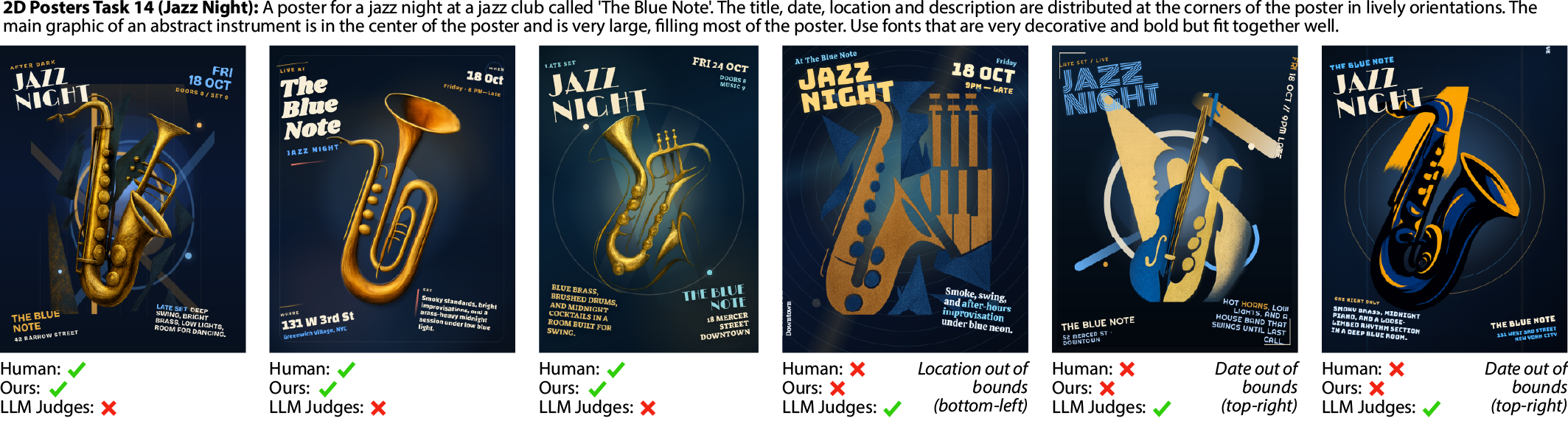}
    \caption{The LLM judges incorrectly reject the two positive layouts on the left and accept the two negative layouts on the right. In the first negative layout, the location is out of bounds. In the second negative layout, the date is both out of bounds and hard to read due to extreme rotation.}
    \label{fig:poster2d-task1}
\end{figure*}

\begin{figure*}
    \centering
    \includegraphics[width=\linewidth]{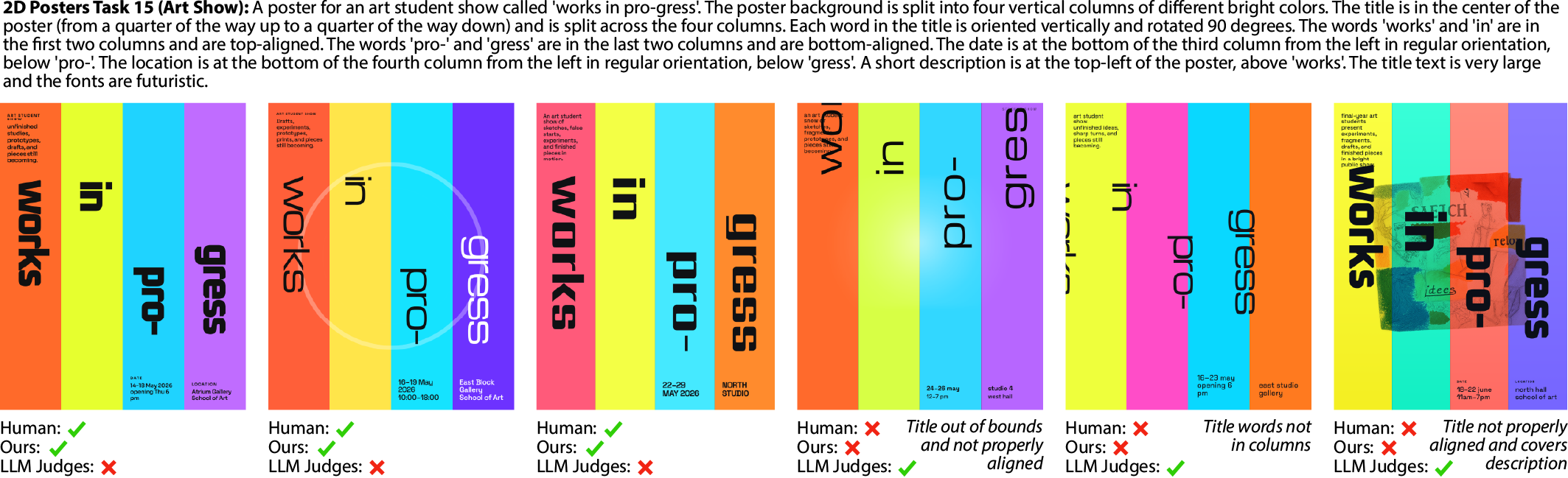}
    \caption{The LLM judge incorrectly reject the two positive layouts on the left and accept the two negative layouts on the right. In the first negative layout the title words are not properly top-aligned and bottom-aligned. In the second negative layout, the title words are not contained in the four columns.}
    \label{fig:poster2d-task2}
\end{figure*}

\begin{figure*}
    \centering
    \includegraphics[width=\linewidth]{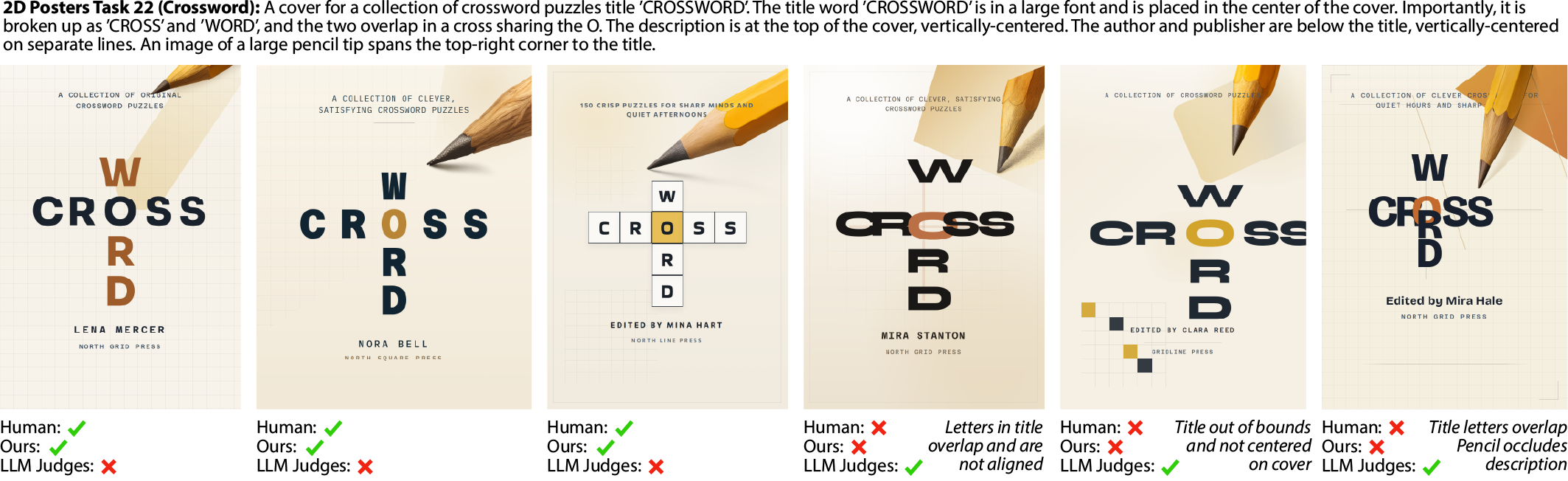}
    \caption{The LLM judges incorrectly reject the two positive layouts on the left and accept the two negative layouts on the right. In the first negative layout, the letters in 'CROSSWORD' are off-center and spaced inconsistently. In the second negative layout, the title is not centered on the cover and extends out of bounds.}
    \label{fig:poster2d-task9}
\end{figure*}

\end{document}


\title{Supplemental Materials: Aggregating LLM-Based Weak Verifiers for Spatial Layout Generation}

\author{Sharon Zhang}
\email{szhang25@stanford.edu}
\orcid{0000-0002-6738-8906}
\affiliation{%
  \institution{Stanford University}
  \city{Stanford}
  \state{California}
  \country{USA}
}

\author{R. Kenny Jones}
\email{rukjones4@gmail.com}
\orcid{0009-0005-1169-0507}
\affiliation{%
  \institution{Stanford University}
  \city{Stanford}
  \state{California}
  \country{USA}
}

\author{Jiajun Wu}
\email{jiajunwu@cs.stanford.edu}
\orcid{0000-0002-4176-343X}
\affiliation{%
  \institution{Stanford University}
  \city{Stanford}
  \state{California}
  \country{USA}
}

\author{Maneesh Agrawala}
\email{maneesh@cs.stanford.edu}
\orcid{0000-0002-8996-7327}
\affiliation{%
  \institution{Stanford University}
  \city{Stanford}
  \state{California}
  \country{USA}
}
\affiliation{%
  \institution{Roblox}
  \city{San Mateo}
  \state{California}
  \country{USA}
}

\renewcommand{\shortauthors}{Zhang et al.}

\maketitle

\section{Overview}

This document provides additional details about our method and evaluations. The sections are organized as follows:
\begin{itemize}
    \item Section\,~\ref{sec:dsl} provides more details about our data model and layout verification DSL, including a descriptions of all the queries in our DSL and a full list of attributes for {\tt elements} in our two domains;
    \item Section\,~\ref{sec:method} provides additional details about our weak verifier generation (Section\,~\ref{subsec:wvg}) and weak verifier aggregation (Section\,~\ref{subsec:wva});
    \item Section\,~\ref{sec:generators} explains the details our generators for 3D room layouts and 2D posters, as well as our setups for Holodeck~\cite{Yang_2024_CVPR} and FlairGPT~\cite{littlefair2025flairgpt}; 
    \item Section\,~\ref{sec:tasks} provides descriptions of all 26 layout verification tasks;
    \item Section\,~\ref{sec:add-results} reports additional layout verification results and analysis;
    \item Section~\ref{sec:vglg} includes additional details and results on verifier-guided layout generation.
\end{itemize}
The remaining sections list code examples and prompts:
\begin{itemize}
    \item Section\,~\ref{sec:code} contains code listings referenced throughout the supplemental materials;
    \item and Section\,~\ref{sec:prompts} includes all the system prompts we use.
\end{itemize}

\section{Layout Verification DSL}\label{sec:dsl}
\begin{table*}[th]
    \centering
    \small
    \captionsetup{justification=raggedright,singlelinecheck=false}
    \caption{The basic data model for an {\tt Element}.
    }\label{tab:data-model}
    \vspace{-\baselineskip}
    \begin{tabularx}{\textwidth}{@{}>{\raggedright\arraybackslash}p{1.5cm}>{\raggedright\arraybackslash}p{2.0cm}>{\raggedright\arraybackslash}X}
    \noalign{\global\arrayrulewidth=1.0pt}
    \hline
    \noalign{\global\arrayrulewidth=0.4pt}
    {Name} & {Syntax} & {Description} \\
    \hline
    \multicolumn{3}{@{}l}{\textit{Data model attributes}} \\
    \hline
    id & $e$.\texttt{id} & A unique identifier for $e$. \\
    class & $e$.\texttt{class} & The class of $e$. \\
    position & $e$.\texttt{position} & The $(x, y, z)$ position of $e$. In 2D layouts, $z$ indicates rendering depth order. \\
    size & $e$.\texttt{size} & The width, height, length of $e$. In 2D layouts, length is ommitted. \\
    orientation & $e$.\texttt{orientation} & In 3D, a set of {\tt up}, {\tt forward} and {\tt right} direction vectors. In 2D, the rotation angle of the element. \\
    bbox & $e$.\texttt{bbox} & The oriented bounding box of $e$. \\
    mesh & $e$.\texttt{mesh} & The mesh geometry of $e$. \\
    \end{tabularx}
    \begin{tabularx}{\textwidth}{@{}>{\raggedright\arraybackslash}p{2.7cm}>{\raggedright\arraybackslash}p{3.2cm}>{\raggedright\arraybackslash}X}
    \noalign{\global\arrayrulewidth=1.0pt}
    \hline
    \end{tabularx}
\end{table*}

\begin{table*}[th]
    \centering
    \small
    \captionsetup{justification=raggedright,singlelinecheck=false}
    \caption{Attributes in our 3D scene layout implementation of the \texttt{Element}  data model.}\label{tab:3d-data-model}
    \vspace{-\baselineskip}
    \begin{tabularx}{\textwidth}{@{}>{\raggedright\arraybackslash}p{2.0cm}>{\raggedright\arraybackslash}X}
    \noalign{\global\arrayrulewidth=1.5pt}
    \hline
    \noalign{\global\arrayrulewidth=0.4pt}
    \multicolumn{2}{@{}l}{\textbf{3D Scene Layouts}} \\
    \hline
    \textit{Attribute} & \textit{Description} \\
    \hline
    \texttt{id} & The unique identifier of the element. \\
    \texttt{class} & The class category of the element. \\
    \texttt{position} & The (x, y, z) position of the element center. \\
    \texttt{size} & The (width, height, depth) of the element. \\
    \texttt{orientation} & We express orientation as three $(x, y, z)$ vectors: \texttt{right}, \texttt{up}, and \texttt{forward}. \\
    \texttt{bbox} & The bounding box of an element. \\
    \texttt{mesh} & The underlying mesh geometry of an element. \\
    \hline
    \end{tabularx}
\end{table*}

\begin{table*}[th]
    \centering
    \small
    \captionsetup{justification=raggedright,singlelinecheck=false}
    \caption{Attributes in our 2D poster layout implementation of the \texttt{Element} data model.}\label{tab:2d-data-model}
    \vspace{-\baselineskip}
    \begin{tabularx}{\textwidth}{@{}>{\raggedright\arraybackslash}p{2.0cm}>{\raggedright\arraybackslash}X}
    \noalign{\global\arrayrulewidth=1.5pt}
    \hline
    \noalign{\global\arrayrulewidth=0.4pt}
    \multicolumn{2}{@{}l}{\textbf{2D Poster Layouts}} \\
    \hline
    \textit{Attribute} & \textit{Description} \\
    \hline
    \texttt{id} & The unique identifier of the element. \\
    \texttt{class} & The category of the element. In our posters we specify five different categories: \{\texttt{title}, \texttt{description}, \texttt{date}, \texttt{location} and \texttt{hero\_graphic}\}. \\
    \texttt{position} & The (x, y) position of the element center. \\
    \texttt{size} & The (width, height) of the element. \\
    \texttt{orientation} & The angle, in degrees, of the orientation of the element. \\
    \texttt{z-index} & The depth ordering of the element. \\
    \texttt{text} & The text string within the text box, may be empty. \\
    \texttt{font\_family} & The font family of the text. \\
    \texttt{font\_weight} & The font weight of the text. \\
    \texttt{justification} & The justification of the text. \\
    \texttt{font\_size} & The font size of the text. \\
    \texttt{text\_color} & The color of the text. \\
    \texttt{image\_path} & The path to the image file, may be empty. \\
    \hline
    \end{tabularx}
\end{table*}

\begin{table*}[th]
    \centering
    \small
    \captionsetup{justification=raggedright,singlelinecheck=false}
    \caption{Our data model represents layouts as collections of elements. Our layout verification DSL includes layout-level and element-level queries about the positions, orientations and intersection properties of elements.
    }\label{tab:api}
    \vspace{-\baselineskip}
    \begin{tabularx}{\textwidth}{@{}>{\raggedright\arraybackslash}p{2.7cm}>{\raggedright\arraybackslash}p{2.4cm}>{\raggedright\arraybackslash}X}
    \noalign{\global\arrayrulewidth=1.0pt}
    \hline
    \noalign{\global\arrayrulewidth=0.4pt}
    {\bf Name} & {\bf Syntax} & {\bf Description} \\
    \hline
    \multicolumn{3}{@{}l}{\textit{Layout-level queries}} \\
    \hline
    \texttt{exists} & $\mathbf{L}$, cls, n & True if there exists at least $n$ instances of elements of a class \texttt{cls} in the layout $\mathbf{L}$. \\
    \texttt{in\_bounds} & $\mathbf{L}$, $e$, $\epsilon$ & True if $e$ is contained entirely within $\epsilon$ of the bounds of layout $\mathbf{L}$. \\
    \hline
    \multicolumn{3}{@{}l}{\textit{Position queries}} \\
    \hline
    \texttt{above} & $e_i$, $e_j$ & True if any part of $e_i$ is above $e_j$ in the global $y$-axis. \\
    \texttt{entirely\_above} & $e_i$, $e_j$ & True if all of $e_i$ is above $e_j$ in the global $y$-axis. \\
    \texttt{below} & $e_i$, $e_j$ & In 3D, true if any part of $e_i$ is below $e_j$ in the global $y$-axis. \\
    \texttt{entirely\_below} & $e_i$, $e_j$ & In 3D, true if all of $e_i$ is below $e_j$ in the global $y$-axis. \\
    \texttt{in\_front} & $e_i$, $e_j$, \texttt{orig}, \texttt{dir} & In 3D, true if any part of $e_i$ is greater than $e_j$ along the $z$-axis in the reference frame of a viewer standing at \texttt{orig} and looking in the direction \texttt{dir}. In 2D, simply $e_i.z > e_j.z$. \\
    \texttt{entirely\_in\_front} & $e_i$, $e_j$, \texttt{orig}, \texttt{dir} & In 3D, true if all of $e_i$ is greater than $e_j$ along the $z$-axis in the reference frame of a viewer standing at \texttt{orig} and looking in the direction \texttt{dir}. In 2D, $e_i.z > e_j.z$ and axis-aligned overlap in $xy$. \\
    \texttt{behind} & $e_i$, $e_j$, \texttt{orig}, \texttt{dir} & In 3D, true if any part of $e_i$ is less than $e_j$ along the $z$-axis in the reference frame of a viewer standing at \texttt{orig} and looking in the direction \texttt{dir}. In 2D, simply $e_i.z < e_j.z$. \\
    \texttt{entirely\_behind} & $e_i$, $e_j$, \texttt{orig}, \texttt{dir} & In 3D, true if all of $e_i$ is less than $e_j$ along the $z$-axis in the reference frame of a viewer standing at \texttt{orig} and looking in the direction \texttt{dir}. In 2D, $e_i.z < e_j.z$ and axis-aligned overlap in $xy$. \\
    \texttt{left\_of} & $e_i$, $e_j$, \texttt{ref} & In 3D, true if any part of $e_i$ is less than $e_j$ along the $x$-axis in the reference frame \texttt{ref}. In 2D, we only use the global reference frame. \\
    \texttt{right\_of} & $e_i$, $e_j$, \texttt{ref} & In 3D, true if any part of $e_i$ is greater than $e_j$ along the $x$-axis in the reference frame \texttt{ref}. In 2D, we only use the global reference frame. \\
    \texttt{between} & $e_i$, $e_j$, $e_k$ & True if $e_i$ intersects the convex hull of the \texttt{bbox} of $e_j$ and $e_k$. \\
    \texttt{distance} & $e_i$, $e_j$ & Returns the shortest distance between any two points on $e_i.{\tt bbox}$ and $e_j.{\tt bbox}$. \\
    \hline
    \multicolumn{3}{@{}l}{\textit{Size queries}} \\
    \hline
    \texttt{larger\_than} & $e_i$, $e_j$ & Checks if $e_i$ has greater volume (3D) or area (2D) than $e_j$. \\
    \texttt{smaller\_than} & $e_i$, $e_j$ & Checks if $e_i$ has smaller volume (3D) or area (2D) than $e_j$. \\
    \hline
    \multicolumn{3}{@{}l}{\textit{Orientation queries (3D only)}} \\
    \hline
    \texttt{facing\_towards} & $e_i$, $e_j$ & True if extruding the forward face of $e_i$ intersects $e_j$. \\
    \texttt{facing\_away\_from} & $e_i$, $e_j$ & True if not \texttt{facing\_towards}($e_i$, $e_j$). \\
    \hline
    \multicolumn{3}{@{}l}{\textit{Bounding box queries}} \\
    \hline
    \texttt{contains} & $e_i$, $e_j$ & Checks if $e_i$ contains $e_j$. \\
    \texttt{inside} & $e_i$, $e_j$ & Checks if $e_i$ is entirely inside $e_j$. \\
    \texttt{iou} & $e_i$, $e_j$ & Returns the intersection over union of the $e_i$.{\tt bbox} and $e_j$.{\tt bbox}. \\
    \texttt{coverage} & $e_i$, $e_j$ & Returns the percentage of $e_j$.{\tt bbox} overlapping with $e_i$.{\tt bbox}. \\
    \hline
    \end{tabularx}
\end{table*}
Table~\ref{tab:data-model} lists the basic attributes in our data model for an {\tt Element}. For each of the 3D rooms and 2D poster designs, we extend this list to include relevant domain-specific attributes that are useful for writing verifiers.
Tables~\ref{tab:3d-data-model} shows the full data model for 3D scenes and and Table~\ref{tab:2d-data-model} shows the full data model 2D poster designs.
Table~\ref{tab:api} contains a comprehensive list of queries and descriptions in our layout verification DSL.

\section{Additional Method Details}\label{sec:method}
This section provides additional details about how we generate weak verifiers (Section\,\ref{subsec:wvg}), how we aggregate them (Section\,\ref{subsec:wva}), and how we use the LLM judges (Section\,\ref{subsec:llm-judge}).

\subsection{Weak Verifier Generation}\label{subsec:wvg}
The LLM-based weak verifier generation is split into three stages: documentation, implementation and verification message addition. The documentation stage prompts the LLM to generate Python function signatures and docstrings for each weak verifier. The implementation stage then prompts the LLM to implement the weak verifier based on the docstring. This two-step process splits the weak verifier generation process into planning and execution subtasks, but also encourages diversity in the weak verifiers.

\subsubsection{Documentation}\label{sec:wva-doc}
In the documentation stage, we provide the LLM with either a holistic or property-specific system prompt, which includes a description of the domain-specific data model. The user prompt contains the task description and the dev set examples expressed in our data model. Each dev set example is annotated with the user-assigned positive/negative label, and any user notes.
We query the LLM 50 times with the same user prompt to generate 50 different function signatures and docstrings. 
The system prompt for this step is in Section~\ref{sec:prompts}, Listing~\ref{lst:3d-scene-holistic-docstring-system-prompt}.

\subsubsection{Implementation}
In this stage, we provide the LLM with the system prompt in Section~\ref{sec:prompts}, Listing~\ref{lst:3d-scene-holistic-implementation-system-prompt} (3D Rooms) and Section~\ref{sec:prompts}, Listing~\ref{lst:2d-poster-holistic-implementation-system-prompt} (2D Posters) that includes our layout verification DSL and the element categories. We give the docstring generated in the documentation stage and the labeled dev set examples in the user prompt. An example (with dev set examples elided for brevity):

\begin{lstlisting}[language={}]
Implement the verifier function in Python from the following stub (signature, docstring, and placeholder body).

def verifier_12(scene: Scene3DLayout) -> int:
    '''
    Holistically verify that the scene depicts a cozy bedroom with a bed and a
    matched pair of nightstands that are identical and symmetrically arranged.

    Expected checks for an implementation:
    - Confirm the scene is a bedroom layout, anchored by a bed.
    - Confirm there are two nightstands associated with the bed.
    - Check the two nightstands are identical or near-identical in class,
      dimensions, and orientation.
    - Check the nightstands are placed symmetrically about the bed's centerline,
      at comparable lateral offsets and similar depth/height.
    - Check the nightstands are appropriately placed beside the bed near the
      headboard wall rather than arbitrarily elsewhere in the room.
    - Reject scenes with clear spatial invalidities that break the intended
      layout, such as severe overlaps/intersections or major out-of-bounds
      placement.

    Args:
        scene: Scene graph / layout.

    Returns:
        1 if the scene satisfies this check, 0 if it fails, -1 if abstaining.
        Abstain only when the DSL lacks enough geometric or semantic information
        to determine symmetry, identity, or placement reliably.
    '''
    # Holistic verifier for "A cozy bedroom with symmetric identical nightstands".
    raise NotImplementedError("This verifier is not implemented")

Keep the implementation straightforward.
If examples are provided below, match their labels using the same **`int`** return convention as the system prompt: return **`1`** on positive examples, **`0`** on negative examples, and **`-1`** only for genuine abstention when the check truly cannot be decided from the available information (not to avoid an incorrect 0/1).

# EXAMPLES
## Positive examples:
...

## Negative examples:
...
\end{lstlisting}

\subsubsection{Verification Message Addition}
In this final stage, we query an LLM to identify every return statement in the weak verifier and inspect the verifier body to identify code that causes this return statement to exit. The LLM is then tasked with writing an output verification message for that return statement which describes this logic. The purpose of this verification message is to provide context on why the weak verifier returned, as a single weak verifier may check for many different conditions that result in a final verification decision. Listing~\ref{lst:verification-message-system-prompt} shows the system prompt for this step.

\subsection{Weak Verifier Aggregation}\label{subsec:wva}

In this section we describe additional details on how we adapt Weaver to our setting, as well as our LLM Judges and VLM Judges baselines.

\subsubsection{Addition Details on Weaver}
Weaver~\cite{2025weaver} aggregates over the responses of LLM judges which  vote {\tt True} or {\tt False} on a given query and response. They adopt a weak supervision framework based on prior work in data programming~\cite{Snorkel}.
Given the ground truth label $Y$, this weak supervision framework defines a weak verifier $S$ as having true positive and true negative rates 
\begin{align*}
\notag\mathbb{P}(S | Y = 1) &\sim\text{Bernoulli}(w_1) \\
\notag\mathbb{P}(S | Y = 0) &\sim\text{Bernoulli}(1 - w_0), 
\end{align*}
respectively. Assuming that the weak verifiers are conditionally independent, i.e.
\[
\mathbb{P}(S_1, \ldots, S_k|Y) = \prod_{m=1}^k \mathbb{P}(S_m | Y),
\]
we can optimize $\{w_{k,0}, w_{k,1}\}_{k=1}^m$ to match the pairwise and marginal voting statistics over our entire dataset, including all the unlabeled data. The ground truth label $Y$ can then be predicted as
\begin{equation*}
\mathbb{P}(Y = y|S_1 = s_1, \ldots, S_m = s_m) \propto \frac{\mathbb{P}(Y = 1)\prod_{k=1}^m\mathbb{P}(S_k = s_k | Y = y)}{\mathbb{P}(S_1 = s_1, \ldots, S_m = s_m)}. 
\end{equation*}
Here, the denominator on the right-hand side is observed from the votes on the data and $\mathbb{P}(Y = y)$ can be estimated from the dev set.

In addition to Section~3.3 of the main paper, we also make the following modifications to adapt Weaver for spatial layout verification:
\begin{itemize}
    \item Since Weaver assumes verifiers which either vote {\tt True} or {\tt False}, we use an abstention-aware variant of their weak learning algorithm \,~\cite{Snorkel, Ratner2019-qe}.
    \item Weaver has an additional class balance filtering step in which it filters out any verifiers which vote {\tt True} or {\tt False} on more than some threshold percentage of the unlabeled data. We update this to also filter out verifiers which {\tt Abstain} on more than some threshold percentage of the unlabeled data. In our experiments, we set this threshold to 95\%, as our data is heavily negative-skewed.
\end{itemize}

\subsection{Aggregating LLM Judges and VLM Judges}\label{subsec:llm-judge}
We use off-the-shelf LLM judges to evaluate whether layouts are positive or negative examples of the task description. These votes are then aggregated with one of our four aggregation methods. In our experiments, we use four different LLMs: \texttt{GPT-5.4} from OpenAI, and the quantized open-source models \texttt{llava:7b-v1.6-mistral}, \texttt{gemma4:e4b} and \texttt{qwen3-vl:8b-thinking} from Ollama.
All four models also have vision capabilities, so we can use them as LLM judges or VLM judges. LLM judges receive the task description and our data model representation of the layout as input. VLM judges receive a top-down rendering for 3D room layouts and an SVG for 2D posters. Each LLM receives the system prompt in Listing~\ref{lst:3d-scene-llm-judge-system-prompt} (3D Rooms) and Listing~\ref{lst:2d-poster-llm-judge-system-prompt} (2D Posters) that describes our layout data model. Each model gets the user message:
\begin{lstlisting}[language={}]
  Determine if the following scene is a good example of the prompt. Output a single line containing only 'True' or 'False'.

  Prompt: <prompt>
  Layout: <layout>
\end{lstlisting}

\section{Layout Generators}\label{sec:generators}
This section provides additional implementation details about the base layout generators for the 3D Rooms and 2D Posters domains.

\subsection{3D Room Generator}
The 3D Room generation pipeline samples uses GPT-5.1 to generate 3D layouts in our data model from an input text description. This process is split into three main stages: {\bf (1) Schema Generation}, {\bf (2) Schema Execution and Asset Population} and {\bf (3) Iterative Refinement}. We describe each of these stages in the following sections.

\subsubsection{Schema Generation}
Our generator produces rectangular rooms with four walls centered at the origin. Given a task description and the system prompt in Section~\ref{sec:prompts}, Listing~\ref{lst:3d-scene-generator-system-prompt}, the LLM initially outputs a schema which consists of some attributes metadata and a list of elements. At the scene level, we specify the {\tt width}, {\tt depth} and {\tt height} dimensions of the scene, as well as an RGB {\tt wall\_color} and {\tt floor\_material\_type} chosen from \{{\tt light\_wood}, {\tt dark\_wood}, {\tt light\_carpet}, {\tt dark\_carpet}, {\tt light\_tile}, {\tt dark\_tile}, {\tt decorative\_tile}, {\tt terrazzo}\}.

An element is instantiated as an {\tt Element} in our data model, but has additional {\tt supported\_by} and {\tt against\_wall} attributes. The {\tt supported\_by} attribute is a string which specifies the {\tt id} of another element that this element must lay on top of, i.e. the bottom face of the {\tt bbox} of this element is fully incident with the top face of the {\tt bbox} of the supporting element. If this entry is empty, then the element is assumed to be placed on the floor. The {\tt against\_wall} attribute optionally specifies a wall \{{\tt N}, {\tt S}, {\tt E}, {\tt W}\} to which this object must be snapped. Note that these attributes operate entirely independently of the actual {\tt position} attribute of the element, which operates in the global space. Any conflicts are resolved in the Schema Execution stage.

\subsubsection{Schema Execution and Asset Population}\label{sec:asset-pop}
We next execute the schema produced in the previous stage to produce a JSON in our data model. First, we resolve any conflicts between the {\tt position}, {\tt supported\_by} and {\tt against\_wall} attributes. We begin by initializing each element at the specified {\tt position}, and then snap it to the specified wall followed by the top surface of the supporting element.

Next, we populate the {\tt mesh} attribute of each element. In our system, we manually curated 350 open-source mesh assets and 19 floor textures from BlenderKit~\cite{BlenderKit2018} and assign each asset into a category. For each mesh asset we also canonicalize the orientation. We first randomly choose a floor texture from the specified category in the schema. Next, we select specific mesh assets for each element in the schema, guided by the element {\tt size}. This is done by assigning a relevance score to all the mesh assets within the target category and randomly picking an asset from the top-$K$ scoring assets. The score is defined as

\begin{equation}
    s(M, {\tt size}) = \alpha s_{\text{prop}}(M, {\tt size}) - \beta s_{\text{vol}}(M, {\tt size}),
\end{equation}
where $s_{\text{prop}}$ checks for matching proportions and $s_{\text{vol}}$ checks for matching global volume. Specifically,  $s_{\text{prop}}(M, {\tt size})$ is defined as the $L_1$ distance between the dimensions of $M$ normalized to the unit cube and {\tt size} and $s_{\text{vol}}(M, {\tt size})$ is the absolute log-ratio of the target volume to the volume of $M$. In our experiments we set $\alpha = 1.0$ and $\beta = 0.15$.
Once an asset is selected, we uniformly resize that asset to fit inside the target {\tt size}, and then update {\tt size} to the new scaled asset dimensions. 

\subsubsection{Iterative Refinement}
The output of the previous stage is a scene in our data model representation. Next, we apply a limited set of refinements to improve any catastrophic errors (i.e., major intersections or out-of-bounds objects). We render this to a top-down view and prompt the same model with the system prompt in Section~\ref{sec:prompts}, Listing~\ref{lst:3d-scene-refinement-system-prompt} to either reject the layout and provide feedback on what is wrong or accept the layout.

\subsection{2D Poster Generator}
The 2D poster generator uses GPT-5.1 to generate poster layouts using HTML/CSS. We then parse the resulting HTML into our 2D poster layout data model.
The task description is passed to the LLM as the user prompt. Each task description contains a single sentence to provide the poster context, followed by the layout specification and the text content for each of the title, description, date and location elements. We also specify possible font families, and color palettes. 
For image assets, the LLM either writes CSS/SVG code directly into the HTML or it can create a placeholder \texttt{<img>} div with an \texttt{alt} attribute. Afterwards, we separately query the LLM to generate images for all such placeholders using the \texttt{alt} text as the prompt.
The LLM receives the system prompt in Section~\ref{sec:prompts}, Listing~\ref{lst:2d-poster-generator-system-prompt}.

\subsection{Holodeck and FlairGPT}
\begin{figure*}[t]
    \centering
    \includegraphics[width=\linewidth]{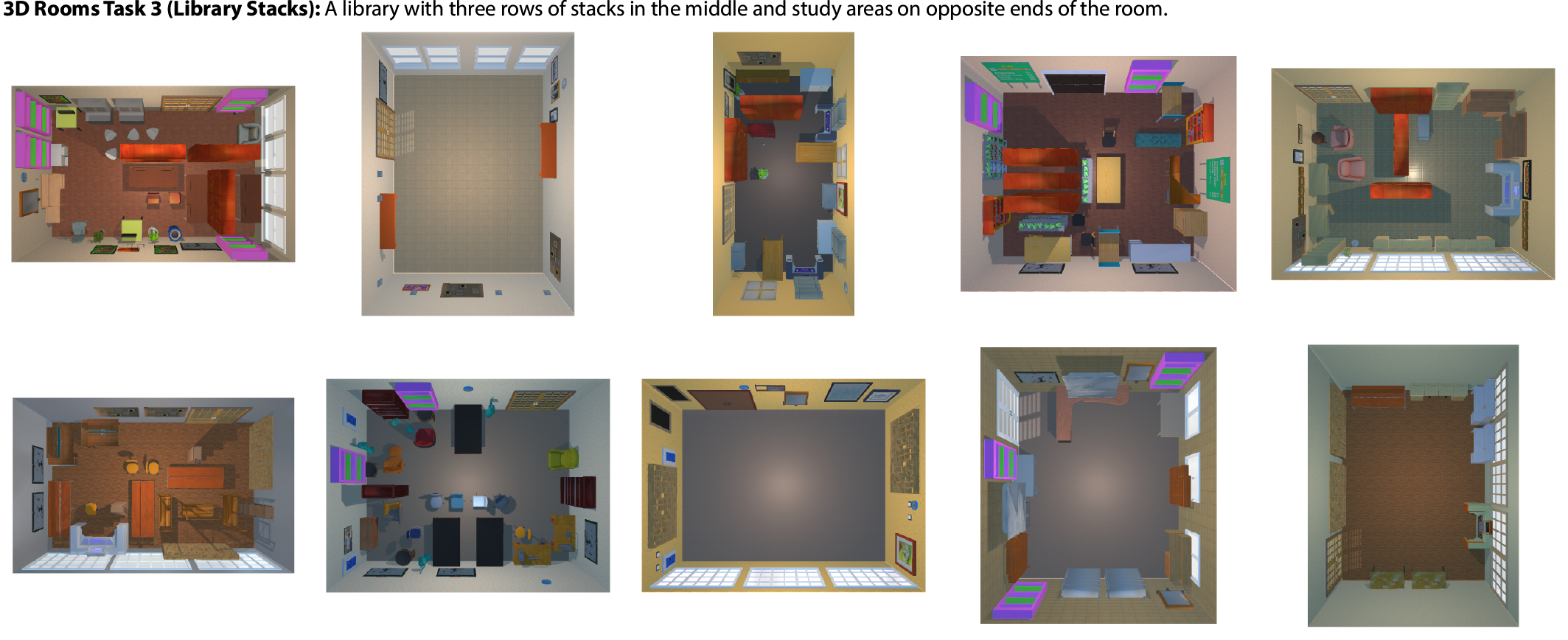}
    \caption{Holodeck generations for the task description in {\bf T3: Library Stacks (3D Rooms)}. Many layouts are missing bookshelves or do not arrange the bookshelves in rows.}
    \label{fig:holodeck-3a}
\end{figure*}

\begin{figure*}[t]
    \centering
    \includegraphics[width=\linewidth]{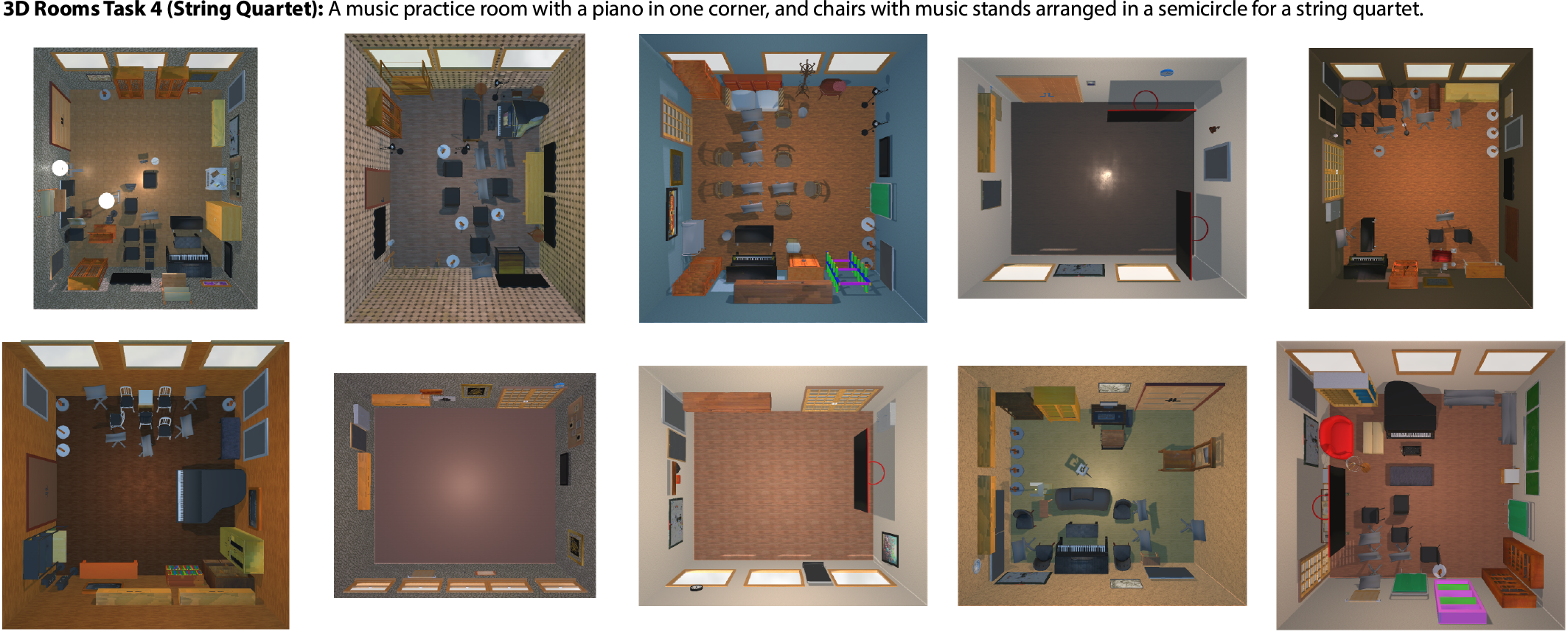}
    \caption{Holodeck generations for the task description in {\bf T4: String Quartet (3D Rooms)}. Many layouts are missing important furniture. None of the layouts have a semicircular arrangement of chairs and music stands.}
    \label{fig:holodeck-4}
\end{figure*}

\begin{figure*}[t]
    \centering
    \includegraphics[width=\linewidth]{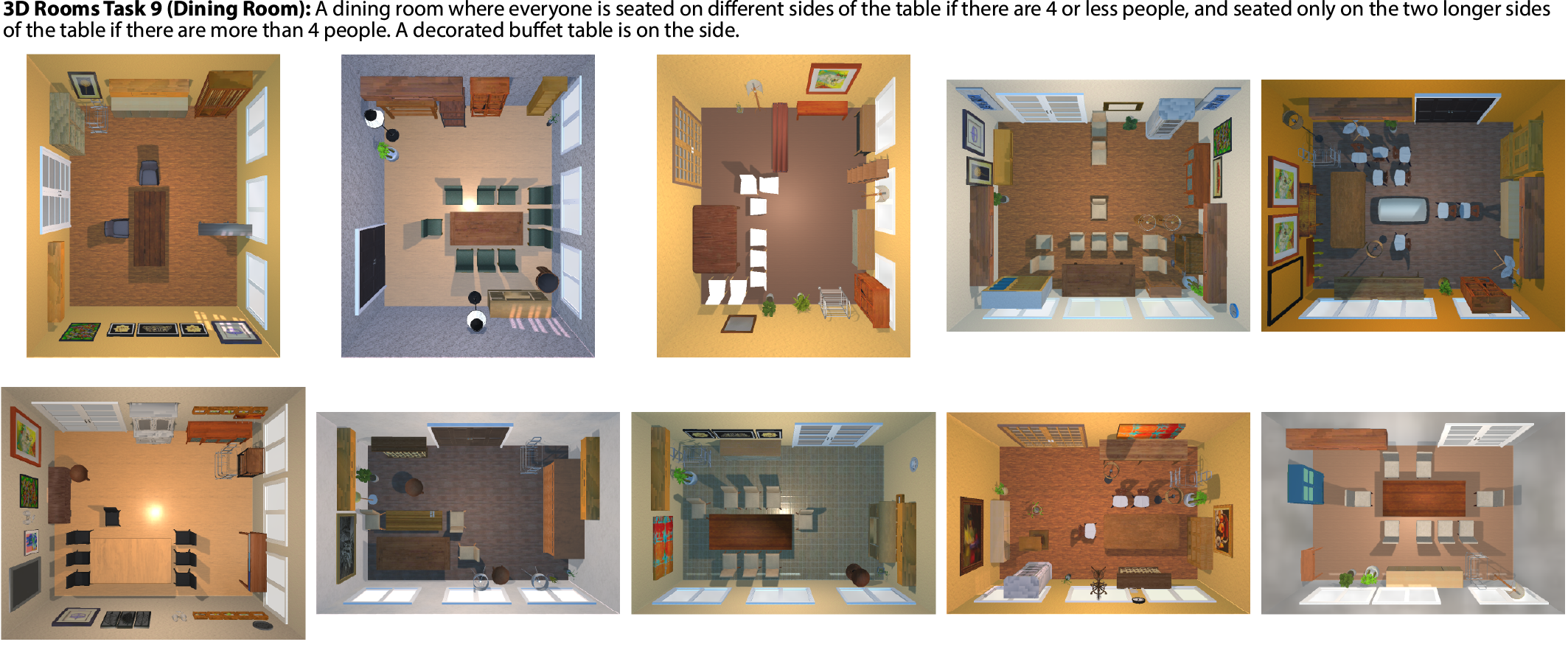}
    \caption{Holodeck generations for the task description in {\bf T9: Dining Room (3D Rooms)}. The top-left image contains a positive layout, but the remaining layouts do not respect the seating rules for more than 4 people.}
    \label{fig:holodeck-9}
\end{figure*}

\begin{figure*}[t]
    \centering
    \includegraphics[width=\linewidth]{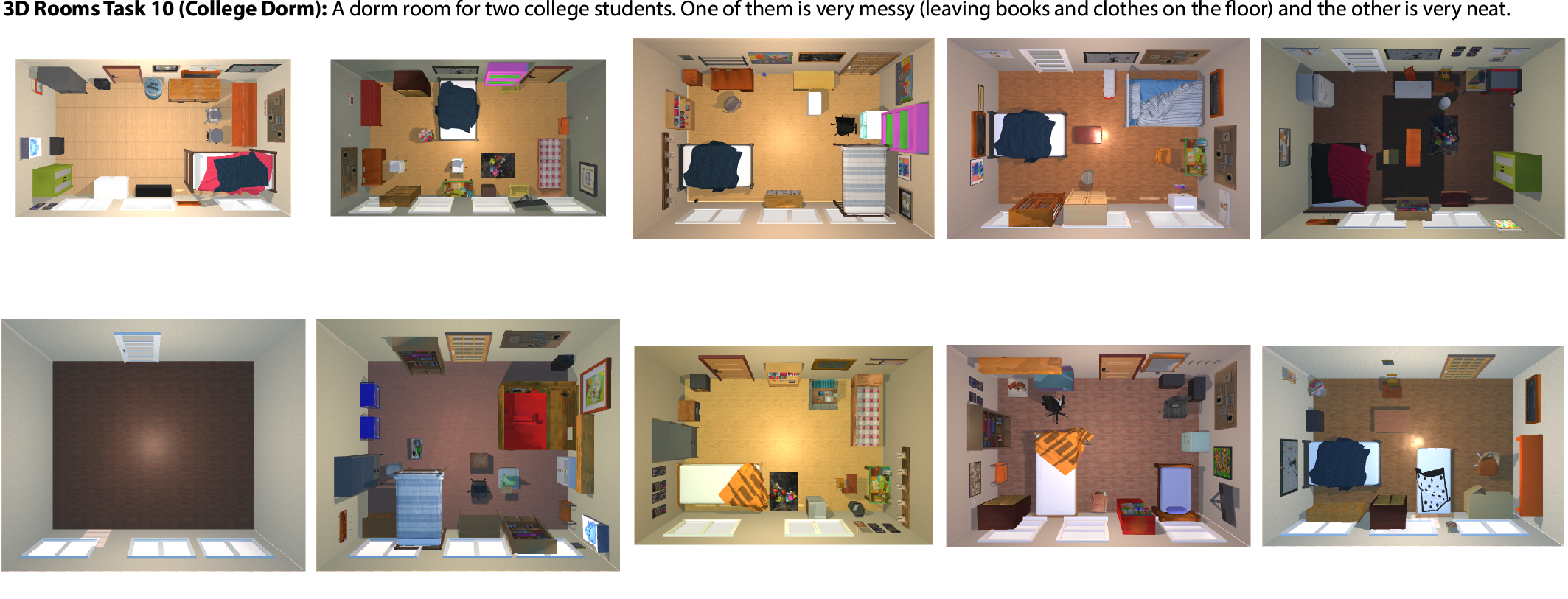}
    \caption{Holodeck generations for the task description in {\bf T10: College Dorm (3D Rooms)}. Several rooms only have a single desk or bed. Many layouts do not have the defined clutter on the floor.}
    \label{fig:holodeck-10a}
\end{figure*}

\begin{figure*}[t]
    \centering
    \includegraphics[width=\linewidth]{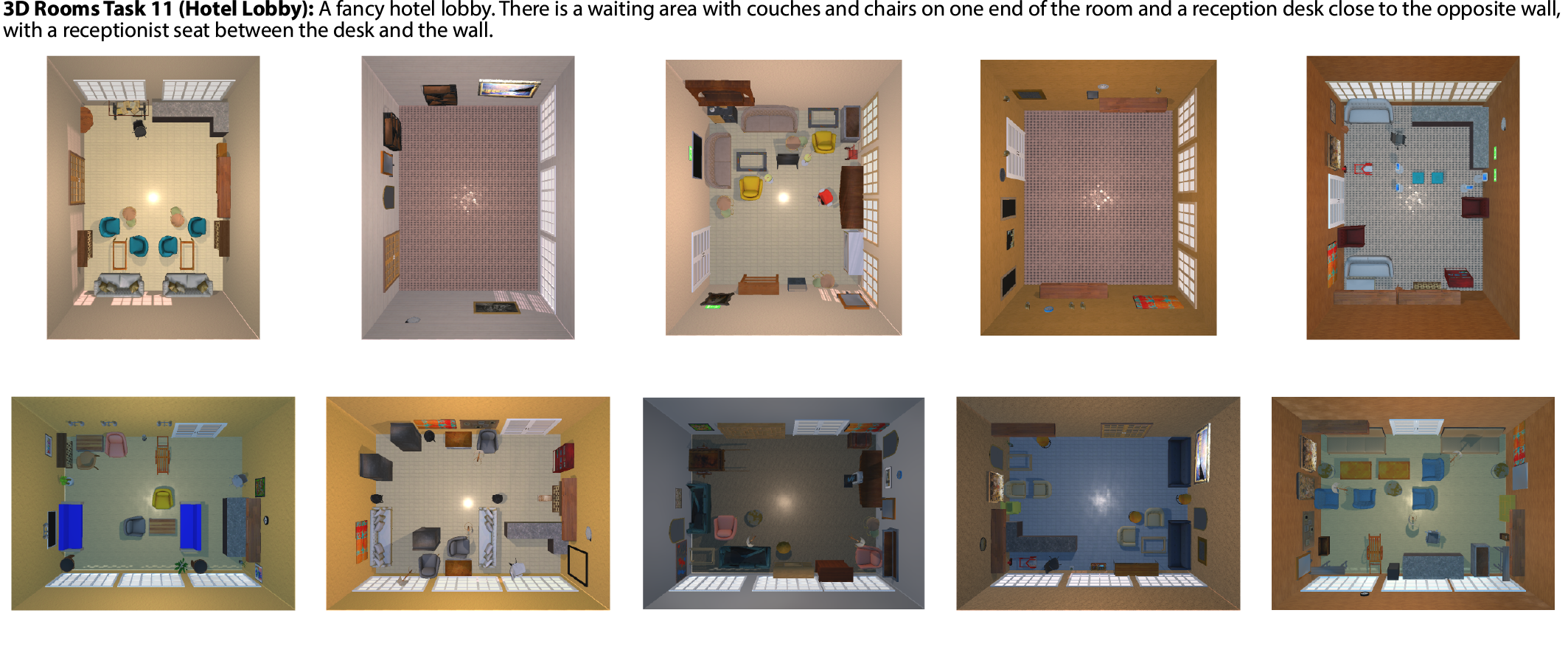}
    \caption{Holodeck generations for the task description in {\bf T11: Hotel Lobby (3D Rooms)}. Several layouts are missing furniture. Many layouts do not satisfy the arrangement of the receptionist seat between the desk and wall.}
    \label{fig:holodeck-11a}
\end{figure*}

\begin{figure*}[t]
    \centering
    \includegraphics[width=\linewidth]{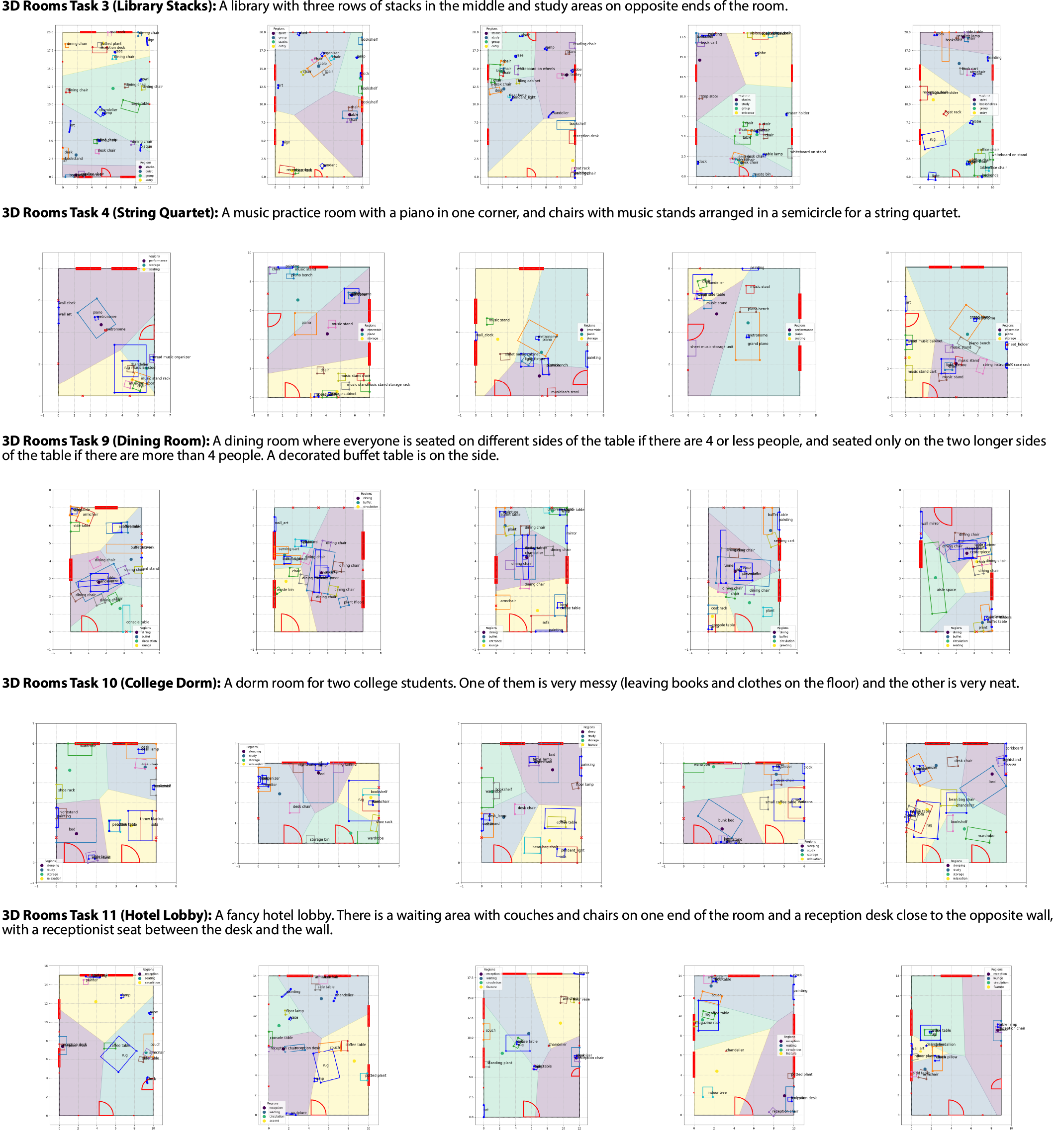}
    \caption{FlairGPT floor plan generations of five task descriptions in the 3D Rooms domain ({\bf T3}, {\bf T4}, {\bf T9}, {\bf T10}, {\bf T11}). Many layouts lack the correct number of furniture and do not have items arranged according to the task description.}
    \label{fig:flairgpt}
\end{figure*}

We experiment with using two existing LLM-based 3D layout generators, Holodeck~\cite{Yang_2024_CVPR} and FlairGPT~\cite{littlefair2025flairgpt}. Holodeck takes in an input prompt and outputs a 3D layout populated with 3D assets from Objaverse~\cite{objaverseXL}. It consists of multiple stages of LLM calls to place large objects, wall-mounted objects and small objects before finally optimizing the positions of all the objects to avoid physical intersections. FlairGPT takes in an input prompt and outputs a 2D floor plan layout, which is not populated with 3D assets. For both methods, we use GPT-5.1 as the LLM (same as in our base 3D layout generator).
%
We find that both generators have significant issues that make them impractical as 3D layout generators for our tasks. Holodeck biases objects towards the edges of the room and often fails to place things in explicitly specified arrangements, such as rows or semicircles. FlairGPT takes a significant amount of time to generate a single layout (about 20 minutes) and often fails before producing a layout. Figures~\ref{fig:holodeck-3a}-\ref{fig:flairgpt} show examples of layouts generated by Holodeck and FlairGPT for the five tasks evaluated in Section~4.4 of the main paper.

\clearpage

\clearpage
\section{Layout Tasks}\label{sec:tasks}
We evaluate a total of 26 layout tasks, 13 for the 3D scene layout domain and 13 for the 2D poster layout domain. Table~\ref{tab:task-descriptions-2d} and Table~\ref{tab:task-descriptions-3d} provides the full task descriptions alongside experiment details for each task, such as the positive sample rate by the corresponding generator and the number of samples used in the entire dataset. For 23 of the 26 tasks, the task description is also the prompt passed to the layout generator. The three 3D Rooms tasks {\bf T1: Symmetric Nighstands}, {\bf T12: Reading Nook} and {\bf T13: Study Area} use a {\em shared} dataset of 100 samples generated from the prompt ``A cozy bedroom."
\begingroup
\small
\captionsetup{justification=raggedright,singlelinecheck=false}
\begin{longtable}{@{}>{\raggedright\arraybackslash}p{3.5cm}>{\raggedleft\arraybackslash}p{1.0cm}>{\raggedright\arraybackslash}p{\dimexpr\textwidth-4.5cm-4\tabcolsep\relax}@{}}
    \caption{{\bf Table of 3D room layout task descriptions.} For each task, we sample 100 layouts for the dataset generation and ask a user to manually label them as positive or negative.}
    \label{tab:task-descriptions-3d} \\

    \noalign{\global\arrayrulewidth=1.5pt}
    \hline
    \noalign{\global\arrayrulewidth=0.4pt}
    \textbf{Name} & \raggedleft\textbf{Pos \%} & \textbf{Layout Task Description} \\
    \hline
    \endfirsthead

    \multicolumn{3}{@{}l}{\small\textit{Table \thetable{} continued from previous page}} \\
    \noalign{\global\arrayrulewidth=1.5pt}
    \hline
    \noalign{\global\arrayrulewidth=0.4pt}
    \textbf{Name} & \raggedleft\textbf{Pos \%} & \textbf{Layout Task Description} \\
    \hline
    \endhead

    \hline
    \multicolumn{3}{@{}r}{\small\textit{Continued on next page}} \\
    \endfoot

    \hline
    \endlastfoot

    \texttt{T1: Symmetric Nightstands} & 26\% & A cozy bedroom with symmetric identical nightstands. \\
    \hline
    \texttt{T2: Two Workstations} & 44\% & An office with two workstations at opposite walls. Each workstation faces the wall. \\
    \hline
    \texttt{T3: Library Stacks} & 18\% & A library with three rows of bookshelves in the middle of a room. There are desks and chairs for studying on opposite ends of the room. \\
    \hline
    \texttt{T4: String Quartet} & 14\% & A music practice room with a piano in one corner, and chairs with music stands arranged in a semicircle for a string quartet. \\
    \hline
    \texttt{T5: Movie Theater} & 23\% & A movie theater with a large screen from floor to ceiling and rows of armchair seats. There is an aisle down the middle of the seats. \\
    \hline
    \texttt{T6: Living Room} & 20\% & A living room with big coffee table and a couch and some armchairs, all angled towards the TV. A small dining area is behind the couch. \\
    \hline
    \texttt{T7: Therapist Office} & 28\% & A therapist's office. The therapist sits behind a desk and the client(s) sit on the other side of the desk. The wall behind the therapist's desk is decorated with a bookshelf, a painting, and/or some plants. \\
    \hline
    \texttt{T8: Arcade Room} & 37\% & An arcade game room that has either a foosball table or a pool table, but not both. People should have enough space around the table to play. There should be a couch with a coffee table or a table with some chairs to play board games on the side. \\
    \hline
    \texttt{T9: Dining Room} & 30\% & A dining room where everyone is seated on different sides of the table if there are 4 or less people, and seated only on the two longer sides of the table if there are more than 4 people. A decorated buffet table is on the side. \\
    \hline
    \texttt{T10: College Dorm} & 21\% & A dorm room for two college students. One of them is very messy (leaving books and clothes on the floor) and the other is very neat. \\
    \hline
    \texttt{T11: Hotel Lobby} & 34\% & A fancy hotel lobby. There is a waiting area with couches and chairs on one end of the room and a reception desk close to the opposite wall, with a receptionist seat between the desk and the wall. \\
    \hline
    \texttt{T12: Reading Nook} & 25\% & A bedroom with a reading nook. \\
    \hline
    \texttt{T13: Study Area} & 21\% & A bedroom with a desk and chair setup for studying. \\
    \hline

\end{longtable}
\endgroup

\begingroup
\small
\captionsetup{justification=raggedright,singlelinecheck=false}
\begin{longtable}{@{}>{\raggedright\arraybackslash}p{3.5cm}>{\raggedleft\arraybackslash}p{1.0cm}>{\raggedright\arraybackslash}p{\dimexpr\textwidth-4.5cm-4\tabcolsep\relax}@{}}
    \caption{{\bf Table of 2D poster layout task descriptions.} For each task, we sample 100 layouts for the dataset generation and ask a user to manually label them as positive or negative.}
    \label{tab:task-descriptions-2d} \\

    \noalign{\global\arrayrulewidth=1.5pt}
    \hline
    \noalign{\global\arrayrulewidth=0.4pt}
    \textbf{Name} & \raggedleft\textbf{Pos \%} & \textbf{Layout Task Description} \\
    \hline
    \endfirsthead

    \multicolumn{3}{@{}l}{\small\textit{Table \thetable{} continued from previous page}} \\
    \noalign{\global\arrayrulewidth=1.5pt}
    \hline
    \noalign{\global\arrayrulewidth=0.4pt}
    \textbf{Name} & \raggedleft\textbf{Pos \%} & \textbf{Layout Task Description} \\
    \hline
    \endhead

    \hline
    \multicolumn{3}{@{}r}{\small\textit{Continued on next page}} \\
    \endfoot

    \hline
    \endlastfoot

    \texttt{T14: Jazz Night} & 82\% & A poster for a jazz night at a jazz club called 'The Blue Note'. The title, date, location and description are distributed at the corners of the poster in lively orientations. The main graphic of an abstract instrument is in the center of the poster and is very large, filling most of the poster. Use fonts that are very decorative and bold but fit together well. \\
    \hline
    \texttt{T15: Art Show} & 27\% & A poster for an art student show called 'works in pro-gress'. The poster background is split into four vertical columns of different bright colors. The title is in the center of the poster (from a quarter of the way up to a quarter of the way down) and is split across the four columns. Each word in the title is oriented vertically and rotated 90 degrees. The words 'works' and 'in' are in the first two columns and are top-aligned. The words 'pro-' and 'gress' are in the last two columns and are bottom-aligned. The date is at the bottom of the third column from the left in regular orientation, below 'pro-'. The location is at the bottom of the fourth column from the left in regular orientation, below 'gress'. A short description is at the top-left of the poster, above 'works'. The title text is very large and the fonts are futuristic. \\
    \hline
    \texttt{T16: Sparrows} & 26\% & A book cover for a book called 'Flight of the Sparrow'. The title fills the entire front cover and each word is arranged in a zigzag pattern. The author is at the bottom right corner and the publisher is at the bottom left corner. Images of sparrows are scattered across the cover in various orientations, interleaving through the title text, some above the text and some below the text. The fonts are serif and the overall design is very traditional and illustrative. \\
    \hline
    \texttt{T17: Modern Home} & 31\% & A landscape-format coffee table book about architecture called 'The Modern Home'. The title is in the top half of the cover on the left, and each word of the title is on a separate line, all left-aligned. Right underneath the title is a large photograph of a modern home that spans the width of the cover. The photograph is left-aligned with the title. The author and publisher are in smaller font at the bottom left corner of the cover, one stacked on the other. There is no description. \\
    \hline
    \texttt{T18: Rock Concert} & 14\% & A rock concert poster for an event called 'EMERALD RIFF FESTIVAL' taking place in Dublin, Ireland. The entire background of the poster is green and all other elements are uniform darker green color. The center of a poster features a transparent carved print style illustration of a guitarist or a drummer with lots of electric energy. The title is in large decorative celtic letters, with the word 'EMERALD' above the hero graphic and the words 'RIFF FESTIVAL' in one line below the hero graphic. The hero graphic covers all the space from 'EMERALD' to 'RIFF FESTIVAL'. The date, description and location are top-aligned and distributed horizontally across the bottom of the poster. The description is a short list of bands playing in the festival. \\
    \hline
    \texttt{T19: OOO} & 83\% & A book cover for a non-fiction book called 'Out Of Office'. The central graphic is a tightly-packed 5x5 grid of solid white people SVG icons, except one cell is empty. The background of the book is blue. The title is in a single line below the central graphic, and the description is in a single line below the title. The author and publisher are at the top two corners of the cover. Everything is very minimalistic and modern. \\
    \hline
    \texttt{T20: Film Memoir} & 71\% & A book cover for a memoir of a filmmaker called 'One Frame At A Time'. The title is center aligned and each word is on a separate line, in total occupying the top 2/3 of the cover. The description is a single line underneath the title in small font. The entire background of the book is a close-up black and white image of messy tangled rolls of film strip. The entire cover is black and white. \\
    \hline
    \texttt{T21: Western Night} & 20\% & A poster for a Western Movie Night. The center of the poster features a transparent woodblock-print style drawing of a cowboy riding a horse. The title is in large western letters above the hero graphic, with the word 'WESTERN' on one line and the words 'MOVIE NIGHT' on the second line. The description is below the hero graphic and the date and location are distributed horizontally across the bottom of the poster. The entire theme of the poster is very western and rustic. All the elements are tightly packed. \\
    \hline
    \texttt{T22: Crossword} & 27\% & A cover for a collection of crossword puzzles title 'CROSSWORD'. The title word 'CROSSWORD' is in a large font and is placed in the center of the cover. Importantly, it is broken up as 'CROSS' and 'WORD', and the two overlap in a cross sharing the O. The description is at the top of the cover, vertically-centered. The author and publisher are below the title, vertically-centered on separate lines. An image of a large pencil tip spans the top-right corner to the title. \\
    \hline
    \texttt{T23: The Thinker} & 48\% & A book cover for a book titled 'The Thinker'. The title is in the top left corner in red script font. The description is below it in a smaller blue serif font. The author and publisher are in the top-right corner in the same blue text. A minimalist red line drawing of a robed Greek sculpture is in the bottom right of the book, extending past the bottom edge of the book. The background of the book is a light scarlet. \\
    \hline
    \texttt{T24: Disappearing} & 28\% & A book cover for a book titled 'Alyssa Mackins is Disappearing Again'. The title occupies the entire front cover, with each word on a separate line that spans the top-left corner to the bottom-right corner. The four words are colored white but the opacity decreases from left to right. The background of the book is blue. The publisher is at the top-center and the author is at the bottom-center. There is no description. \\
    \hline
    \texttt{T25: Craft Fair} & 46\% & A poster for a craft fair. The entire background of the poster is a deep red-orange. The top 2/3 of the poster is a large detailed illustration of a flower. The title is in decorate script font right below the hero graphic. The location, date and description are distributed in three columns from left to right below the title, in that order. All the text is white. \\
    \hline
    \texttt{T26: Whirlpool} & 31\% & A movie poster for a movie called 'Whirlpool'. The title 'Whirlpool' is repeated three times, with the entire string arranged in a large spiral at the center of the poster. The outermost iteration of the word 'Whirlpool' is solid white, and the rest are at 25\% opacity. The date, location and description are in a triangle around the spiral, and the text is all in random orientations. A subtle psychedelic abstract hero graphic is directly behind the title text and covers a large part of the poster. \\
    \hline
\end{longtable}
\endgroup

\clearpage
\section{Additional Results}\label{sec:add-results}
\begin{table*}[t]
    \centering
    \small
    \caption{
    F1-scores of our verifiers with four aggregation methods. The top performing verifier per-task is bolded in each column, though in the case of ties we bold Weaver. 
    Overall, our Weaver verifier performs strongest on average and is the top-performing verifier on the most number of tasks (18 out of 26 tasks).
    } \label{tab:claim2}
\begin{tabularx}{\textwidth}{@{}>{\raggedright\arraybackslash}p{2.8cm}|*{13}{>{\centering\arraybackslash}X}|>{\raggedleft\arraybackslash}X@{}}
\multicolumn{15}{@{}l}{\textbf{3D Rooms (F1-Score $\uparrow$)}} \\
\noalign{\global\arrayrulewidth=1.2pt}\hline\noalign{\global\arrayrulewidth=0.4pt}
\textbf{Method} & T1 & T2 & T3 & T4 & T5 & T6 & T7 & T8 & T9 & T10 & T11 & T12 & T13 & Avg \\
\hline
Naive Majority & 0.82 & \textbf{0.92} & 0.94 & 0.66 & 0.86 & 0.75 & 0.83 & 0.80 & 0.56 & 0.80 & 0.82 & \textbf{0.85} & 0.70 & 0.79 \\
Logistic Regression & 0.82 & 0.91 & \textbf{0.95} & 0.59 & 0.84 & 0.46 & 0.80 & 0.80 & 0.73 & 0.80 & 0.81 & 0.81 & 0.68 & 0.77 \\
Top-1 & 0.80 & 0.91 & \textbf{0.95} & 0.67 & 0.85 & 0.68 & 0.77 & 0.79 & \textbf{0.75} & 0.83 & 0.79 & 0.78 & 0.73 & 0.79 \\
Weaver & \textbf{0.84} & 0.91 & 0.94 & \textbf{0.74} & \textbf{0.86} & \textbf{0.76} & \textbf{0.86} & \textbf{0.82} & 0.74 & \textbf{0.90} & \textbf{0.84} & 0.82 & \textbf{0.91} & \textbf{0.84} \\
\hline
\end{tabularx}
\vskip 1em
\begin{tabularx}{\textwidth}{@{}>{\raggedright\arraybackslash}p{2.8cm}|*{13}{>{\centering\arraybackslash}X}|>{\raggedleft\arraybackslash}X@{}}
\multicolumn{15}{@{}l}{\textbf{2D Posters (F1-Score $\uparrow$)}} \\
\noalign{\global\arrayrulewidth=1.2pt}\hline\noalign{\global\arrayrulewidth=0.4pt}
\textbf{Method} & T14 & T15 & T16 & T17 & T18 & T19 & T20 & T21 & T22 & T23 & T24 & T25 & T26 & Avg \\
\hline
Naive Majority & 0.94 & 0.87 & 0.91 & 0.06 & 0.92 & 0.91 & 0.20 & \textbf{0.83} & 0.29 & \textbf{0.85} & 0.22 & 0.72 & 0.29 & 0.62 \\
Logistic Regression & 0.94 & 0.84 & 0.91 & 0.93 & 0.59 & 0.90 & 0.95 & 0.77 & 0.21 & \textbf{0.85} & 0.13 & 0.66 & 0.28 & 0.69 \\
Top-1 & 0.93 & 0.81 & 0.89 & \textbf{0.95} & 0.90 & 0.87 & \textbf{0.96} & 0.67 & 0.38 & 0.81 & 0.30 & 0.69 & 0.46 & 0.74 \\
Weaver & \textbf{0.94} & \textbf{0.90} & \textbf{0.91} & 0.93 & \textbf{0.94} & \textbf{0.91} & 0.94 & 0.77 & \textbf{0.38} & 0.84 & \textbf{0.46} & \textbf{0.74} & \textbf{0.59} & \textbf{0.79} \\
\hline
\end{tabularx}
\end{table*}

\begin{table*}
    \centering
    \small
    \caption{
    Precision of our verifiers with four aggregation methods.
    } \label{tab:prec}
\begin{tabularx}{\textwidth}{@{}>{\raggedright\arraybackslash}p{2.8cm}|*{13}{>{\centering\arraybackslash}X}|>{\raggedleft\arraybackslash}X@{}}
\multicolumn{15}{@{}l}{\textbf{3D Rooms (Precision $\uparrow$)}} \\
\noalign{\global\arrayrulewidth=1.2pt}\hline\noalign{\global\arrayrulewidth=0.4pt}
\textbf{Method} & T1 & T2 & T3 & T4 & T5 & T6 & T7 & T8 & T9 & T10 & T11 & T12 & T13 & Avg \\
\hline
Naive Majority & 0.79 & \textbf{0.91} & \textbf{1.00} & 0.68 & \textbf{0.88} & \textbf{0.85} & 0.82 & 0.83 & \textbf{0.66} & \textbf{0.98} & 0.80 & \textbf{0.87} & 0.90 & \textbf{0.84} \\
Logistic Regression & 0.78 & 0.89 & \textbf{1.00} & 0.70 & 0.86 & 0.40 & \textbf{0.84} & \textbf{0.84} & 0.59 & 0.96 & 0.76 & 0.84 & 0.91 & 0.80 \\
Top-1 & 0.72 & 0.89 & \textbf{1.00} & \textbf{0.71} & 0.78 & 0.70 & 0.78 & 0.82 & 0.61 & 0.95 & \textbf{0.84} & 0.73 & \textbf{0.94} & 0.81 \\
Weaver & \textbf{0.79} & 0.89 & 0.97 & 0.67 & 0.85 & 0.71 & 0.83 & 0.82 & 0.61 & 0.92 & 0.78 & 0.77 & 0.91 & 0.81 \\
\hline
\end{tabularx}
\vskip 1em
\begin{tabularx}{\textwidth}{@{}>{\raggedright\arraybackslash}p{2.8cm}|*{13}{>{\centering\arraybackslash}X}|>{\raggedleft\arraybackslash}X@{}}
\multicolumn{15}{@{}l}{\textbf{2D Posters (Precision $\uparrow$)}} \\
\noalign{\global\arrayrulewidth=1.2pt}\hline\noalign{\global\arrayrulewidth=0.4pt}
\textbf{Method} & T14 & T15 & T16 & T17 & T18 & T19 & T20 & T21 & T22 & T23 & T24 & T25 & T26 & Avg \\
\hline
Naive Majority & 0.96 & \textbf{0.98} & \textbf{0.99} & \textbf{1.00} & 0.92 & 0.92 & 0.96 & \textbf{0.96} & \textbf{0.88} & \textbf{0.91} & \textbf{1.00} & \textbf{0.77} & \textbf{1.00} & \textbf{0.94} \\
Logistic Regression & 0.91 & \textbf{0.98} & 0.98 & \textbf{1.00} & \textbf{0.97} & 0.88 & 0.97 & 0.94 & 0.49 & 0.90 & 0.67 & 0.72 & 0.58 & 0.85 \\
Top-1 & 0.96 & 0.96 & 0.96 & 0.99 & 0.87 & 0.92 & 0.98 & 0.68 & 0.78 & \textbf{0.91} & \textbf{1.00} & \textbf{0.77} & 0.82 & 0.89 \\
Weaver & \textbf{0.96} & 0.96 & 0.95 & 0.94 & 0.92 & \textbf{0.92} & \textbf{0.98} & 0.73 & 0.40 & 0.90 & 0.65 & 0.76 & 0.92 & 0.85 \\
\hline
\end{tabularx}
\end{table*}

\begin{table*}
    \centering
    \small
    \caption{
    Recall of our verifiers with four aggregation methods.
    } \label{tab:rec}
\begin{tabularx}{\textwidth}{@{}>{\raggedright\arraybackslash}p{2.8cm}|*{13}{>{\centering\arraybackslash}X}|>{\raggedleft\arraybackslash}X@{}}
\multicolumn{15}{@{}l}{\textbf{3D Rooms (Recall $\uparrow$)}} \\
\noalign{\global\arrayrulewidth=1.2pt}\hline\noalign{\global\arrayrulewidth=0.4pt}
\textbf{Method} & T1 & T2 & T3 & T4 & T5 & T6 & T7 & T8 & T9 & T10 & T11 & T12 & T13 & Avg \\
\hline
Naive Majority & 0.85 & 0.92 & 0.89 & 0.68 & 0.84 & 0.67 & 0.87 & 0.80 & 0.68 & 0.68 & 0.84 & 0.85 & 0.70 & 0.79 \\
Logistic Regression & 0.86 & 0.93 & 0.90 & 0.53 & 0.82 & 0.55 & 0.82 & 0.80 & 0.95 & 0.72 & 0.88 & 0.82 & 0.66 & 0.79 \\
Top-1 & \textbf{0.91} & 0.92 & 0.91 & 0.65 & \textbf{0.93} & 0.73 & 0.80 & 0.78 & \textbf{0.96} & 0.76 & 0.75 & 0.85 & 0.70 & 0.82 \\
Weaver & 0.89 & \textbf{0.93} & \textbf{0.93} & \textbf{0.84} & 0.88 & \textbf{0.85} & \textbf{0.92} & \textbf{0.84} & 0.94 & \textbf{0.89} & \textbf{0.92} & \textbf{0.89} & \textbf{0.92} & \textbf{0.90} \\
\hline
\end{tabularx}
\vskip 1em
\begin{tabularx}{\textwidth}{@{}>{\raggedright\arraybackslash}p{2.8cm}|*{13}{>{\centering\arraybackslash}X}|>{\raggedleft\arraybackslash}X@{}}
\multicolumn{15}{@{}l}{\textbf{2D Posters (Recall $\uparrow$)}} \\
\noalign{\global\arrayrulewidth=1.2pt}\hline\noalign{\global\arrayrulewidth=0.4pt}
\textbf{Method} & T14 & T15 & T16 & T17 & T18 & T19 & T20 & T21 & T22 & T23 & T24 & T25 & T26 & Avg \\
\hline
Naive Majority & 0.91 & 0.78 & 0.85 & 0.03 & 0.93 & 0.90 & 0.11 & 0.73 & 0.22 & 0.80 & 0.13 & 0.67 & 0.18 & 0.56 \\
Logistic Regression & \textbf{0.97} & 0.73 & 0.85 & 0.87 & 0.46 & \textbf{0.93} & \textbf{0.93} & 0.65 & 0.19 & 0.80 & 0.08 & 0.61 & 0.19 & 0.64 \\
Top-1 & 0.91 & 0.70 & 0.83 & 0.91 & 0.95 & 0.84 & \textbf{0.93} & 0.68 & 0.38 & 0.72 & 0.21 & 0.63 & 0.32 & 0.70 \\
Weaver & 0.93 & \textbf{0.85} & \textbf{0.88} & \textbf{0.94} & \textbf{0.98} & 0.90 & 0.90 & \textbf{0.83} & \textbf{0.43} & \textbf{0.80} & \textbf{0.36} & \textbf{0.72} & \textbf{0.47} & \textbf{0.77} \\
\hline
\end{tabularx}
\end{table*}

\begin{table*}
    \centering
    \small
    \caption{
    Accuracy of our verifiers with four aggregation methods.
    } \label{tab:acc}
\begin{tabularx}{\textwidth}{@{}>{\raggedright\arraybackslash}p{2.8cm}|*{13}{>{\centering\arraybackslash}X}|>{\raggedleft\arraybackslash}X@{}}
\multicolumn{15}{@{}l}{\textbf{3D Rooms (Accuracy $\uparrow$)}} \\
\noalign{\global\arrayrulewidth=1.2pt}\hline\noalign{\global\arrayrulewidth=0.4pt}
\textbf{Method} & T1 & T2 & T3 & T4 & T5 & T6 & T7 & T8 & T9 & T10 & T11 & T12 & T13 & Avg \\
\hline
Naive Majority & 0.92 & \textbf{0.93} & 0.98 & 0.90 & 0.94 & \textbf{0.91} & 0.90 & 0.86 & 0.78 & 0.93 & 0.87 & \textbf{0.92} & 0.91 & 0.90 \\
Logistic Regression & 0.92 & 0.92 & 0.98 & 0.90 & 0.93 & 0.84 & 0.89 & 0.86 & 0.79 & 0.93 & 0.86 & 0.91 & 0.90 & 0.89 \\
Top-1 & 0.90 & 0.92 & 0.98 & 0.91 & 0.92 & 0.88 & 0.87 & 0.85 & \textbf{0.81} & 0.94 & 0.86 & 0.88 & 0.92 & 0.90 \\
Weaver & \textbf{0.92} & 0.92 & \textbf{0.98} & \textbf{0.91} & \textbf{0.94} & 0.89 & \textbf{0.91} & \textbf{0.87} & 0.80 & \textbf{0.96} & \textbf{0.88} & 0.90 & \textbf{0.96} & \textbf{0.91} \\
\hline
\end{tabularx}
\vskip 1em
\begin{tabularx}{\textwidth}{@{}>{\raggedright\arraybackslash}p{2.8cm}|*{13}{>{\centering\arraybackslash}X}|>{\raggedleft\arraybackslash}X@{}}
\multicolumn{15}{@{}l}{\textbf{2D Posters (Accuracy $\uparrow$)}} \\
\noalign{\global\arrayrulewidth=1.2pt}\hline\noalign{\global\arrayrulewidth=0.4pt}
\textbf{Method} & T14 & T15 & T16 & T17 & T18 & T19 & T20 & T21 & T22 & T23 & T24 & T25 & T26 & Avg \\
\hline
Naive Majority & 0.90 & 0.94 & 0.96 & 0.70 & 0.98 & 0.85 & 0.37 & \textbf{0.94} & \textbf{0.77} & 0.86 & 0.76 & 0.75 & 0.75 & 0.81 \\
Logistic Regression & 0.89 & 0.93 & 0.96 & 0.96 & 0.92 & 0.83 & 0.93 & 0.93 & 0.73 & 0.86 & 0.73 & 0.71 & 0.74 & 0.86 \\
Top-1 & 0.89 & 0.91 & 0.95 & \textbf{0.97} & 0.97 & 0.81 & \textbf{0.94} & 0.87 & 0.76 & 0.83 & 0.78 & 0.74 & 0.77 & 0.86 \\
Weaver & \textbf{0.91} & \textbf{0.95} & \textbf{0.96} & 0.96 & \textbf{0.98} & \textbf{0.85} & 0.92 & 0.90 & 0.51 & \textbf{0.86} & \textbf{0.82} & \textbf{0.77} & \textbf{0.82} & \textbf{0.86} \\
\hline
\end{tabularx}
\end{table*}

\begin{table*}[t]
    \centering
    \small
    \caption{%
    F1-scores and accuracy with four aggregation methods on the 3D-FRONT dataset~\cite{3dfront}.
    Higher is better ($\uparrow$).
    } \label{tab:3d-front-f1-acc}
    \vspace{-\baselineskip}
    \noindent
    \begin{minipage}[t]{0.48\textwidth}
    \begin{tabularx}{\linewidth}{@{}>{\raggedright\arraybackslash}p{2.3cm}|*{5}{>{\centering\arraybackslash}X}|>{\raggedleft\arraybackslash}X@{}}
    \multicolumn{7}{@{}l@{}}{\textbf{3D-FRONT (F1-Score $\uparrow$})} \\
    \noalign{\global\arrayrulewidth=1.5pt}
    \hline
    \noalign{\global\arrayrulewidth=0.4pt}
    \textbf{Method} & T27 & T28 & T29 & T30 & T31 & Avg \\
    \hline
    Naive Majority & 0.88 & \textbf{0.73} & 0.15 & 0.00 & 0.04 & 0.36 \\
    Logistic Regression & 0.86 & 0.55 & 0.37 & 0.52 & 0.20 & 0.50 \\
    Top-1 & 0.85 & 0.18 & 0.77 & 0.50 & 0.15 & 0.49 \\
    Weaver & \textbf{0.88} & 0.70 & \textbf{0.80} & \textbf{0.57} & \textbf{0.24} & \textbf{0.64} \\
    \hline
    \end{tabularx}
    \end{minipage}\hfill
    \begin{minipage}[t]{0.48\textwidth}
    \begin{tabularx}{\linewidth}{@{}>{\raggedright\arraybackslash}p{2.3cm}|*{5}{>{\centering\arraybackslash}X}|>{\raggedleft\arraybackslash}X@{}}
    \multicolumn{7}{@{}l@{}}{\textbf{3D-FRONT (Accuracy $\uparrow$)}} \\
    \noalign{\global\arrayrulewidth=1.5pt}
    \hline
    \noalign{\global\arrayrulewidth=0.4pt}
    \textbf{Method} & T27 & T28 & T29 & T30 & T31 & Avg \\
    \hline
    Naive Majority & 0.88 & 0.90 & 0.87 & 0.86 & 0.92 & 0.89 \\
    Logistic Regression & 0.87 & 0.87 & 0.89 & 0.84 & 0.93 & 0.88 \\
    Top-1 & 0.87 & 0.96 & 0.94 & 0.87 & 0.94 & 0.92 \\
    Weaver & 0.88 & 0.85 & 0.95 & 0.81 & 0.67 & 0.83 \\
    \hline
    \end{tabularx}
    \end{minipage}
    \vspace{-1em}
\end{table*}

\begin{figure*}
    \centering
    \includegraphics[width=\linewidth]{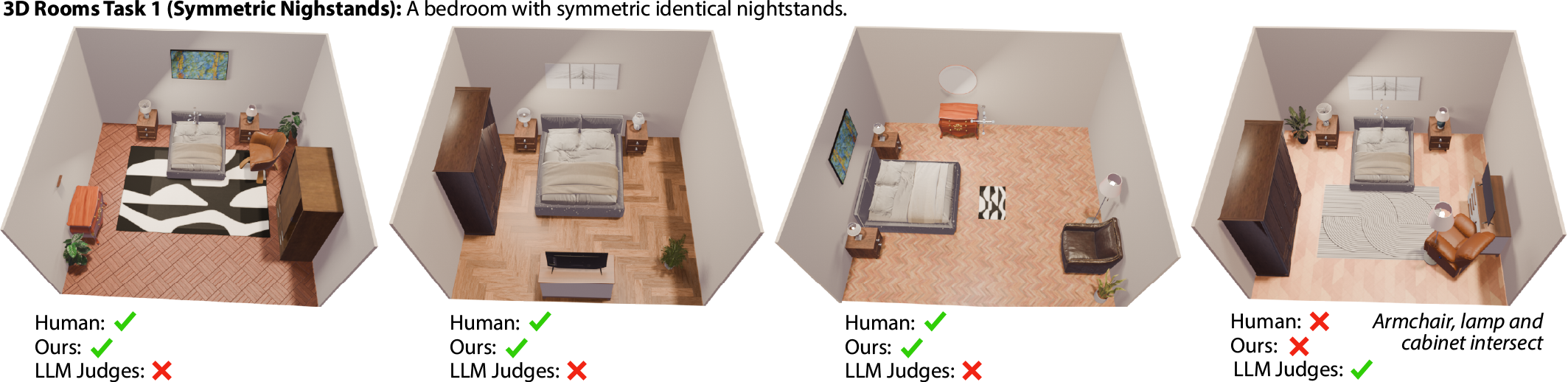}
    \caption{The LLM judges incorrectly reject the three positive layouts on the left and accept the negative layouts on the right. In negative layout, three major pieces of furniture in the bottom-right corner intersect.}
    \label{fig:scene3d-task1}
\end{figure*}

\begin{figure*}
    \centering
    \includegraphics[width=\linewidth]{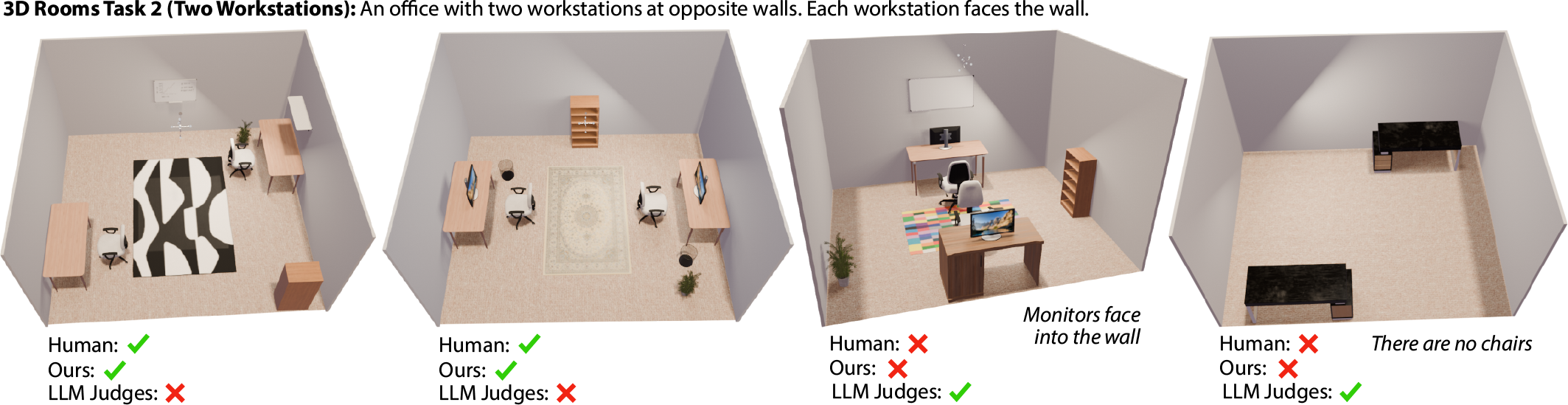}
    \caption{The LLM judges incorrectly reject the two positive layouts on the left and accept the two negative layouts on the right. In the first negative layout, the workstations are missing crucial items like a chair pointed at the desk. In the second negative layout, the workstation furniture fails to follow common sense arrangements, monitors pointed towards the chair.}
    \label{fig:scene3d-task2}
\end{figure*}

\begin{figure*}
    \centering
    \includegraphics[width=\linewidth]{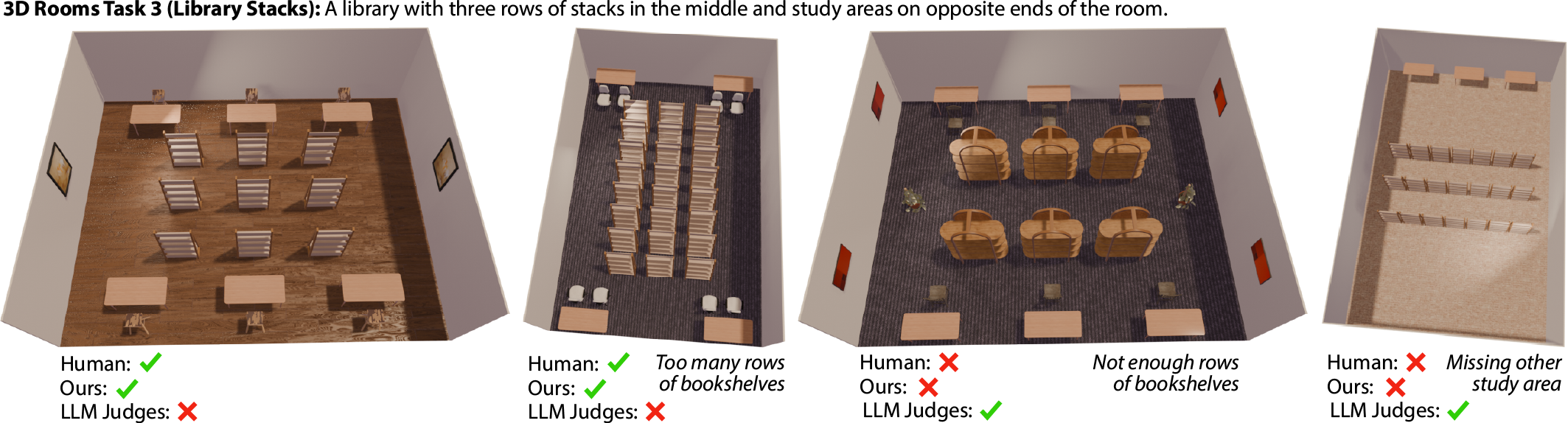}
    \caption{The LLM judges incorrectly reject the two positive layouts on the left and accept the two negative layouts on the right. In the first two negative layouts, the number of rows of stacks in the middle of the room is too many or too few. In the first negative layout, there are not enough rows of bookshelves. In the second negative layout, the study area is missing chairs to sit in and there is no study area on the opposing end of the room.}
    \label{fig:scene3d-task3}
\end{figure*}

\begin{figure*}
    \centering
    \includegraphics[width=\linewidth]{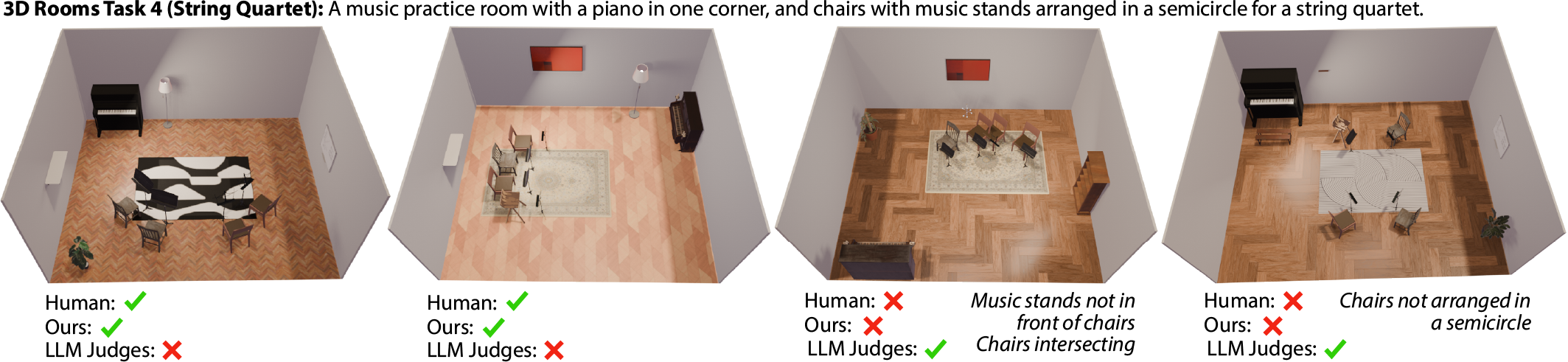}
    \caption{The LLM judges incorrectly reject the two positive layouts on the left and accept the two negative layouts on the right. In the first negative layout, the music stands are not placed in front of the chairs so that musicians can use them. There are also two intersecting chairs. In the second layout, the chairs are not arranged in the requested semicircle.}
    \label{fig:scene3d-task4}
\end{figure*}

\begin{figure*}
    \centering
    \includegraphics[width=\linewidth]{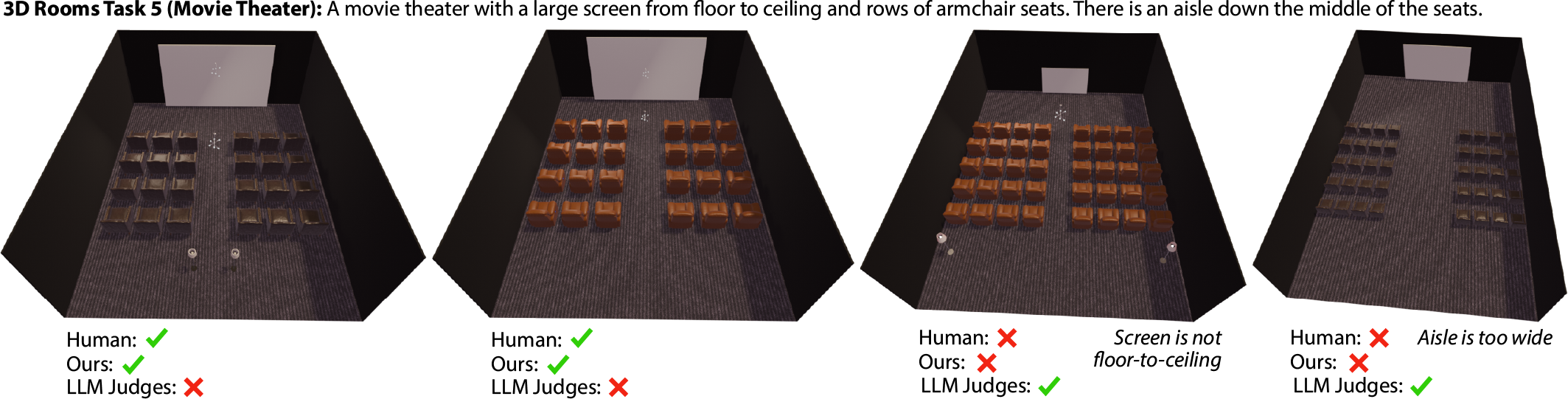}
    \caption{The LLM judges incorrectly reject the two positive layouts on the left and accept the two negative layouts on the right. In the first negative layout, the screen does not go from floor-to-ceiling. In the second negative layout, the aisle is too wide, so the majority of the seats are facing into the wall.}
    \label{fig:scene3d-task5}
\end{figure*}

\begin{figure*}
    \centering
    \includegraphics[width=\linewidth]{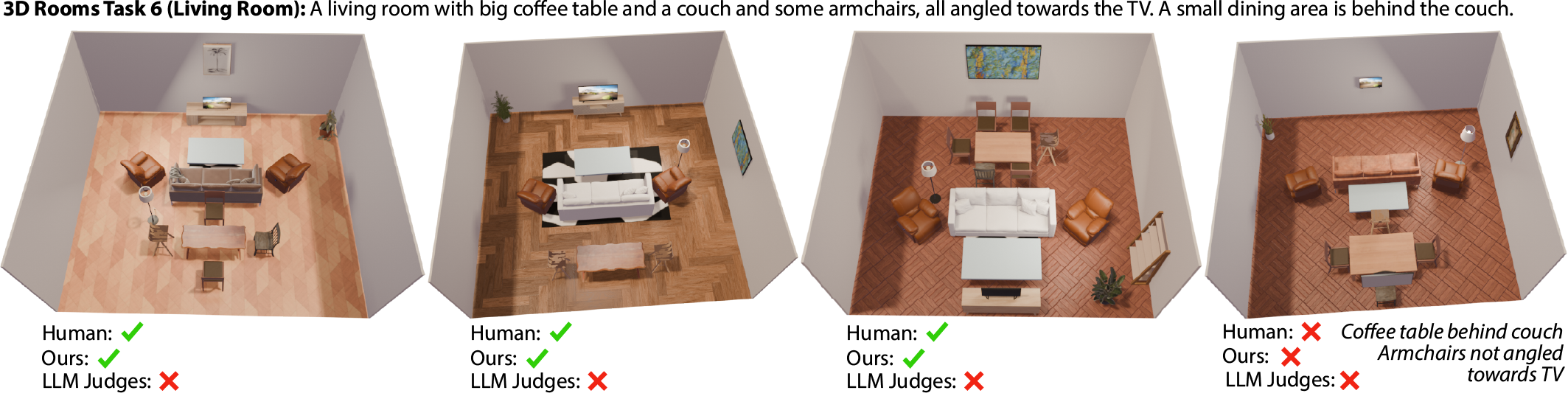}
    \caption{The LLM judges incorrectly reject the three positive layouts on the left and accept the negative layout on the right. In the negative layout, the armchairs are not angled towards the TV, and the coffee table is behind the couch, which does not follow a common sense placement.}
    \label{fig:scene3d-task6}
\end{figure*}

\begin{figure*}
    \centering
    \includegraphics[width=\linewidth]{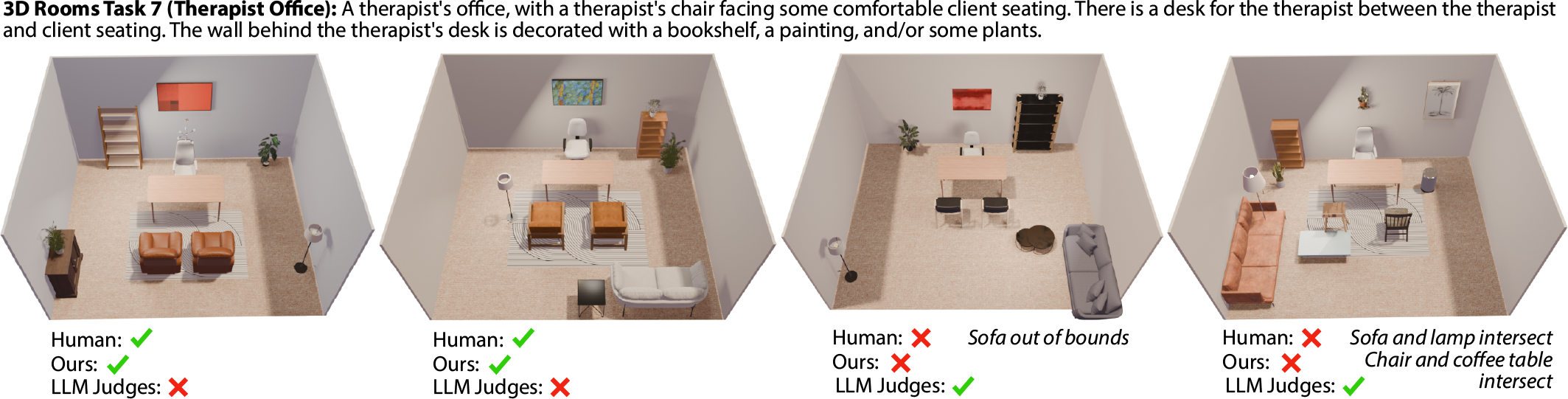}
    \caption{The LLM judges incorrectly reject the two positive layouts on the left and accept the two negative layouts on the right. In the first negative layout, the sofa is out of bounds. In the second negative layout, there are major furniture intersections.}
    \label{fig:scene3d-task7}
\end{figure*}

\begin{figure*}
    \centering
    \includegraphics[width=\linewidth]{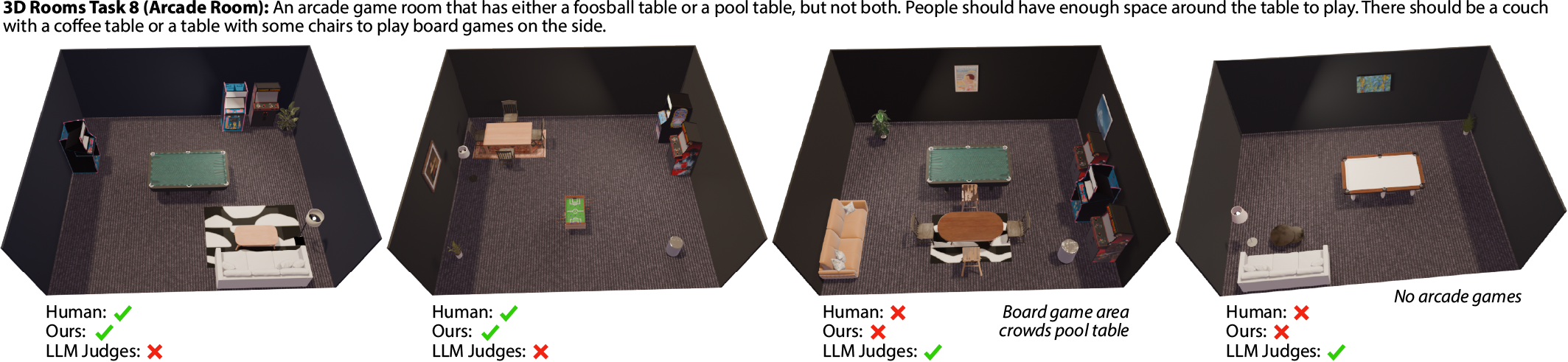}
    \caption{The LLM judges incorrectly reject the two positive layouts on the left and accept the two negative layouts on the right. In the first negative layout, the dining table area crowd the pool table and do not leave enough space to play. In the second negative layout, there is no indication that the room is an arcade game room.}
    \label{fig:scene3d-task8}
\end{figure*}

\begin{figure*}
    \centering
    \includegraphics[width=\linewidth]{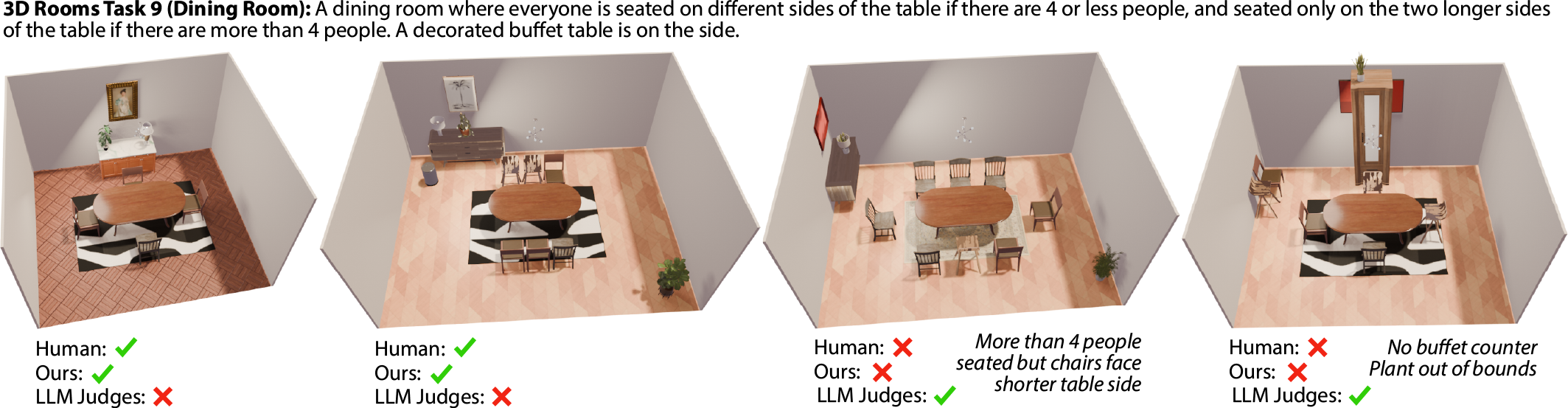}
    \caption{The LLM judges incorrectly reject the two positive layouts on the left and accept the two negative layouts on the right. In the first negative layout, there are more than 4 people seated but two chairs are pointed at the shorter sides of the table. In the second negative layout, there is no buffet counter but instead an extremely tall cabinet.}
    \label{fig:scene3d-task9}
\end{figure*}

\begin{figure*}
    \centering
    \includegraphics[width=\linewidth]{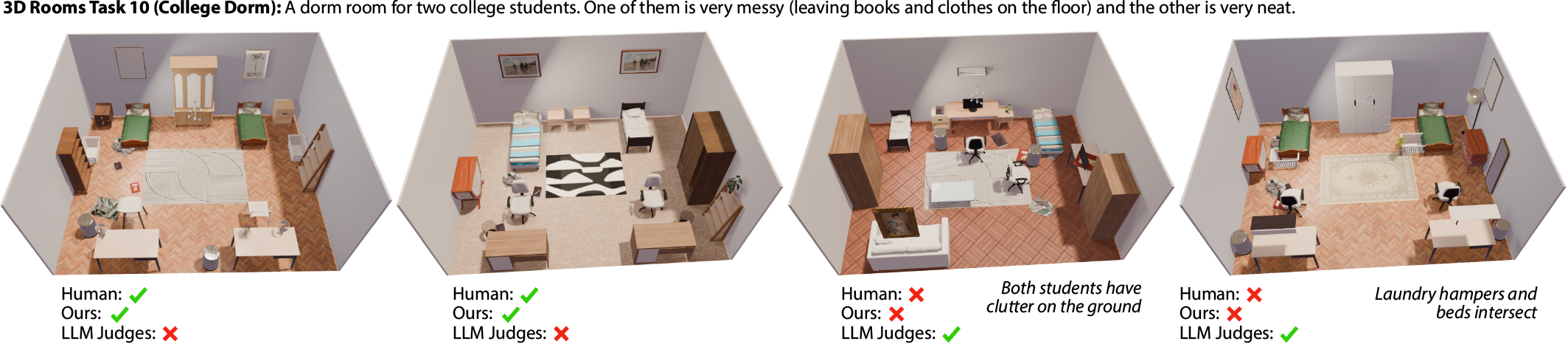}
    \caption{The LLM judges incorrectly reject the two positive layouts on the left and accept the two negative layouts on the right. In both negative examples there are many major object intersections, such as between the wardrobes and chairs or the desk and nightstands. In the first negative example, the paintings are also oriented perpendicularly to the walls, which is not physically plausible. In the second negative example, it is not entirely clear that one student is very neat, as the clutter on the floor is close to both student desks. These requirements are expressed through the task description and the user's dev set notes (see supplemental).}
    \label{fig:scene3d-task10}
\end{figure*}

\begin{figure*}
    \centering
    \includegraphics[width=\linewidth]{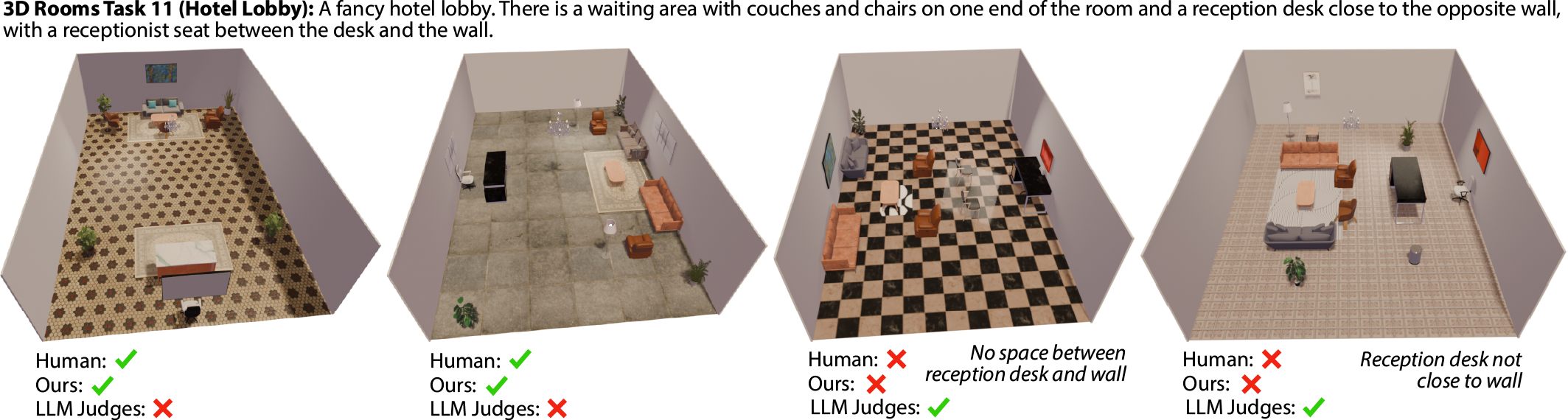}
    \caption{The LLM judges incorrectly reject the two positive layouts on the left and accept the two negative layouts on the right. In the first negative layout the receptionist seat is not between the wall and the reception desk, rather it is completely overlapped by the reception desk. In the second negative layout the receptionist desk is far from the wall, and the waiting area and the reception area are not clearly on opposite ends of the room.}
    \label{fig:scene3d-task11}
\end{figure*}

\begin{figure*}
    \centering
    \includegraphics[width=\linewidth]{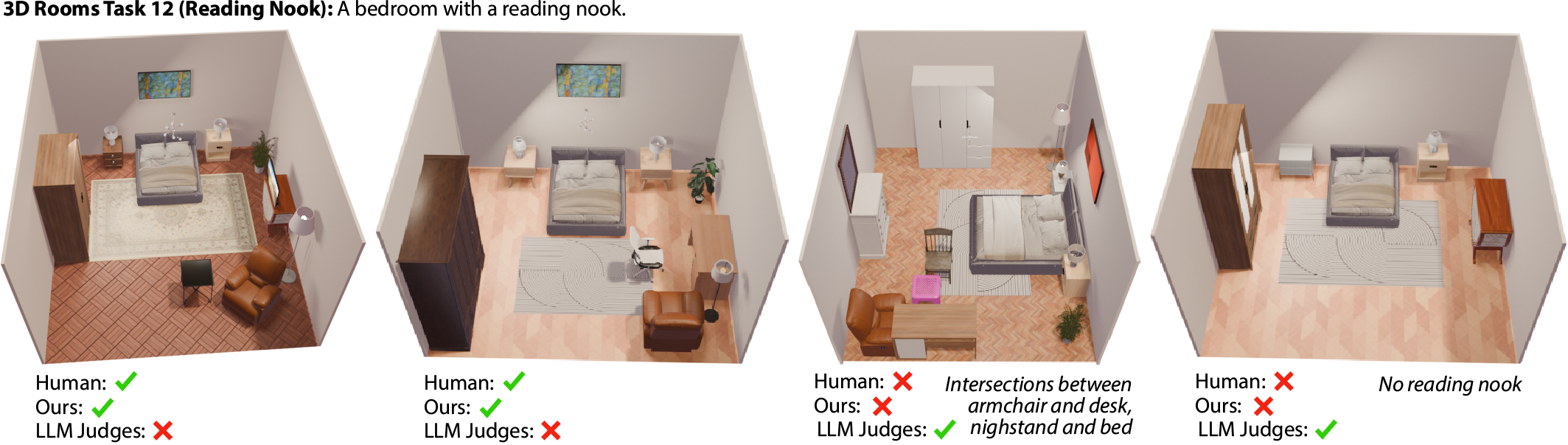}
    \caption{The LLM judges incorrectly reject the two positive layouts on the left and accept the two negative layouts on the right. In the first negative example are major furniture intersections. In the second example, there is no reading nook (e.g., an armchair with a light by it).}
    \label{fig:scene3d-task12}
\end{figure*}

\begin{figure*}
    \centering
    \includegraphics[width=\linewidth]{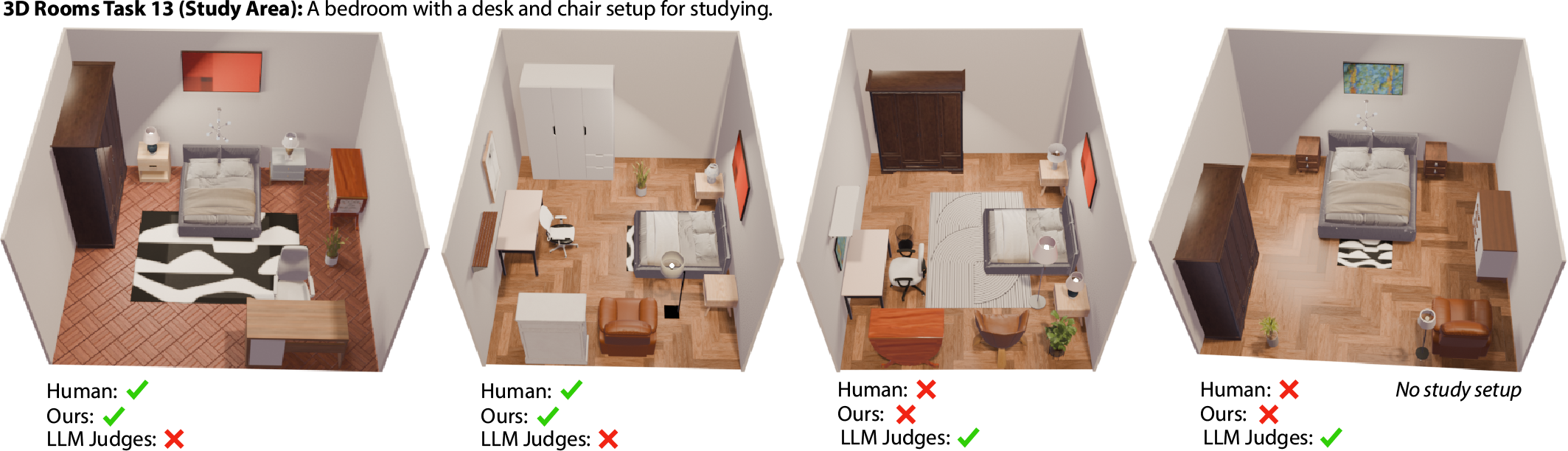}
    \caption{The LLM judges incorrectly reject all three positive layouts on the left, even though they clearly have desk and chairs set up.}
    \label{fig:scene3d-task13}
\end{figure*}

\begin{figure*}
    \centering
    \includegraphics[width=\linewidth]{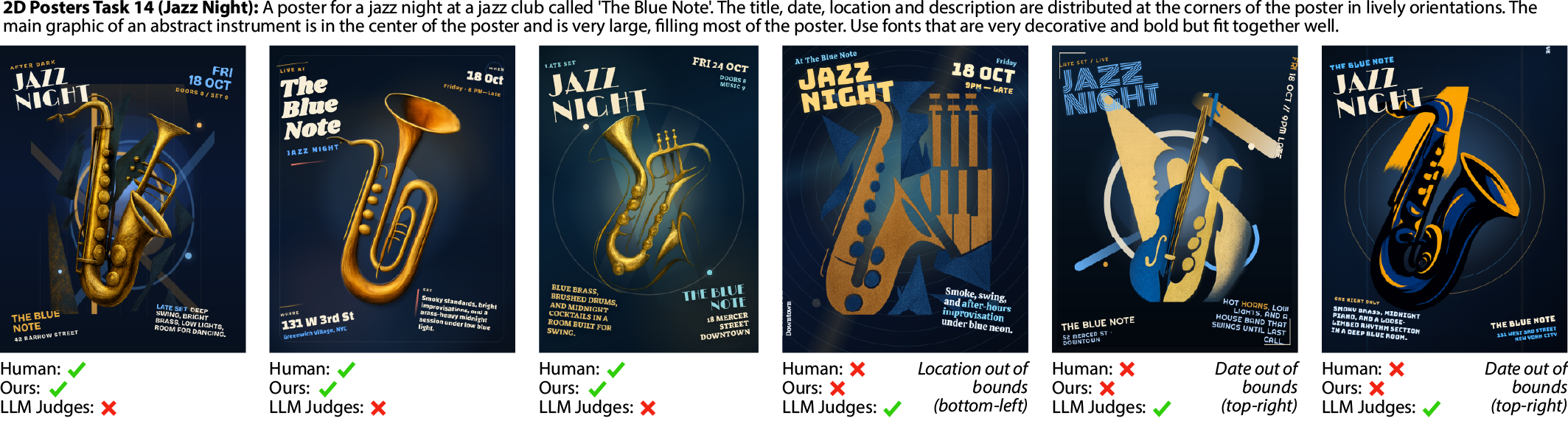}
    \caption{The LLM judges incorrectly reject the two positive layouts on the left and accept the two negative layouts on the right. In the first negative layout, the location is out of bounds. In the second negative layout, the date is both out of bounds and hard to read due to extreme rotation.}
    \label{fig:poster2d-task1}
\end{figure*}

\begin{figure*}
    \centering
    \includegraphics[width=\linewidth]{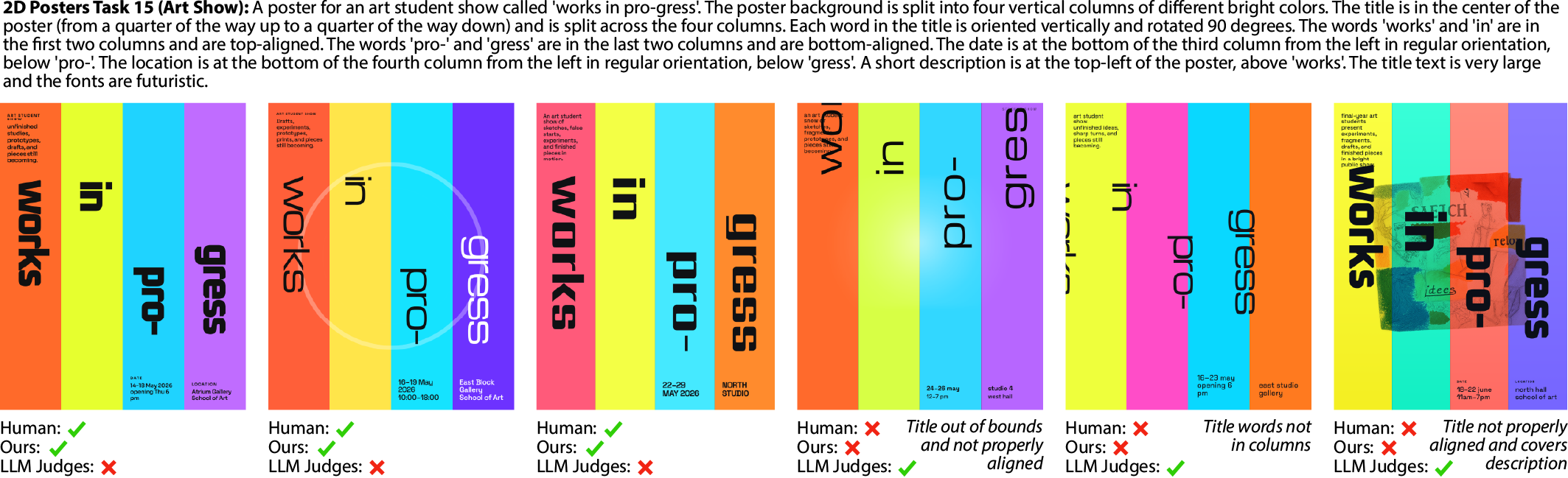}
    \caption{The LLM judge incorrectly reject the two positive layouts on the left and accept the two negative layouts on the right. In the first negative layout the title words are not properly top-aligned and bottom-aligned. In the second negative layout, the title words are not contained in the four columns.}
    \label{fig:poster2d-task2}
\end{figure*}

\begin{figure*}
    \centering
    \includegraphics[width=\linewidth]{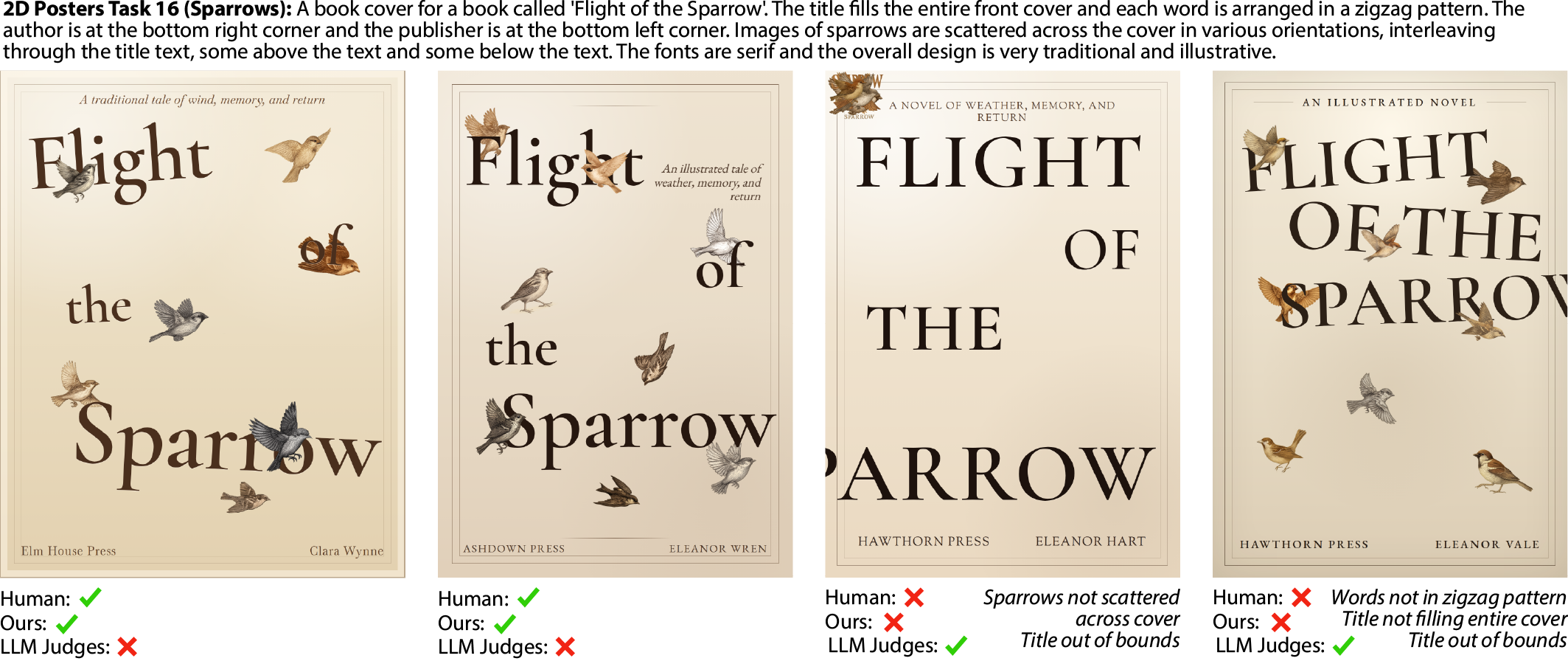}
    \caption{The LLM judges incorrectly reject the two positive layouts on the left and accept the two negative layouts on the right. In both negative layouts the title is our of bounds. In the first negative layout, the sparrows are not scattered across the cover. In the second negative layout, the title words are not individually arranged in a zig-zag pattern, and they do not fill the entire cover.}
    \label{fig:poster2d-task3}
\end{figure*}

\begin{figure*}
    \centering
    \includegraphics[width=\linewidth]{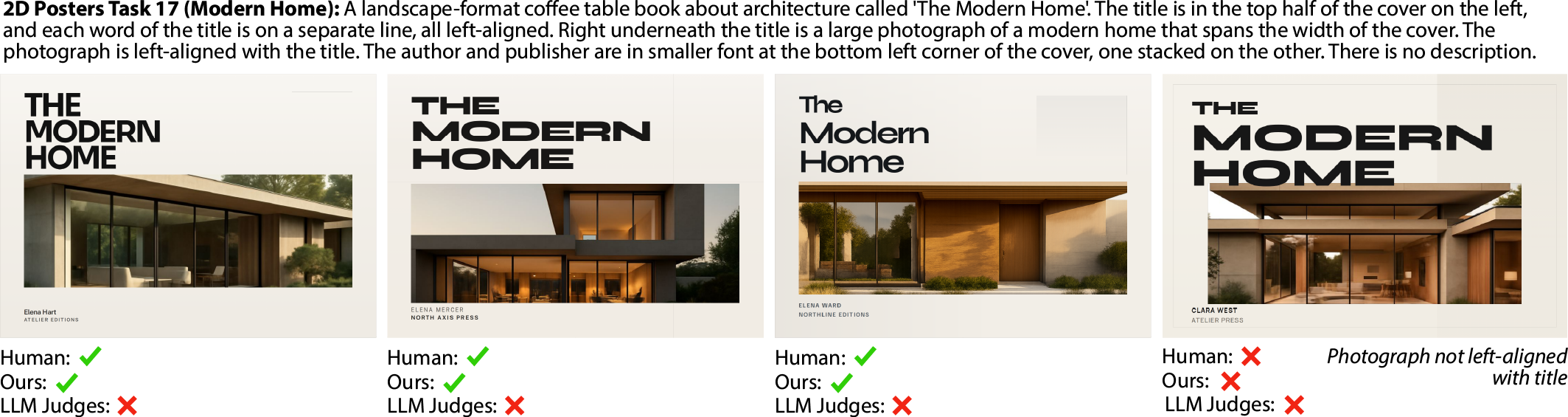}
    \caption{The LLM judges incorrectly reject the three positive layouts on the left, even though they satisfy all layout criteria.}
    \label{fig:poster2d-task4}
\end{figure*}

\begin{figure*}
    \centering
    \includegraphics[width=\linewidth]{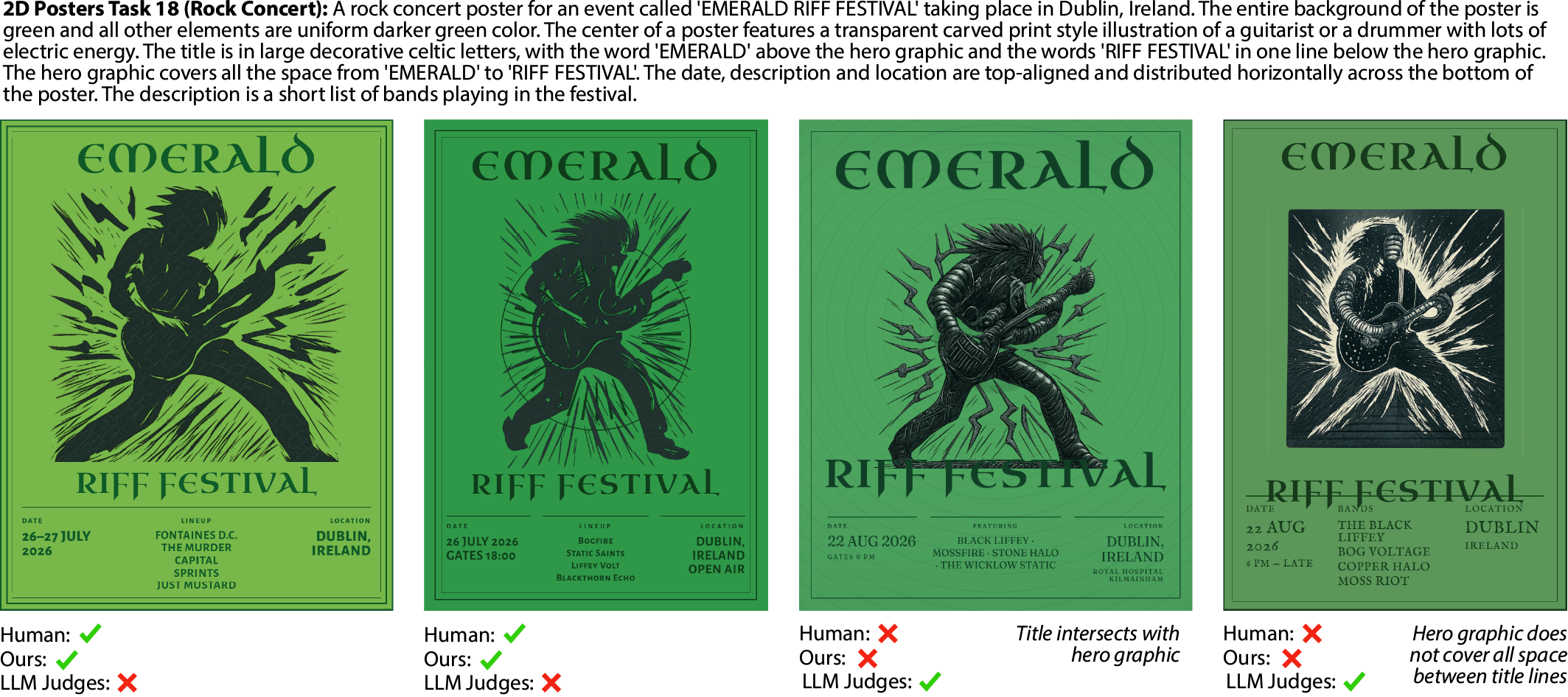}
    \caption{The LLM judges incorrectly reject the two positive layouts on the left and accept the two negative layouts on the right. In the first negative layout, the title and hero graphic overlap. In the second negative layout, the hero graphic does not fill the space between the two title lines.}
    \label{fig:poster2d-task5}
\end{figure*}

\begin{figure*}
    \centering
    \includegraphics[width=\linewidth]{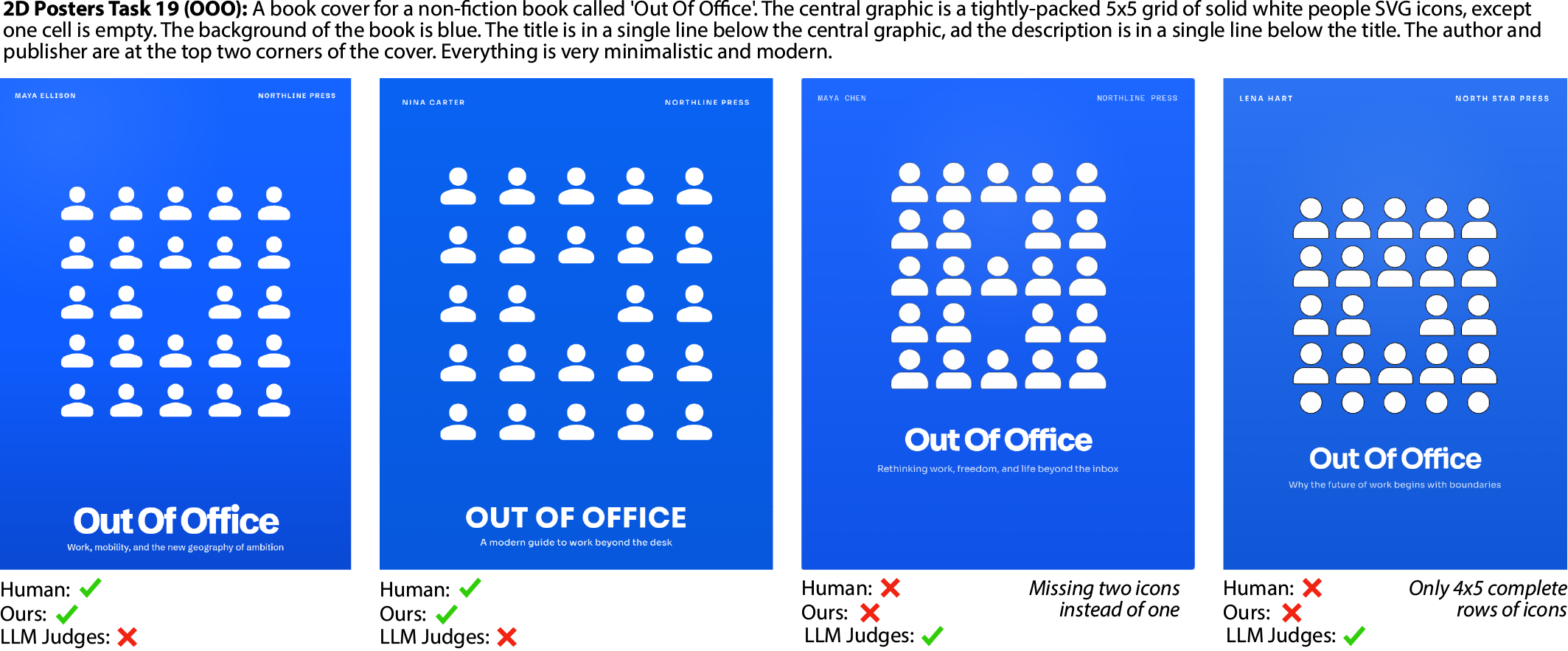}
    \caption{The LLM judges incorrectly reject the two positive layouts on the left and accept the two negative layouts on the right. In the first negative layout, there are two missing icons in the 5x5 grid rather than one. In the second negative layout, there are only 4x5 complete icon grids.}
    \label{fig:poster2d-task6}
\end{figure*}

\begin{figure*}
    \centering
    \includegraphics[width=\linewidth]{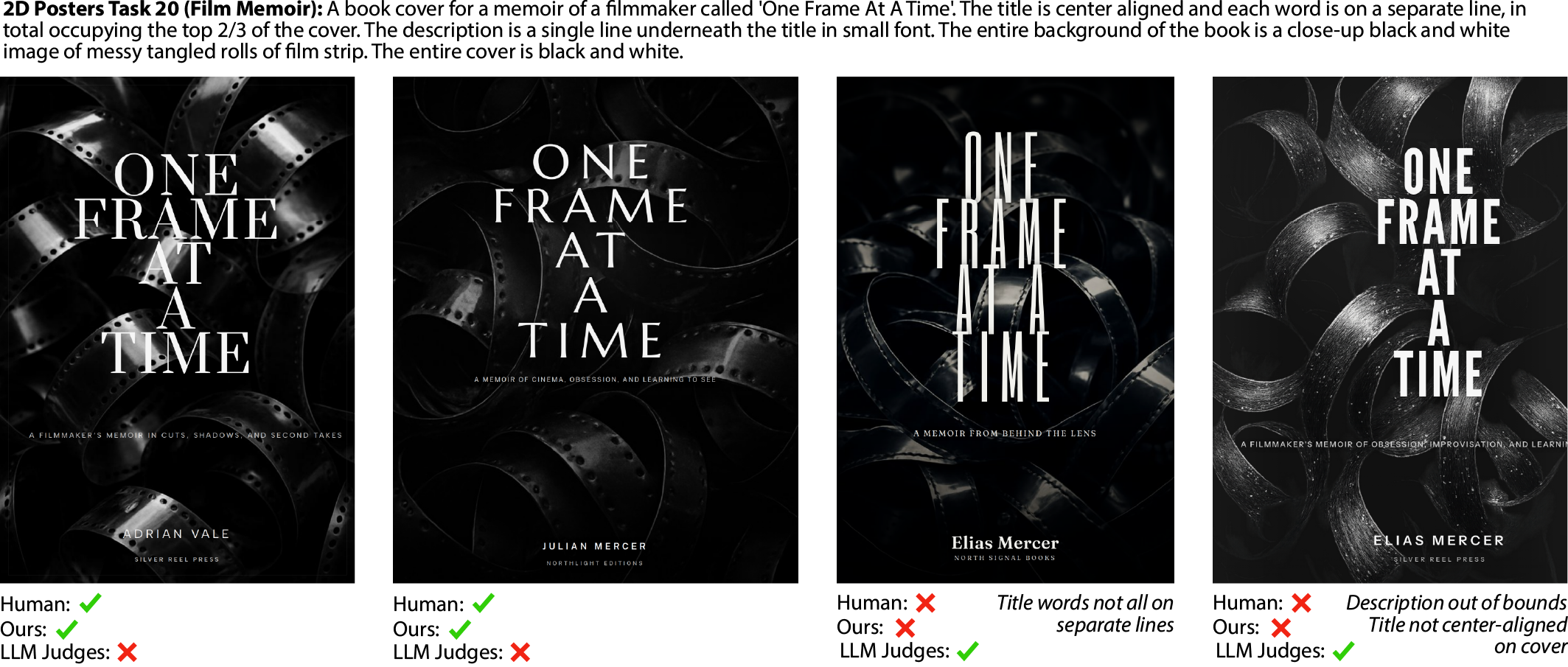}
    \caption{The LLM judges incorrectly reject the two positive layouts on the left and incorrectly accept the two negative layouts on the right. In the first negative layout, the title words 'At' and 'A' are not on separate lines. These requirements are expressed in the task description. In the second negative layout, the title is not centered-aligned with the cover and the description goes out of bounds.}
    \label{fig:poster2d-task7}
\end{figure*}

\begin{figure*}
    \centering
    \includegraphics[width=\linewidth]{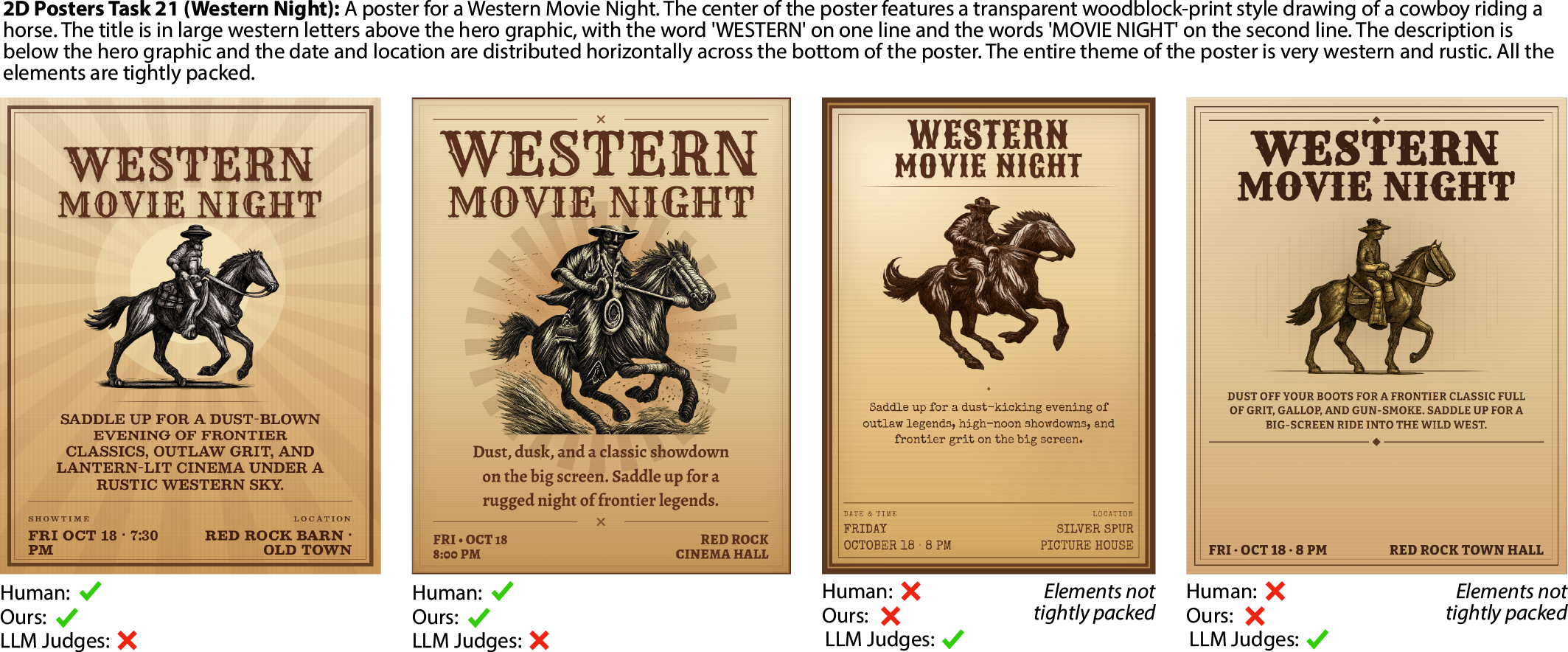}
    \caption{The LLM judges incorrectly reject the two positive layouts on the left and accept the two negative layouts on the right. In both negative layouts, the text elements are not packed tightly enough.}
    \label{fig:poster2d-task8}
\end{figure*}

\begin{figure*}
    \centering
    \includegraphics[width=\linewidth]{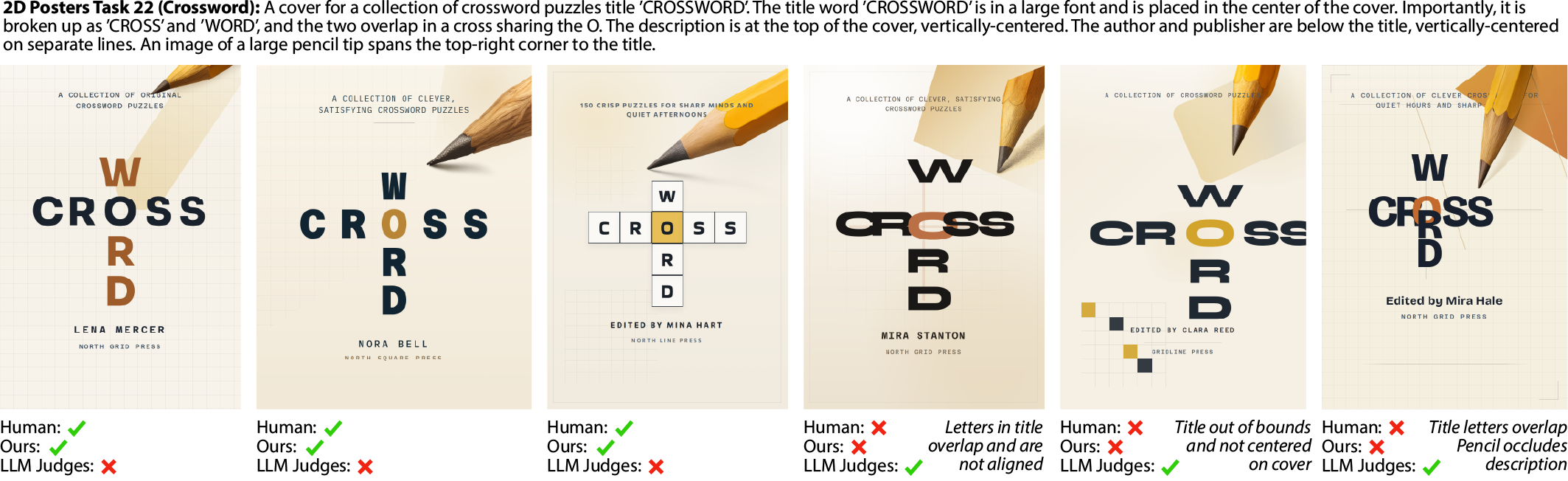}
    \caption{The LLM judges incorrectly reject the two positive layouts on the left and accept the two negative layouts on the right. In the first negative layout, the letters in 'CROSSWORD' are off-center and spaced inconsistently. In the second negative layout, the title is not centered on the cover and extends out of bounds.}
    \label{fig:poster2d-task9}
\end{figure*}

\begin{figure*}
    \centering
    \includegraphics[width=\linewidth]{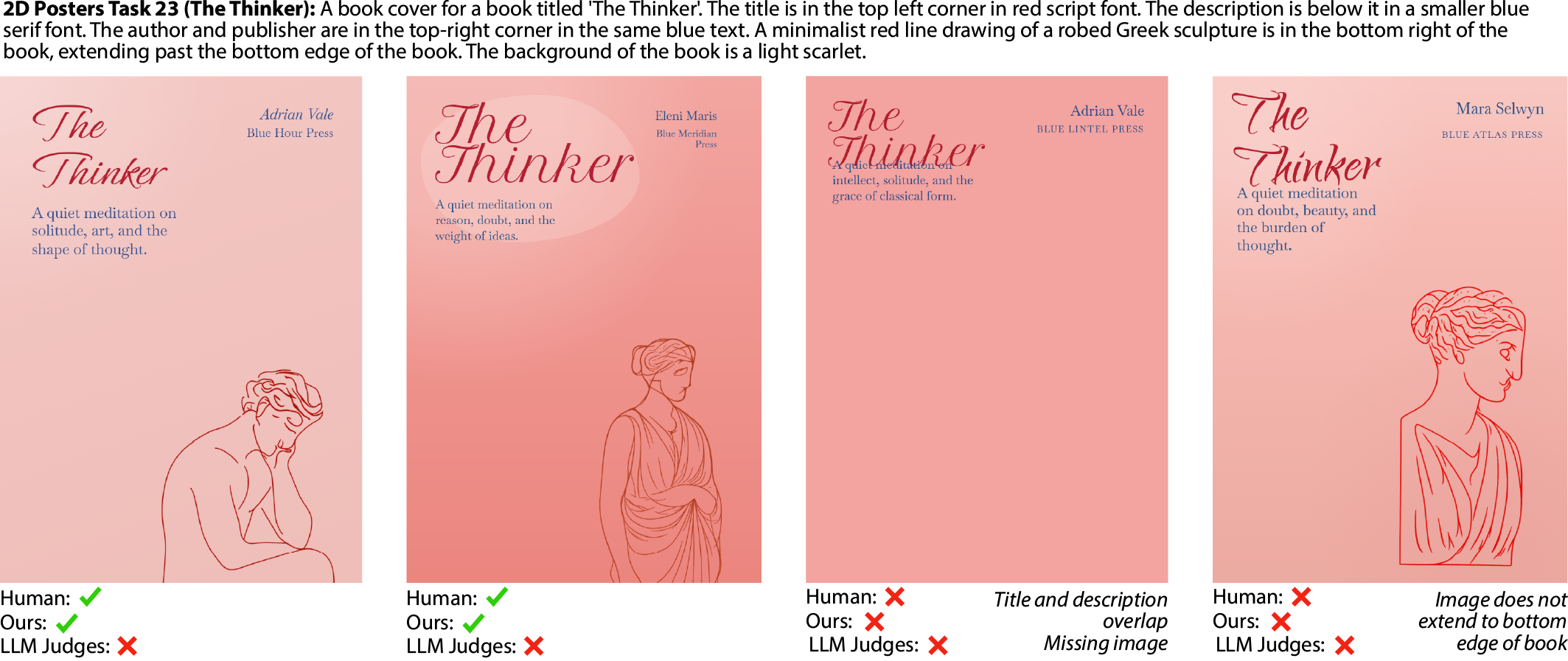}
    \caption{The LLM judges vote overly negative, incorrectly rejecting the two positive layouts on the left, even though they satisfy all layout requirements.}
    \label{fig:poster2d-task10}
\end{figure*}

\begin{figure*}
    \centering
    \includegraphics[width=\linewidth]{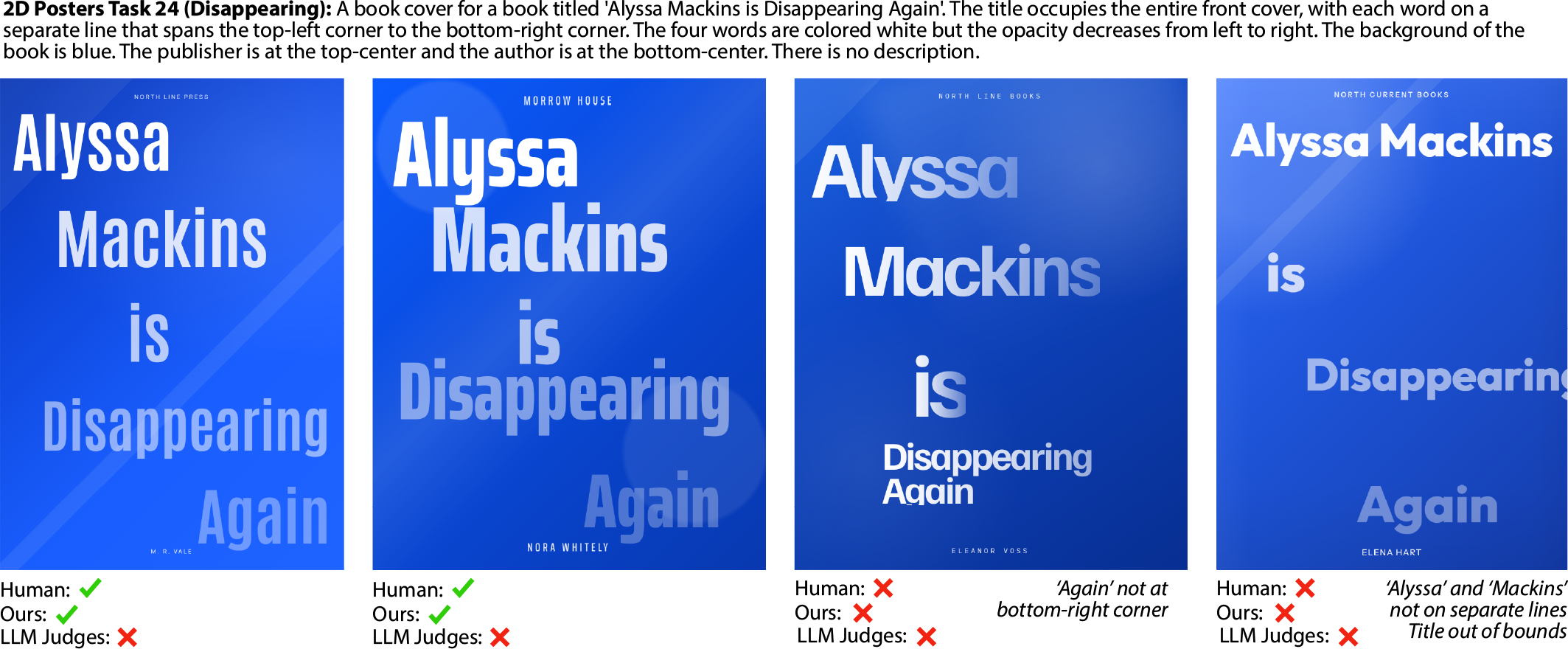}
    \caption{The LLM judges vote overly negative, incorrectly rejecting the two positive layouts on the left, even though they satisfy all layout requirements.}
    \label{fig:poster2d-task11}
\end{figure*}

\begin{figure*}
    \centering
    \includegraphics[width=\linewidth]{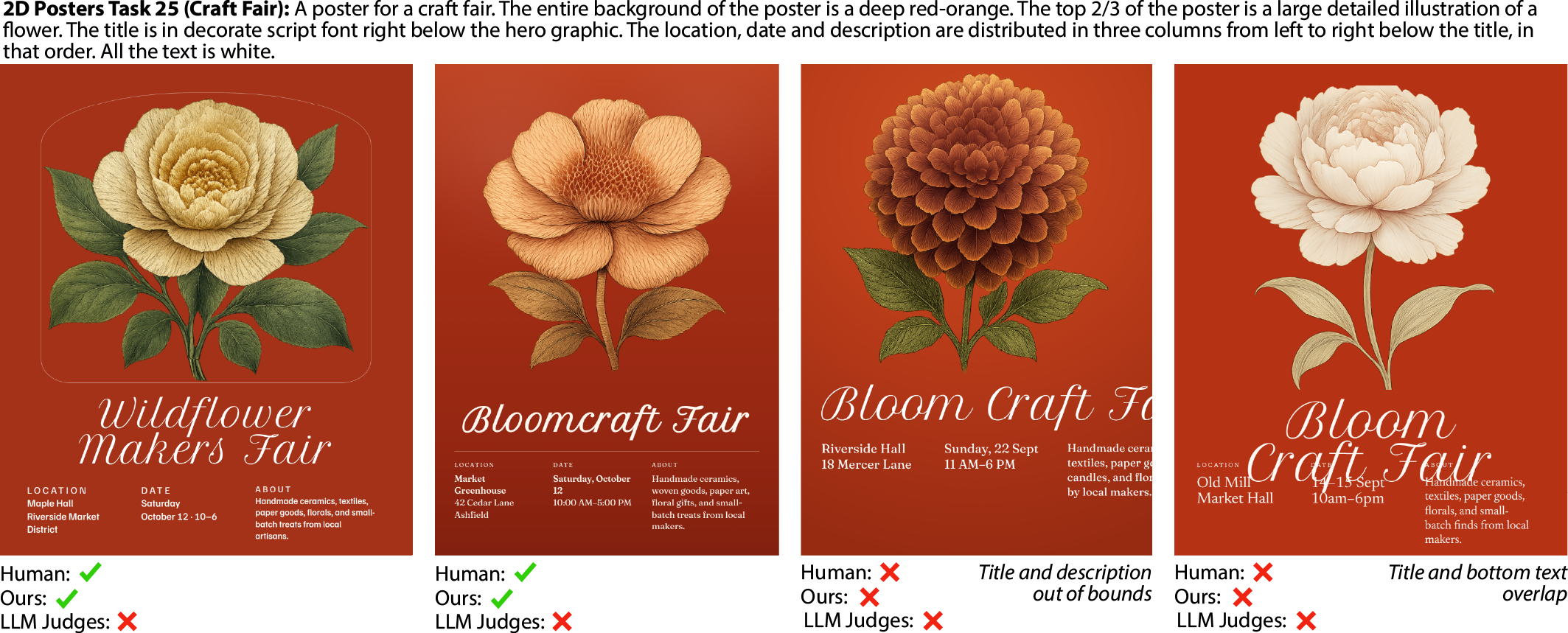}
    \caption{The LLM judges vote overly negative, incorrectly rejecting the two positive layouts on the left, even though they satisfy all layout requirements.}
    \label{fig:poster2d-task12}
\end{figure*}

\begin{figure*}
    \centering
    \includegraphics[width=\linewidth]{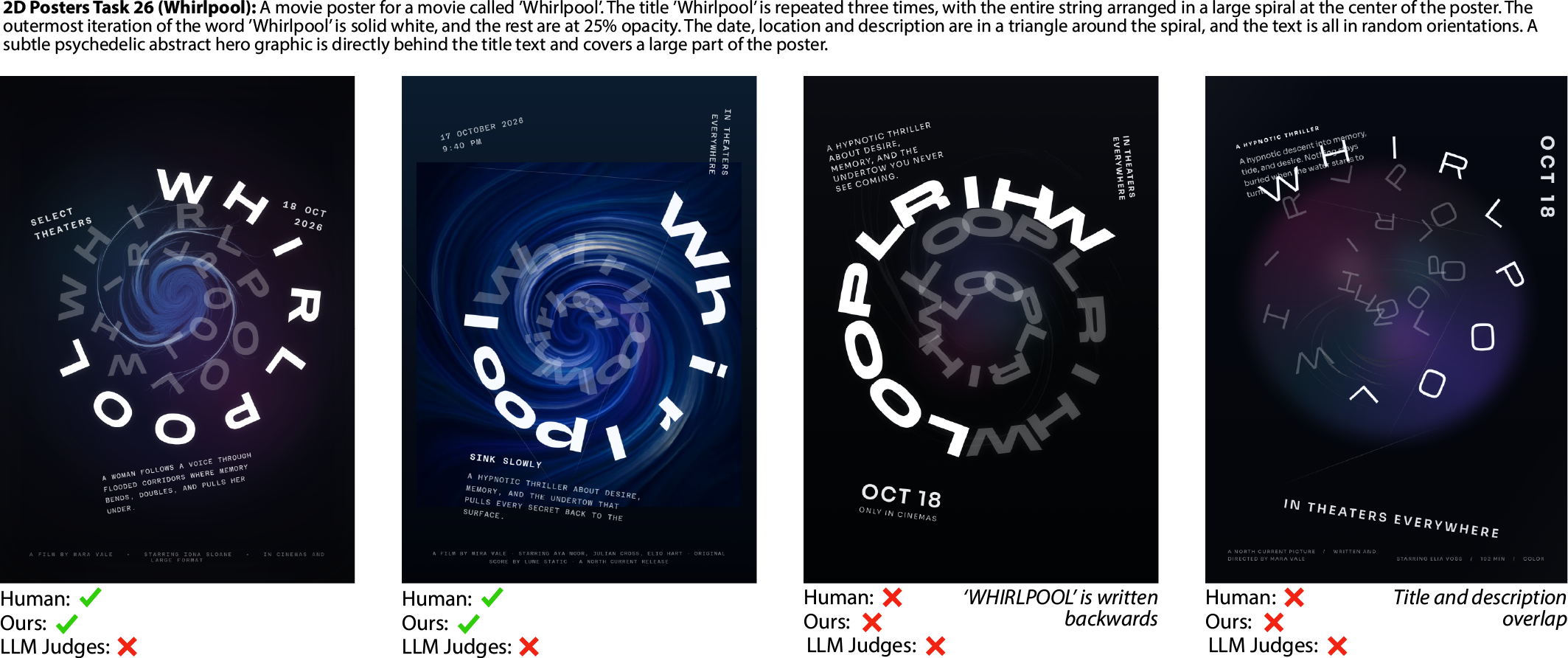}
    \caption{The LLM judges vote overly negative, incorrectly rejecting the two positive layouts on the left, even though they satisfy all layout requirements.}
    \label{fig:poster2d-task13}
\end{figure*}

\begin{table*}[t]
    \centering
    \small
    \caption{%
    Precision and recall with four aggregation methods on the 3D-FRONT dataset~\cite{3dfront}.
    Higher is better ($\uparrow$).
    } \label{tab:3d-front-prec-rec}
    \vspace{-\baselineskip}
    \noindent
    \begin{minipage}[t]{0.48\textwidth}
    \begin{tabularx}{\linewidth}{@{}>{\raggedright\arraybackslash}p{2.3cm}|*{5}{>{\centering\arraybackslash}X}|>{\raggedleft\arraybackslash}X@{}}
    \multicolumn{7}{@{}l@{}}{\textbf{3D-FRONT (Precision $\uparrow$)}} \\
    \noalign{\global\arrayrulewidth=1.5pt}
    \hline
    \noalign{\global\arrayrulewidth=0.4pt}
    \textbf{Method} & T27 & T28 & T29 & T30 & T31 & Avg \\
    \hline
    Naive Majority      & 0.78 & 0.78 & 0.96 & 0.00 & 0.50 & 0.60 \\
    Logistic Regression & 0.80 & 0.74 & 0.92 & 0.48 & 0.40 & 0.67 \\
    Top-1               & 0.84 & 0.17 & 0.85 & 0.52 & 0.12 & 0.50 \\
    Weaver              & 0.78 & 0.57 & 0.88 & 0.46 & 0.19 & 0.58 \\
    \hline
    \end{tabularx}
    \end{minipage}\hfill
    \begin{minipage}[t]{0.48\textwidth}
    \begin{tabularx}{\linewidth}{@{}>{\raggedright\arraybackslash}p{2.3cm}|*{5}{>{\centering\arraybackslash}X}|>{\raggedleft\arraybackslash}X@{}}
    \multicolumn{7}{@{}l@{}}{\textbf{3D-FRONT (Recall $\uparrow$)}} \\
    \noalign{\global\arrayrulewidth=1.5pt}
    \hline
    \noalign{\global\arrayrulewidth=0.4pt}
    \textbf{Method} & T27 & T28 & T29 & T30 & T31 & Avg \\
    \hline
    Naive Majority      & 1.00 & 0.70 & 0.08 & 0.00 & 0.02 & 0.54 \\
    Logistic Regression & 0.94 & 0.55 & 0.26 & 0.76 & 0.15 & 0.71 \\
    Top-1               & 0.89 & 0.19 & 0.74 & 0.60 & 0.19 & 0.71 \\
    Weaver              & 1.00 & 0.90 & 0.76 & 0.84 & 0.32 & 0.93 \\
    \hline
    \end{tabularx}
    \end{minipage}
    \vspace{-1em}
\end{table*}
Tables~\ref{tab:prec},~\ref{tab:rec} and~\ref{tab:acc} provides additional precision, recall and accuracy results for each of the tasks.
Figures~\ref{fig:scene3d-task1}--\ref{fig:scene3d-task13} show positive and negative layout examples for our 3D Rooms tasks and Figures~\ref{fig:poster2d-task1}--\ref{fig:poster2d-task13} show positive and negative layout examples for our 2D Posters tasks. 
Sections~\ref{subsec:lrt1} and~\ref{subsec:weak-verifiers} provide additional results from our analysis of Logistic Regression, Top-1 and Naive Majority discussed in Section~4.3 of the main paper.
Section~\ref{subsec:3dfront} discusses additional tasks in which we are able to amortize dataset generation (Stage 1 of our verification pipeline) by taking advantage of a pre-existing dataset.

\clearpage
\subsection{Logistic Regression and Top-1 are unreliable on small dev sets}\label{subsec:lrt1}
\begin{figure*}
    \centering
    \includegraphics[width=0.9\linewidth]{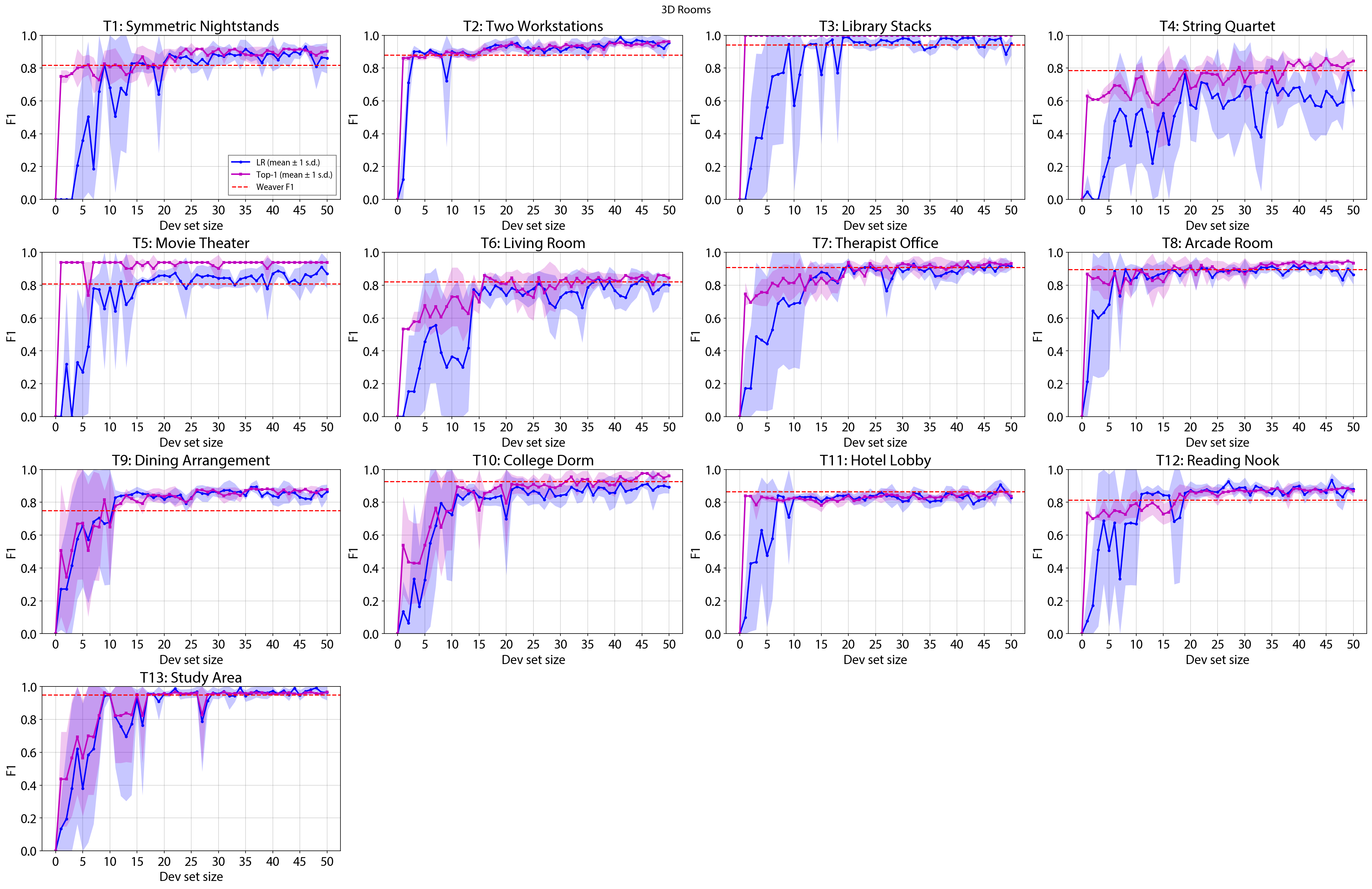}
    \includegraphics[width=0.9\linewidth]{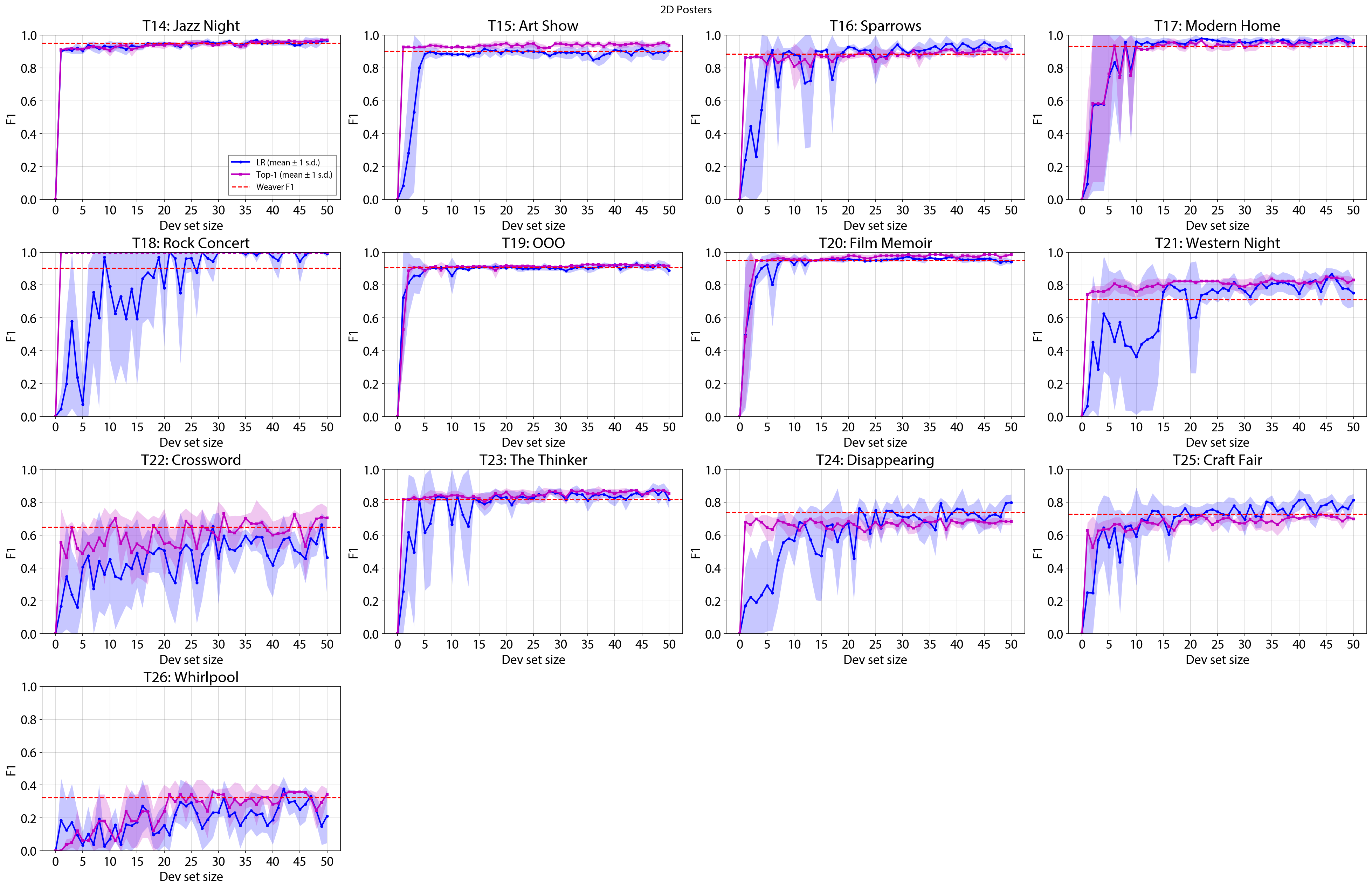}
    \caption{Performance of Logistic Regression (blue) and Top-1 (purple) as dev set size increases.  In most cases, increasing the dev set size allows Logistic Regression and Top-1 to match the performance of Weaver. In several cases, the performance of these two aggregation methods still exhibits high variance with respect to the dev set composition even with large dev set sizes (e.g., {\bf T4: String Quartet}, {\bf T6: Living Room}, {\bf T10: College Dorm}, {\bf T22: Crossword}, {\bf T26: Whirlpool}). In those tasks, even 50 dev set examples is not enough for Logistic Regression to outperform Weaver.}
    \label{fig:lrt1}
\end{figure*}
Since Logistic Regression and Top-1 use the dev set to learn a weighting of the weak verifiers, we investigate how these two methods perform as we increase the size of the dev set in Stage 3: Weak verifier aggregation. For dev set sizes from $0, \ldots, 50$, we randomly sample five different dev sets compute aggregation weights for each dev set. Figure~\ref{fig:lrt1} reports the mean/std F1-score of the resulting aggregate verifier and compares them against the F1-score of our strong Weaver verifier. We see that in many cases, the performance of these two aggregation methods still exhibits high variance with respect to the dev set composition even with large dev set sizes (e.g., {\bf T4: String Quartet}, {\bf T6: Living Room}, {\bf T10: College Dorm}, {\bf T22: Crossword}, {\bf T26: Whirlpool}). In those tasks, even 50 dev set examples is not enough for Logistic Regression to outperform Weaver.

\subsection{Naive Majority is susceptible to low recall weak verifiers}\label{subsec:weak-verifiers}
\begin{figure}
    \centering
    \includegraphics[width=0.5\linewidth]{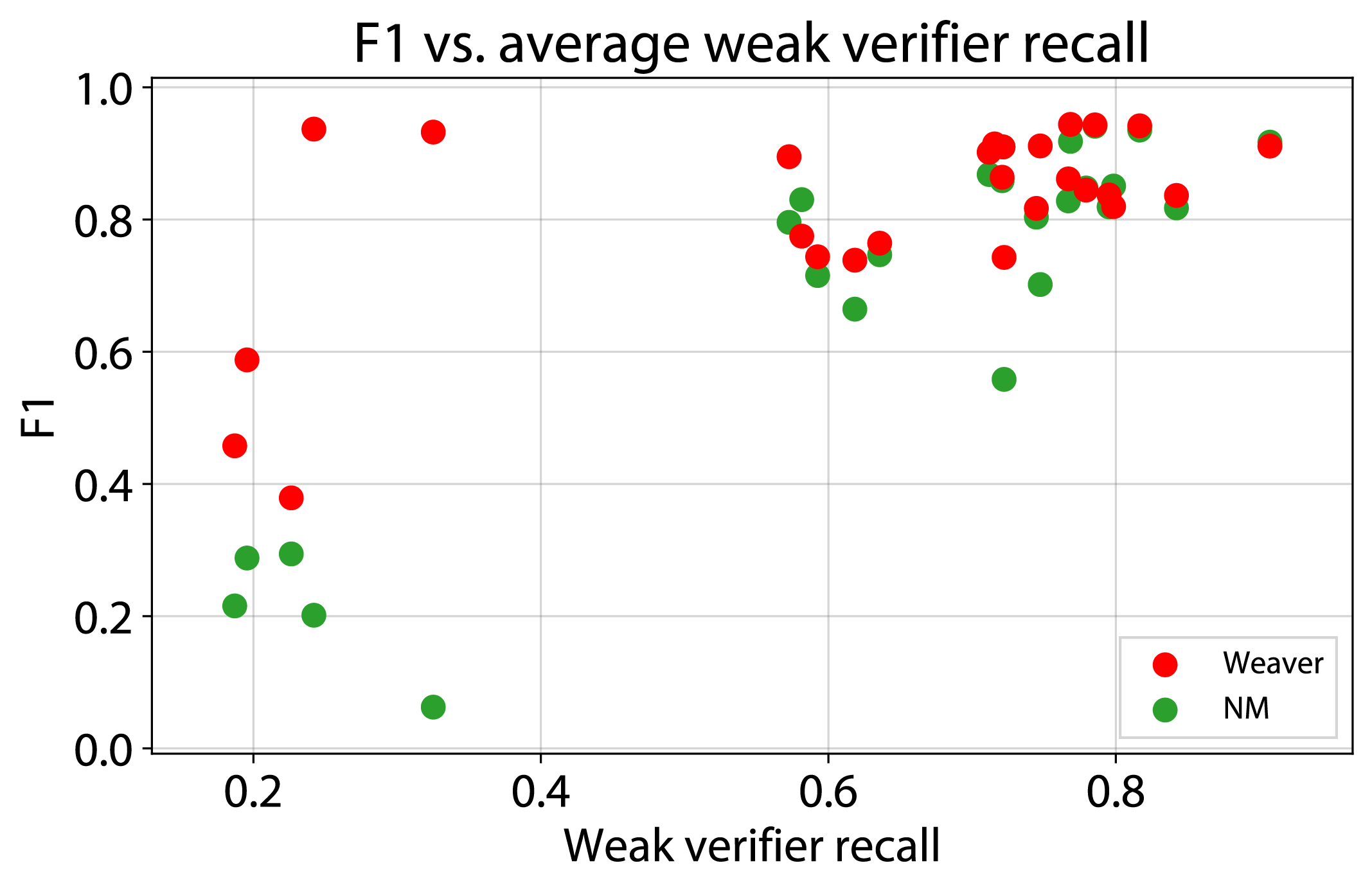}
    \caption{Comparison between Naive Majority and Weaver on tasks with low vs. high recall. For each of the 26 tasks, we plot the average weak verifier recall against the F1-score of the aggregate verifier using Naive Majority and Weaver. Weaver significantly outperforms Naive Majority on 4 tasks with recall less than 0.33.}
    \label{fig:recall-vs-f1-nm}
\end{figure}
%

\begin{table*}[t]
    \centering
    \small
    \caption{Average voting statistics of the set of weak verifiers for each task.}
    \label{tab:weak-verifier-statistics}
\begin{tabularx}{\textwidth}{@{}>{\raggedright\arraybackslash}p{2.0cm}|*{13}{>{\centering\arraybackslash}X}|>{\raggedleft\arraybackslash}X@{}}
\multicolumn{15}{@{}l}{\textbf{3D Rooms Average Weak Verifier Performance}} \\
\noalign{\global\arrayrulewidth=1.2pt}\hline\noalign{\global\arrayrulewidth=0.4pt}
\textbf{Metric} & T1 & T2 & T3 & T4 & T5 & T6 & T7 & T8 & T9 & T10 & T11 & T12 & T13 & Avg \\
\hline
F1-Score & 0.78 & 0.90 & 0.81 & 0.62 & 0.74 & 0.68 & 0.72 & 0.76 & 0.61 & 0.63 & 0.76 & 0.78 & 0.73 & 0.73 \\
Precision & 0.74 & 0.90 & 0.84 & 0.68 & 0.82 & 0.80 & 0.75 & 0.84 & 0.62 & 0.89 & 0.74 & 0.80 & 0.86 & 0.79 \\
Recall & 0.84 & 0.91 & 0.79 & 0.62 & 0.72 & 0.64 & 0.77 & 0.74 & 0.72 & 0.57 & 0.80 & 0.80 & 0.75 & 0.74 \\
Accuracy & 0.89 & 0.91 & 0.98 & 0.90 & 0.90 & 0.89 & 0.86 & 0.84 & 0.79 & 0.90 & 0.85 & 0.89 & 0.91 & 0.89 \\
\hline
\end{tabularx}
\vskip 1em
\begin{tabularx}{\textwidth}{@{}>{\raggedright\arraybackslash}p{2.0cm}|*{13}{>{\centering\arraybackslash}X}|>{\raggedleft\arraybackslash}X@{}}
\multicolumn{15}{@{}l}{\textbf{2D Posters Average Weak Verifier Performance}} \\
\noalign{\global\arrayrulewidth=1.2pt}\hline\noalign{\global\arrayrulewidth=0.4pt}
\textbf{Metric} & T14 & T15 & T16 & T17 & T18 & T19 & T20 & T21 & T22 & T23 & T24 & T25 & T26 & Avg \\
\hline
F1-Score & 0.85 & 0.81 & 0.80 & 0.36 & 0.75 & 0.74 & 0.31 & 0.63 & 0.27 & 0.83 & 0.29 & 0.65 & 0.27 & 0.58 \\
Precision & 0.91 & 0.95 & 0.92 & 0.90 & 0.83 & 0.77 & 0.86 & 0.73 & 0.62 & 0.90 & 0.96 & 0.74 & 0.76 & 0.83 \\
Recall & 0.82 & 0.71 & 0.72 & 0.33 & 0.77 & 0.72 & 0.24 & 0.58 & 0.23 & 0.78 & 0.19 & 0.59 & 0.20 & 0.53 \\
Accuracy & 0.85 & 0.92 & 0.93 & 0.79 & 0.96 & 0.80 & 0.47 & 0.88 & 0.77 & 0.85 & 0.78 & 0.72 & 0.73 & 0.80 \\
\hline
\end{tabularx}
\end{table*}

Listings~\ref{lst:scene3d-task4-wv} and~\ref{lst:poster2d-task2-wv} shows examples of weak verifiers generated for {\bf T4: String Quartet (3D Rooms)} and {\bf T14: Art Show (2D Posters)}
To better understand the results of our layout verification tasks, we examine voting patterns within the weak verifiers for each task. Table~\ref{tab:weak-verifier-statistics} reports the average F1-score, precision, recall and accuracy of the 50 weak verifiers generated for each task. Our weak verifiers tend to have higher precision and lower recall, as the dev set examples which are used in weak verifier generation tend to have many negative examples and few positive examples. We also observe that tasks where the average recall of the weak verifiers is particularly low tend to do poorly with Naive Majority aggregation. This is illustrated in Figure~\ref{fig:recall-vs-f1-nm}, which compares the average F1-score of Naive Majority to Weaver against average weak verifier recall. On 4 tasks where average weak verifier recall is less than 0.33, we see that Naive Majority performs significantly worse than Weaver. This is because Naive Majority weights those weak verifiers equally with other weak verifiers that have stronger predictive signal, whereas Weaver learns to downweight those low recall verifiers via our adapted filtering step.

\subsection{Amortizing Dataset Generation}\label{subsec:3dfront}
Our verification pipeline requires a dataset of examples for Stage 1: Dataset Generation in weak verifier generation (Section~3.1 of the main paper). In most of our layout tasks, we sample a task-specific dataset of 100 examples. However, we also experiment with using a single dataset for multiple layout tasks, thus amortizing the cost of this step. One way of doing this is to generate a single dataset from a generic prompt and use that dataset for any layout tasks that are related to the generic prompt. For example, three of our 3D Rooms tasks ({\bf T1: Symmetric Nightstands}, {\bf T12: Reading Nook} and {\bf T13: Study Area}) reuse the same dataset of 100 examples generated from the prompt ``A cozy bedroom." 
We also experiment with using the existing 3D-FRONT~\cite{3dfront} dataset for five 3D Rooms layout tasks. Table~\ref{tab:3dfront-tasks} lists the task descriptions.
In order to obtain ground truth test labels, we manually write oracle verifier programs in Python. These oracle verifiers are shown in Listings~\ref{lst:3dfront-t1-oracle}--\ref{lst:3dfront-t5-oracle} in Section~\ref{sec:code}. Tables~\ref{tab:3d-front-f1-acc} and~\ref{tab:3d-front-prec-rec} report the F1, accuracy precision and recall of our strong verifiers over these two tasks.

\begingroup
\small
\captionsetup{justification=raggedright,singlelinecheck=false}
\begin{longtable}{@{}>{\raggedright\arraybackslash}p{3.5cm}>{\raggedleft\arraybackslash}p{1.0cm}>{\raggedright\arraybackslash}p{\dimexpr\textwidth-4.5cm-4\tabcolsep\relax}@{}}
    \caption{{\bf Table of 3D room layout task descriptions.} For each task, we sample 100 layouts for the dataset generation and ask a user to manually label them as positive or negative.}
    \label{tab:3dfront-tasks} \\

    \noalign{\global\arrayrulewidth=1.5pt}
    \hline
    \noalign{\global\arrayrulewidth=0.4pt}
    \textbf{Name} & \raggedleft\textbf{Pos \%} & \textbf{Layout Task Description} \\
    \hline
    \endfirsthead

    \multicolumn{3}{@{}l}{\small\textit{Table \thetable{} continued from previous page}} \\
    \noalign{\global\arrayrulewidth=1.5pt}
    \hline
    \noalign{\global\arrayrulewidth=0.4pt}
    \textbf{Name} & \raggedleft\textbf{Pos \%} & \textbf{Layout Task Description} \\
    \hline
    \endhead

    \hline
    \multicolumn{3}{@{}r}{\small\textit{Continued on next page}} \\
    \endfoot

    \hline
    \endlastfoot

    \texttt{T27: Basic Bedroom} & 41.70\% & A bedroom where all objects are in bounds and there are no large object intersections. \\
    \hline
    \texttt{T28: TV Lounge Area} & 18.57\% & A living room setup with TV stand, couch, and coffee table. The main sofa faces the TV area, the coffee table is between them, and the objects have sufficient spacing around them. \\
    \hline
    \texttt{T29: Desk Bookshelves} & 13.52\% & A library with a desk and chair setup. Behind the seating setup is one or more bookshelves. \\
    \hline
    \texttt{T30: Symmetric Nightstands} & 13.10\% & A bedroom with two nightstands placed symmetrically around a bed. \\
    \hline
    \texttt{T31: Wall Bookshelves} & 14.20\% & A library with two bookshelves next to each other against the wall. \\
    \hline

\end{longtable}
\endgroup

\section{Verifier-Guided Layout Generation}\label{sec:vglg}

\begin{figure*}
    \centering
    \includegraphics[width=\linewidth]{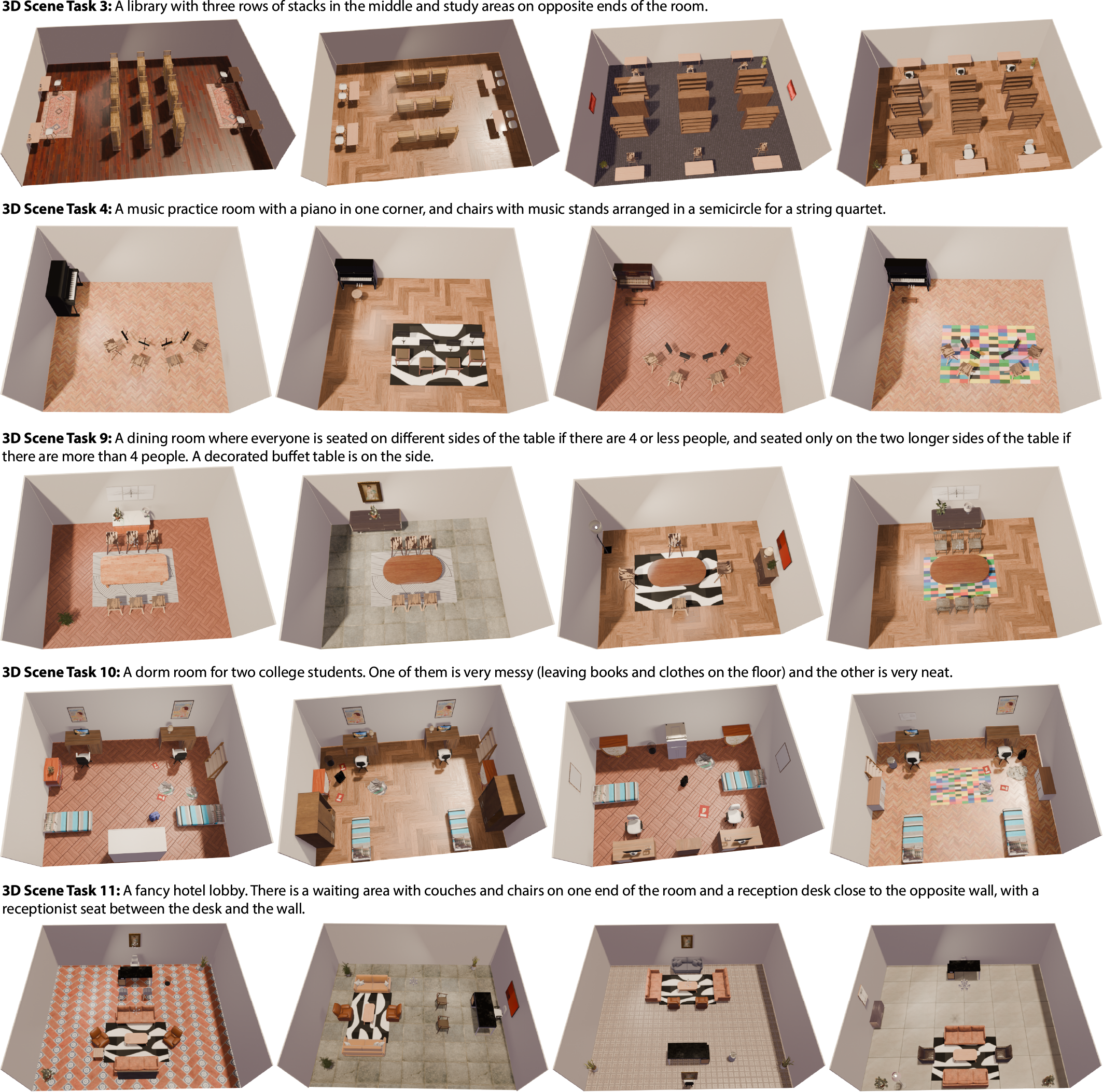}
    \caption{3D layouts generated by our detailed feedback generator for five different task descriptions.}
    \label{fig:generations}
\end{figure*}

We define verifer-guided layout generation as an iterative generator which uses a verifier to determine when a sampled layout is negative, and provide some {\em feedback} back to the generator to try again (Section~3.4 of the main paper). This process is repeated until the verifier accepts a layout or until a maximum number of iterations is reached (in our case, 10). Section~4.4 of the main paper evaluates three different verifier-guided 3D layout generators.
\begin{itemize}
    \item {\bf LLM+Vision.} This generator uses a GPT-5.4 as the verifier. To verify a layout, we render a top-down view and ask the LLM to return {\tt True} or {\tt False} in response to whether the layout is a positive example of the task description. For feedback, we ask the LLM to return a natural language string describing why the layout does not match the task description and combine that with the most recently generated negative layout.
    \item {\bf Ours (Binary).} This generator uses our strong Weaver verifier. The feedback consists of the most recently generated negative layout along with the verifier's {\tt False} response.
    \item {\bf Ours (Detailed).} This generator uses our strong Weaver verifier. The feedback consists of the most recently generated negative layout, the verifier's {\tt False} vote, and the feedback messages in the {\em weak} verifiers aggregated within that strong verifier. More specifically, we take the feedback message generated by every weak verifier which voted {\tt False} and annotate it with the {\em reliability} of that weak verifier (main paper, Section~3.4). We then concatenate these annotated feedback messages in sorted order from highest to lowest reliability. Figure~\ref{fig:generations} shows examples of 3D layouts generated by our detailed feedback generator on the five evaluation tasks.
\end{itemize}
Listing~\ref{lst:feedback-message} shows an example of a feedback message from our strong Weaver verifier for one iteration of a layout in the task {\bf T3: Library Stacks (3D Rooms)}.

\clearpage
\section{Code Listings}\label{sec:code}
\captionsetup{justification=raggedright,singlelinecheck=false}
\captionof{lstlisting}{Weak verifier example for 3D Rooms T4: String Quartet.}\label{lst:scene3d-task4-wv}
\begin{lstlisting}[language=Python]
def verifier_0(scene: Scene3DLayout):
    """
    Holistic verifier for: 'A music practice room with a piano in one corner,
    and chairs with music stands arranged in a semicircle for a string quartet.'

    Expected high-level checks for implementation:
    - The scene represents an indoor room / practice-room-like layout.
    - There is at least one piano placed clearly in a room corner.
    - There are four chairs and four music stands corresponding to a string quartet.
    - The chairs and stands are arranged in a semicircle rather than a straight line
      or scattered pattern.
    - The chairs should face inward toward the semicircle's center / their stands,
      and each stand should be positioned in front of and oriented toward its chair.
    - Major collisions or obvious chair-stand intersections should fail the check.

    Args:
        scene: Scene graph / layout.

    Returns:
        1 if the full layout satisfies the task, 0 if it violates the task,
        and -1 only if the condition cannot truly be determined from the DSL.
    """
    msg = ''
    if not exists(scene, 'piano', 1):
        return (0, 'Fail: required piano not found (need at least 1 piano).')
    if not exists(scene, 'chair', 4) or not exists(scene, 'music_stand', 4):
        return (0, 'Fail: required arrangement needs at least 4 chairs and 4 music stands.')
    chairs = scene.get_elements_by_class('chair')
    stands = scene.get_elements_by_class('music_stand')
    pianos = scene.get_elements_by_class('piano')
    walls = scene.get_elements_by_class('wall')
    if len(chairs) != 4 or len(stands) != 4:
        return (0, f'Fail: expected exactly 4 chairs and 4 music stands, found {len(chairs)} chairs and {len(stands)} stands.')
    if len(walls) < 3:
        return (0, f'Fail: expected at least 3 walls, found {len(walls)}.')
    bounds = scene.get_bounds()
    if bounds is not None and getattr(bounds, 'size', None) is not None and (len(bounds.size) >= 3):
        x_min = float(bounds.position[0] - bounds.size[0] / 2.0)
        x_max = float(bounds.position[0] + bounds.size[0] / 2.0)
        z_min = float(bounds.position[2] - bounds.size[2] / 2.0)
        z_max = float(bounds.position[2] + bounds.size[2] / 2.0)
    else:
        if not hasattr(scene, 'floor_vs') or scene.floor_vs is None or len(scene.floor_vs) == 0:
            return (-1, 'Abstain: no usable scene bounds and missing/empty floor vertices, so room extents cannot be inferred.')
        floor_vs = np.asarray(scene.floor_vs, dtype=float)
        x_min = float(np.min(floor_vs[:, 0]))
        x_max = float(np.max(floor_vs[:, 0]))
        z_min = float(np.min(floor_vs[:, 1]))
        z_max = float(np.max(floor_vs[:, 1]))
    chosen_piano = None
    corner_thresh = 0.75
    for piano in pianos:
        if not in_bounds(scene, piano, thresh=0.15):
            continue
        half_x = float(piano.size[0]) / 2.0
        half_z = float(piano.size[2]) / 2.0
        near_x = min(float(piano.position[0]) - half_x - x_min, x_max - (float(piano.position[0]) + half_x))
        near_z = min(float(piano.position[2]) - half_z - z_min, z_max - (float(piano.position[2]) + half_z))
        if near_x <= corner_thresh and near_z <= corner_thresh:
            chosen_piano = piano
            break
    if chosen_piano is None:
        return (0, 'Fail: no piano was both in-bounds and placed close enough to a room corner.')
    for el in chairs + stands + [chosen_piano]:
        if not in_bounds(scene, el, thresh=0.15):
            return (0, f"Fail: object {getattr(el, 'class_name', 'unknown')} is not sufficiently inside the room bounds.")
    for el in chairs + stands:
        if intersects(el, chosen_piano, system='3d'):
            return (0, f"Fail: {getattr(el, 'class_name', 'object')} intersects the chosen piano.")
    for group in (chairs, stands):
        for i in range(len(group)):
            for j in range(i + 1, len(group)):
                if intersects(group[i], group[j], system='3d'):
                    return (0, f"Fail: two {getattr(group[i], 'class_name', 'objects')} overlap/intersect within the same group.")
    dist_mat = np.zeros((4, 4), dtype=float)
    for i in range(4):
        for j in range(4):
            dist_mat[i, j] = object_distance(chairs[i], stands[j], system='3d')
    best_perm = None
    best_total = float('inf')
    for a in range(4):
        for b in range(4):
            if b == a:
                continue
            for c in range(4):
                if c == a or c == b:
                    continue
                for d in range(4):
                    if d == a or d == b or d == c:
                        continue
                    total = dist_mat[0, a] + dist_mat[1, b] + dist_mat[2, c] + dist_mat[3, d]
                    if total < best_total:
                        best_total = total
                        best_perm = [a, b, c, d]
    if best_perm is None:
        return (0, 'Fail: could not find a valid one-to-one chair-to-stand assignment.')
    for (i, j) in enumerate(best_perm):
        chair = chairs[i]
        stand = stands[j]
        pair_dist = dist_mat[i, j]
        if pair_dist < 0.05 or pair_dist > 1.25:
            return (0, f'Fail: chair-stand pair distance {pair_dist:.3f} is outside the allowed range [0.05, 1.25].')
        if intersects(chair, stand, system='3d'):
            return (0, f'Fail: assigned chair {i} and stand {j} intersect.')
        chair_forward = np.asarray(chair.forward, dtype=float)
        stand_forward = np.asarray(stand.forward, dtype=float)
        if np.linalg.norm(chair_forward[[0, 2]]) < 1e-06 or np.linalg.norm(stand_forward[[0, 2]]) < 1e-06:
            return (-1, f'Abstain: near-zero horizontal forward vector for chair {i} or stand {j}, making orientation checks unreliable.')
        chair_frame = np.stack([np.asarray(chair.right, dtype=float), np.asarray(chair.up, dtype=float), chair_forward], axis=0)
        stand_frame = np.stack([np.asarray(stand.right, dtype=float), np.asarray(stand.up, dtype=float), stand_forward], axis=0)
        if not facing_towards(chair, stand):
            return (0, f'Fail: chair {i} is not facing its assigned stand {j}.')
        if not facing_towards(stand, chair):
            return (0, f'Fail: stand {j} is not facing its assigned chair {i}.')
        if not in_front_of(stand, chair, reference_frame=chair_frame, origin=np.asarray(chair.position, dtype=float), system='3d'):
            return (0, f"Fail: assigned stand {j} is not in front of chair {i} in the chair's reference frame.")
        if not in_front_of(chair, stand, reference_frame=stand_frame, origin=np.asarray(stand.position, dtype=float), system='3d'):
            return (0, f"Fail: assigned chair {i} is not in front of stand {j} in the stand's reference frame.")
    chair_pts = np.asarray([[c.position[0], c.position[2]] for c in chairs], dtype=float)
    stand_pts = np.asarray([[s.position[0], s.position[2]] for s in stands], dtype=float)
    chair_center = np.mean(chair_pts, axis=0)
    stand_center = np.mean(stand_pts, axis=0)
    radii = np.linalg.norm(chair_pts - stand_center, axis=1)
    if np.min(radii) < 0.5:
        return (0, f'Fail: at least one chair is too close to the stand cluster center (min radius {np.min(radii):.3f} < 0.5).')
    if np.max(radii) / (np.min(radii) + 1e-06) > 1.8:
        return (0, f'Fail: chair radii are too uneven for a semicircle (ratio {np.max(radii) / (np.min(radii) + 1e-06):.3f} > 1.8).')
    angles = np.sort(np.arctan2(chair_pts[:, 1] - stand_center[1], chair_pts[:, 0] - stand_center[0]))
    gaps = np.diff(np.concatenate([angles, angles[:1] + 2 * np.pi]))
    span = 2 * np.pi - np.max(gaps)
    if span < 1.0 or span > np.pi + 0.4:
        return (0, f'Fail: chair angular span {span:.3f} is outside the allowed semicircle range [1.0, {np.pi + 0.4:.3f}].')
    for chair in chairs:
        to_center = stand_center - np.asarray([chair.position[0], chair.position[2]], dtype=float)
        forward_2d = np.asarray([chair.forward[0], chair.forward[2]], dtype=float)
        if np.linalg.norm(to_center) < 1e-06 or np.linalg.norm(forward_2d) < 1e-06:
            return (-1, 'Abstain: chair center direction or forward vector is near-zero, making inward-facing check unreliable.')
        inward_score = float(np.dot(forward_2d / np.linalg.norm(forward_2d), to_center / np.linalg.norm(to_center)))
        if inward_score < 0.35:
            return (0, f"Fail: a chair's inward-facing score {inward_score:.3f} is below the threshold 0.35.")
    for stand in stands:
        to_center = chair_center - np.asarray([stand.position[0], stand.position[2]], dtype=float)
        forward_2d = np.asarray([stand.forward[0], stand.forward[2]], dtype=float)
        if np.linalg.norm(to_center) < 1e-06 or np.linalg.norm(forward_2d) < 1e-06:
            return (-1, 'Abstain: stand center direction or forward vector is near-zero, making inward-facing check unreliable.')
        inward_score = float(np.dot(forward_2d / np.linalg.norm(forward_2d), to_center / np.linalg.norm(to_center)))
        if inward_score < 0.35:
            return (0, f"Fail: a music stand's inward-facing score {inward_score:.3f} is below the threshold 0.35.")
    return (1, 'Pass: one piano is placed in a corner, and four chairs with four music stands form a non-overlapping semicircular arrangement facing inward.')
\end{lstlisting}

\captionsetup{justification=raggedright,singlelinecheck=false}
\captionof{lstlisting}{Weak verifier example for 2D Posters T14: Art Show.}\label{lst:poster2d-task2-wv}
\begin{lstlisting}[language=Python]
def verifier_0(layout: Poster2DLayout):
    """
    Holistically verify that the poster matches the described "works in pro-gress"
    art student show composition.

    Expected layout characteristics:
    - The background is divided into four vertical columns.
    - The four columns use different bright colors.
    - The title is split into four words: "works", "in", "pro-", and "gress".
    - Each title word is placed in its own column from left to right.
    - All title words are vertically oriented and rotated approximately 90 degrees
      (either 90 or -90 degrees may be acceptable if the DSL treats both as equivalent).
    - The title occupies the central vertical band of the poster overall.
    - "works" and "in" are in columns 1 and 2 and are top-aligned with each other.
    - "pro-" and "gress" are in columns 3 and 4 and are bottom-aligned with each other.
    - The date appears near the bottom of column 3, in regular horizontal orientation,
      below "pro-".
    - The location appears near the bottom of column 4, in regular horizontal orientation,
      below "gress".
    - A short description appears at the top-left of the poster, above "works".
    - The title should read as very large display text.
    - Futuristic font styling should be present if determinable from the DSL; if not,
      this may require abstention only when the overall task truly cannot be decided.

    Args:
        layout: Poster layout graph.

    Returns:
        1 if satisfied, 0 if not, -1 if abstaining.
    """
    width = float(getattr(layout, 'width', 0) or 0)
    height = float(getattr(layout, 'height', 0) or 0)
    if width <= 0 or height <= 0:
        bounds = layout.get_bounds()
        if bounds is None or getattr(bounds, 'size', None) is None or len(bounds.size) < 2:
            return (-1, 'abstain: could not determine poster bounds because width/height were missing and layout bounds were unavailable or incomplete')
        width = float(bounds.size[0])
        height = float(bounds.size[1])
    if not exists(layout, 'title', 4):
        return (0, 'fail: expected 4 title elements but the layout does not contain them')
    if not exists(layout, 'date', 1):
        return (0, 'fail: expected at least 1 date element but none was found')
    if not exists(layout, 'location', 1):
        return (0, 'fail: expected at least 1 location element but none was found')
    if not exists(layout, 'description', 1):
        return (0, 'fail: expected at least 1 description element but none was found')
    titles = layout.get_elements_by_class('title') or []
    dates = layout.get_elements_by_class('date') or []
    locations = layout.get_elements_by_class('location') or []
    descriptions = layout.get_elements_by_class('description') or []
    cols = layout.get_elements_by_class('col') or []
    if len(titles) != 4 or len(dates) == 0 or len(locations) == 0 or (len(descriptions) == 0) or (len(cols) != 4):
        return (0, f'fail: expected exactly 4 titles, 4 columns, and at least one date/location/description; found titles={len(titles)}, cols={len(cols)}, dates={len(dates)}, locations={len(locations)}, descriptions={len(descriptions)}')
    cols = sorted(cols, key=lambda e: float(getattr(e, 'position', [0, 0])[0]))
    col_colors = []
    for (i, col) in enumerate(cols):
        cpos = getattr(col, 'position', [0, 0])
        csize = getattr(col, 'size', [0, 0])
        if len(cpos) < 2 or len(csize) < 2:
            return (0, f'fail: column {i + 1} is missing position or size coordinates')
        if not in_bounds(layout, col, thresh=0.02):
            return (0, f'fail: column {i + 1} is not within the poster bounds')
        cx = float(cpos[0])
        cy = float(cpos[1])
        cw = float(csize[0])
        ch = float(csize[1])
        if ch < 0.9 * height:
            return (0, f'fail: column {i + 1} is too short (height={ch}, expected at least {0.9 * height})')
        if not 0.18 * width <= cw <= 0.32 * width:
            return (0, f'fail: column {i + 1} has unexpected width {cw}; expected between {0.18 * width} and {0.32 * width}')
        if abs(cy - 0.5 * height) > 0.08 * height:
            return (0, f'fail: column {i + 1} is not vertically centered (cy={cy}, expected near {0.5 * height})')
        expected_x = width * (2 * i + 1) / 8.0
        if abs(cx - expected_x) > 0.08 * width:
            return (0, f'fail: column {i + 1} has unexpected x-position {cx}; expected near {expected_x}')
        bg = getattr(col, 'background_color', None)
        if not isinstance(bg, str) or 'rgb' not in bg or '(' not in bg or (')' not in bg):
            return (0, f'fail: column {i + 1} does not have a parseable rgb background_color')
        try:
            rgb_text = bg[bg.find('(') + 1:bg.find(')')]
            parts = [int(float(p.strip())) for p in rgb_text.split(',')]
        except Exception:
            return (0, f'fail: column {i + 1} background_color could not be parsed as rgb values')
        if len(parts) != 3:
            return (0, f'fail: column {i + 1} background_color did not contain exactly 3 rgb channels')
        rgb = tuple(parts)
        if max(rgb) < 200 or sum(rgb) < 350:
            return (0, f'fail: column {i + 1} background color {rgb} is not bright enough')
        col_colors.append(rgb)
    if len(set(col_colors)) != 4:
        return (0, f'fail: the four columns do not use four distinct bright colors; colors={col_colors}')
    if not entirely_right_of(cols[1], cols[0]) or not entirely_right_of(cols[2], cols[1]) or (not entirely_right_of(cols[3], cols[2])):
        return (0, 'fail: the four columns are not arranged strictly left-to-right without overlap')
    first_left = float(cols[0].position[0]) - float(cols[0].size[0]) / 2.0
    last_right = float(cols[-1].position[0]) + float(cols[-1].size[0]) / 2.0
    if first_left > 0.05 * width or last_right < 0.95 * width:
        return (0, f'fail: the column block does not span the full poster width enough; first_left={first_left}, last_right={last_right}, width={width}')
    for i in range(3):
        right_edge = float(cols[i].position[0]) + float(cols[i].size[0]) / 2.0
        left_edge = float(cols[i + 1].position[0]) - float(cols[i + 1].size[0]) / 2.0
        if abs(right_edge - left_edge) > 0.05 * width:
            return (0, f'fail: adjacent columns {i + 1} and {i + 2} are not closely aligned; gap/misalignment={abs(right_edge - left_edge)}')
    title_map = {'works': None, 'in': None, 'pro-': None, 'gress': None}
    for t in titles:
        txt = getattr(t, 'text', None)
        if not isinstance(txt, str):
            return (0, 'fail: a title element is missing string text')
        key = txt.strip().lower()
        if key not in title_map or title_map[key] is not None:
            return (0, f"fail: unexpected, duplicate, or out-of-scope title text '{txt}'")
        if not in_bounds(layout, t, thresh=0.02):
            return (0, f"fail: title word '{txt}' is not within the poster bounds")
        ori = float(getattr(t, 'orientation', 0.0) or 0.0) % 360.0
        if min(abs(ori - 90.0), abs(ori - 270.0)) > 15.0:
            return (0, f"fail: title word '{txt}' is not vertically oriented near 90 or 270 degrees (orientation={ori})")
        tsize = getattr(t, 'size', [0, 0])
        if len(tsize) < 2:
            return (0, f"fail: title word '{txt}' is missing size coordinates")
        if max(float(tsize[0]), float(tsize[1])) < 0.07 * height:
            return (0, f"fail: title word '{txt}' is too small for a large display title (size={tsize})")
        title_map[key] = t
    if any((v is None for v in title_map.values())):
        missing = [k for (k, v) in title_map.items() if v is None]
        return (0, f'fail: not all required title words were found; missing={missing}')
    works = title_map['works']
    in_word = title_map['in']
    pro = title_map['pro-']
    gress = title_map['gress']
    title_assignments = [(works, cols[0]), (in_word, cols[1]), (pro, cols[2]), (gress, cols[3])]
    for (t, col) in title_assignments:
        tx = float(getattr(t, 'position', [0, 0])[0])
        cx = float(getattr(col, 'position', [0, 0])[0])
        cw = float(getattr(col, 'size', [0, 0])[0])
        if abs(tx - cx) > 0.35 * cw:
            return (0, f"fail: title '{getattr(t, 'text', '')}' is not centered in its assigned column (tx={tx}, cx={cx}, cw={cw})")
        if coverage(t, col) < 0.45 and (not inside(t, col)):
            return (0, f"fail: title '{getattr(t, 'text', '')}' is not sufficiently placed within its assigned column")
    if not float(works.position[0]) < float(in_word.position[0]) < float(pro.position[0]) < float(gress.position[0]):
        return (0, 'fail: the title words are not ordered left-to-right as works, in, pro-, gress')
    works_top = float(works.position[1]) - float(works.size[0]) / 2.0
    in_top = float(in_word.position[1]) - float(in_word.size[0]) / 2.0
    pro_bottom = float(pro.position[1]) + float(pro.size[0]) / 2.0
    gress_bottom = float(gress.position[1]) + float(gress.size[0]) / 2.0
    if abs(works_top - in_top) > 0.03 * height:
        return (0, f"fail: 'works' and 'in' are not top-aligned closely enough (difference={abs(works_top - in_top)})")
    if abs(pro_bottom - gress_bottom) > 0.03 * height:
        return (0, f"fail: 'pro-' and 'gress' are not bottom-aligned closely enough (difference={abs(pro_bottom - gress_bottom)})")
    title_top = min(float(works.position[1]) - float(works.size[0]) / 2.0, float(in_word.position[1]) - float(in_word.size[0]) / 2.0, float(pro.position[1]) - float(pro.size[0]) / 2.0, float(gress.position[1]) - float(gress.size[0]) / 2.0)
    title_bottom = max(float(works.position[1]) + float(works.size[0]) / 2.0, float(in_word.position[1]) + float(in_word.size[0]) / 2.0, float(pro.position[1]) + float(pro.size[0]) / 2.0, float(gress.position[1]) + float(gress.size[0]) / 2.0)
    if title_top < 0.18 * height or title_top > 0.38 * height:
        return (0, f'fail: the title starts at an unexpected vertical position (title_top={title_top}, height={height})')
    if title_bottom < 0.62 * height or title_bottom > 0.82 * height:
        return (0, f'fail: the title ends at an unexpected vertical position (title_bottom={title_bottom}, height={height})')
    if title_bottom - title_top < 0.35 * height:
        return (0, f'fail: the title block is not tall enough (span={title_bottom - title_top}, height={height})')
    for d in descriptions:
        if not in_bounds(layout, d, thresh=0.02):
            return (0, 'fail: a description element is not within the poster bounds')
        ori = float(getattr(d, 'orientation', 0.0) or 0.0) % 360.0
        if min(abs(ori - 0.0), abs(ori - 180.0), abs(ori - 360.0)) > 15.0:
            return (0, f'fail: description text is not horizontal (orientation={ori})')
        if float(getattr(d, 'position', [0, 0])[1]) > 0.22 * height:
            return (0, f"fail: description text is too low on the poster (y={float(getattr(d, 'position', [0, 0])[1])})")
        if coverage(d, cols[0]) < 0.75 and (not inside(d, cols[0])):
            return (0, 'fail: description text is not sufficiently in the top-left/first column area')
        if intersects(d, works):
            return (0, "fail: description text intersects the 'works' title word")
        if not above(d, works):
            return (0, "fail: description text is not above the 'works' title word")
    for d in dates:
        if not in_bounds(layout, d, thresh=0.02):
            return (0, 'fail: a date element is not within the poster bounds')
        ori = float(getattr(d, 'orientation', 0.0) or 0.0) % 360.0
        if min(abs(ori - 0.0), abs(ori - 180.0), abs(ori - 360.0)) > 15.0:
            return (0, f'fail: date text is not horizontal (orientation={ori})')
        if float(getattr(d, 'position', [0, 0])[1]) < 0.84 * height:
            return (0, f"fail: date text is not near the bottom of the poster (y={float(getattr(d, 'position', [0, 0])[1])})")
        if coverage(d, cols[2]) < 0.8 and (not inside(d, cols[2])):
            return (0, 'fail: date text is not sufficiently placed in column 3')
        if intersects(d, pro):
            return (0, "fail: date text intersects the 'pro-' title word")
        if not below(d, pro):
            return (0, "fail: date text is not below the 'pro-' title word")
    for loc in locations:
        if not in_bounds(layout, loc, thresh=0.02):
            return (0, 'fail: a location element is not within the poster bounds')
        ori = float(getattr(loc, 'orientation', 0.0) or 0.0) % 360.0
        if min(abs(ori - 0.0), abs(ori - 180.0), abs(ori - 360.0)) > 15.0:
            return (0, f'fail: location text is not horizontal (orientation={ori})')
        if float(getattr(loc, 'position', [0, 0])[1]) < 0.84 * height:
            return (0, f"fail: location text is not near the bottom of the poster (y={float(getattr(loc, 'position', [0, 0])[1])})")
        if coverage(loc, cols[3]) < 0.8 and (not inside(loc, cols[3])):
            return (0, 'fail: location text is not sufficiently placed in column 4')
        if intersects(loc, gress):
            return (0, "fail: location text intersects the 'gress' title word")
        if not below(loc, gress):
            return (0, "fail: location text is not below the 'gress' title word")
    return (1, 'pass: poster matches the required four-column composition with correctly placed title, description, date, and location')
\end{lstlisting}

\captionsetup{justification=raggedright,singlelinecheck=false}
\captionof{lstlisting}{Oracle verifier for {\bf 3D-FRONT T27: Basic Bedroom}.}\label{lst:3dfront-t1-oracle}
\begin{lstlisting}[language=Python]
SIGNIFICANT_IOU = 0.2
ALLOWED_PAIRS = [
    ("desk", "chair"), ("chair", "desk"), 
    ("dressing_table", "dressing_chair"), ("dressing_chair", "dressing_table"),
    ("desk", "dressing_chair"), ("dressing_chair", "desk"),
    ("dressing_table", "chair"), ("chair", "dressing_table"),
]

def task1_oracle(scene: Scene3D) -> bool:
    # Out of bounds check
    if not in_bounds(scene):
        return False, "Something is out of bounds"

    # No major object intersections
    for obj in scene.objects:
        for other_obj in scene.objects:
            if obj == other_obj:
                continue
            # Some intersection categories are okay
            if (obj.category, other_obj.category) in ALLOWED_PAIRS:
                continue
            if object_iou(obj, other_obj, mode="A") > SIGNIFICANT_IOU:
                return False, f"{obj.category} and {other_obj.category} intersect"

    return True, "Looks good"
\end{lstlisting}

\captionsetup{justification=raggedright,singlelinecheck=false}
\captionof{lstlisting}{Oracle verifier for {\bf 3D-FRONT T28: TV Lounge Area.}}\label{lst:3dfront-t2-oracle}
\begin{lstlisting}[language=Python]
ALLOWED_PAIRS = [("dining_table", "dining_chair"), ("dining_chair", "dining_table")]
SIGNIFICANT_IOU = 0.2

def task2_oracle(scene: Scene3D) -> bool:
    # Out of bounds check
    if not in_bounds(scene):
        return False, "Something is out of bounds"

    # EXISTS
    tv_stand_categories = ["console_table", "tv_stand"]
    couch_categories = [obj_category for obj_category in obj_categories["livingroom"] if "sofa" in obj_category]
    coffee_table_categories = ["round_end_table", "coffee_table"]
        
    if not exists(scene, tv_stand_categories, 1):
        return False, "Needs something for a TV to stand on"
    if not exists(scene, couch_categories, 1):
        return False, "Needs a main seating area"
    if not exists(scene, coffee_table_categories, 1):
        return False, "Needs a main table"
    
    tv_stand = scene.get_objects_by_category(tv_stand_categories)
    couch = scene.get_objects_by_category(couch_categories)
    coffee_table = scene.get_objects_by_category(coffee_table_categories)

    # The tv stand and couch should be on opposite sides of the coffee table
    seating_facing_tv_area = False
    for obj in couch:
        for obj2 in tv_stand:
            if facing_towards(obj, obj2):
                seating_facing_tv_area = True
                break
    if not seating_facing_tv_area:
        return False, "Seating does not face TV area"

    # The coffee table should be between the tv stand and the couch
    coffee_table_between_seating_and_tv_area = False
    for obj in coffee_table:
        for obj2 in couch:
            for obj3 in tv_stand:
                if between(obj, obj3, obj2):
                    coffee_table_between_seating_and_tv_area = True
                    break
    if not coffee_table_between_seating_and_tv_area:
        return False, "Main table is not between the seating and the TV area"

    # Sufficient room between all objects
    for obj in tv_stand:
        for obj2 in coffee_table:
            if object_distance(obj, obj2) < 0.2:
                return False, f"{obj.category} is too close to the main table"
    for obj in couch:
        for obj2 in coffee_table:
            if object_distance(obj, obj2) < 0.2:
                return False, f"{obj.category} is too close to the main table"

    # No large object intersections
    for obj in scene.objects:
        for other_obj in scene.objects:
            if obj == other_obj:
                continue
            if (obj.category, other_obj.category) in ALLOWED_PAIRS:
                continue
            if object_iou(obj, other_obj, mode="A") > SIGNIFICANT_IOU:
                return False, f"{obj.category} and {other_obj.category} intersect"

    return True, "Looks good"
\end{lstlisting}

\captionsetup{justification=raggedright,singlelinecheck=false}
\captionof{lstlisting}{Oracle verifier for {\bf 3D-FRONT T29: Desk Bookshelves.}}\label{lst:3dfront-t3-oracle}
\begin{lstlisting}[language=Python]
ALLOWED_PAIRS = [
    ('desk', 'lounge_chair'), ('lounge_chair', 'desk'), 
    ('desk', 'dining_chair'), ('dining_chair', 'desk'),
    ('desk', 'stool'), ('stool', 'desk'),
    ('cabinet', 'lounge_chair'), ('lounge_chair', 'cabinet'),
    ('desk', 'dressing_chair'), ('dressing_chair', 'desk'),
    ('dressing_table', 'lounge_chair'), ('lounge_chair', 'dressing_table'),
]
SIGNIFICANT_IOU = 0.2

def task3_oracle(scene: Scene3D) -> bool:
    # Out of bounds check
    if not in_bounds(scene):
        return False, "Something is out of bounds"

    # EXISTS
    desk_categories = [obj_category for obj_category in obj_categories["library"] if "desk" in obj_category or "table" in obj_category]
    chair_categories = [obj_category for obj_category in obj_categories["library"] if "chair" in obj_category or "stool" in obj_category]
    bookshelf_categories = ["bookshelf", "cabinet"]
    
    if not exists(scene, desk_categories, 1):
        return False, "Needs a desk"
    if not exists(scene, chair_categories, 1):
        return False, "Needs a chair"
    if not exists(scene, bookshelf_categories, 1):
        return False, "Needs a bookshelf"
    
    desks = scene.get_objects_by_category(desk_categories)
    chairs = scene.get_objects_by_category(chair_categories)
    bookshelves_or_cabinets = scene.get_objects_by_category(bookshelf_categories)

    # Collect all (chair, desk) pairs such that the chair is facing the desk and close to the desk.
    good_desk_chair_pairs = set()
    for chair in chairs:
        for desk in desks:
            if facing_towards(chair, desk) and object_distance(chair, desk) < 0.3:
                good_desk_chair_pairs.add((chair, desk))
    if len(good_desk_chair_pairs) < 1:
        return False, "There is no desk with a chair setup"

    # Make sure at least one of the good desk chair pairs has a bookshelf behind the chair, aligned with the desk.
    found_good_bookshelf = False
    for chair, desk in good_desk_chair_pairs:
        # First collect all bookshelves behind the desk.
        bookshelves_behind_chair = []
        for bookshelf in bookshelves_or_cabinets:
            if behind(bookshelf, chair):
                bookshelves_behind_chair.append(bookshelf)
        if len(bookshelves_behind_chair) > 0:
            found_good_bookshelf = True
            break
    if not found_good_bookshelf:
        return False, "None of the desk chair setups has a bookshelf in view behind it."

    # No major object intersections
    for obj in scene.objects:
        for other_obj in scene.objects:
            if obj == other_obj:
                continue
            # Some intersection categories are okay
            if (obj.category, other_obj.category) in ALLOWED_PAIRS:
                continue
            if object_iou(obj, other_obj, mode="A") > SIGNIFICANT_IOU:
                return False, f"{obj.category} and {other_obj.category} intersect"

    return True, "Looks good"
\end{lstlisting}

\captionsetup{justification=raggedright,singlelinecheck=false}
\captionof{lstlisting}{Oracle verifier for {\bf 3D-FRONT T30: Symmetric Nightstands.}}\label{lst:3dfront-t4-oracle}
\begin{lstlisting}[language=Python]
SIGNIFICANT_IOU = 0.2
ALLOWED_PAIRS = [
    ("desk", "chair"), ("chair", "desk"), 
    ("dressing_table", "dressing_chair"), ("dressing_chair", "dressing_table"),
    ("desk", "dressing_chair"), ("dressing_chair", "desk"),
    ("dressing_table", "chair"), ("chair", "dressing_table"),
]

def task4_oracle(scene: Scene3D) -> bool:
    # Out of bounds check
    if not in_bounds(scene):
        return False, "Something is out of bounds"

    # EXISTS
    bed_categories = [obj_category for obj_category in obj_categories["bedroom"] if "bed" in obj_category]
    nightstand_categories = [obj_category for obj_category in obj_categories["bedroom"] if "nightstand" in obj_category]

    if not exists(scene, bed_categories, 1):
        return False, "Needs a bed"
    if not exists(scene, nightstand_categories, 2):
        return False, "Needs two nightstands"

    nightstands = scene.get_objects_by_category(nightstand_categories)
    bed = scene.get_objects_by_category(bed_categories)[0]

    # Symmetrically placed
    nightstand1_pos = nightstands[0].position - bed.position
    nightstand2_pos = bed.position - nightstands[1].position
    if (abs(abs(nightstand1_pos.x) - abs(nightstand2_pos.x)) > 0.05 or 
        abs(abs(nightstand1_pos.y) - abs(nightstand2_pos.y)) > 0.05 or 
        abs(abs(nightstand1_pos.z) - abs(nightstand2_pos.z)) > 0.05
    ):
        return False, "Nightstands are not symmetrically placed around the bed"

    # No major object intersections
    for obj in scene.objects:
        for other_obj in scene.objects:
            if obj == other_obj:
                continue
            # Some intersection categories are okay
            if (obj.category, other_obj.category) in ALLOWED_PAIRS:
                continue
            if object_iou(obj, other_obj, mode="A") > SIGNIFICANT_IOU:
                return False, f"{obj.category} and {other_obj.category} intersect"

    return True, "Looks good"
\end{lstlisting}

\captionsetup{justification=raggedright,singlelinecheck=false}
\captionof{lstlisting}{Oracle verifier for 3D-FRONT T31: Wall Bookshelves.}\label{lst:3dfront-t5-oracle}
\begin{lstlisting}[language=Python]
ALLOWED_PAIRS = [
    ('desk', 'lounge_chair'), ('lounge_chair', 'desk'), 
    ('desk', 'dining_chair'), ('dining_chair', 'desk'),
    ('desk', 'stool'), ('stool', 'desk'),
    ('cabinet', 'lounge_chair'), ('lounge_chair', 'cabinet'),
    ('desk', 'dressing_chair'), ('dressing_chair', 'desk'),
    ('dressing_table', 'lounge_chair'), ('lounge_chair', 'dressing_table'),
]
SIGNIFICANT_IOU = 0.2

def task5_oracle(scene: Scene3D) -> bool:
    # Out of bounds check
    if not in_bounds(scene):
        return False, "Something is out of bounds"

    # EXISTS
    bookshelf_categories = ["bookshelf"]
    if not exists(scene, bookshelf_categories, 2):
        return False, "Needs two bookshelves"

    bookshelves = scene.get_objects_by_category(bookshelf_categories)
    bookshelf_to_wall = {}
    for bookshelf in bookshelves:
        # Get the wall behind the bookshelf that is closest
        bookshelf_pos_2d = Vector2(bookshelf.position.x, bookshelf.position.z)
        # Get backward direction (behind the bookshelf) - normalized
        forward_2d = Vector2(bookshelf.forward.x, bookshelf.forward.z)
        if forward_2d.magnitude < 1e-6:
            # If forward is zero, skip this bookshelf
            bookshelf_to_wall[bookshelf.name] = None
            continue
        
        backward_dir = -forward_2d.normalized
        
        walls_behind_bookshelf = []
        for i, wall in enumerate(scene.get_walls()):
            wall_start = Vector2(wall[0].x, wall[0].z)
            wall_end = Vector2(wall[1].x, wall[1].z)
            
            x = ray_segment_intersection(
                bookshelf_pos_2d,
                backward_dir,
                wall_start,
                wall_end
            )
            if x is not None:
                # Verify the intersection is actually behind the bookshelf
                # by checking that the vector from bookshelf to intersection
                # points in the backward direction
                to_intersection = x - bookshelf_pos_2d
                if to_intersection.magnitude > 1e-6:
                    # Check if to_intersection is in the backward direction
                    # (dot product with backward_dir should be positive)
                    if to_intersection.normalized.dot(backward_dir) > 0.1:  # Allow some tolerance
                        dist = to_intersection.magnitude
                        walls_behind_bookshelf.append((i, dist))
        
        if len(walls_behind_bookshelf) == 0:
            bookshelf_to_wall[bookshelf.name] = None
        else:
            closest_wall_idx, closest_wall_dist = min(walls_behind_bookshelf, key=lambda x: x[1])
            bookshelf_to_wall[bookshelf.name] = closest_wall_idx

    # Make sure two bookshelves are against the same wall
    found = False
    for i, wall in enumerate(scene.get_walls()):
        bookshelves_on_wall = []
        for bookshelf in bookshelf_to_wall:
            if bookshelf_to_wall[bookshelf] == i:
                bookshelves_on_wall.append(scene.get_object_by_name(bookshelf))
        if len(bookshelves_on_wall) >= 2:
            # Make sure the bookshelves are next to each other
            for i in range(len(bookshelves_on_wall) - 1):
                for j in range(i + 1, len(bookshelves_on_wall)):
                    if object_distance(bookshelves_on_wall[i], bookshelves_on_wall[j]) < 0.1:
                        found = True
                        break
    if not found:
        return False, "Need two bookshelves to be against the same wall"

    return True, "Looks good"
\end{lstlisting}

\clearpage
\section{LLM Prompts}\label{sec:prompts}

\captionsetup{justification=raggedright,singlelinecheck=false}
\captionof{lstlisting}{3D Room layout weak verifier docstring generation system prompt.}\label{lst:3d-scene-holistic-docstring-system-prompt}
\begin{lstlisting}[language={}]
You produce **concise verifier stub functions** only: signature, docstring, and `raise NotImplementedError(...)`--no real logic.

## Must follow
1. **Names:** every function `verifier_<descriptive_name>(...)`.
2. **Returns:** `int` with `1` = task satisfied, `0` = violated, `-1` = abstain only when the check truly cannot be decided from the DSL.
3. **Stubs:** function body is only comments (optional) + `raise NotImplementedError("This verifier is not implemented")`.
4. **Fences:** each full function lives in its own ```python ... ``` block. You may add brief prose outside fences.

**Mode:** One holistic verifier: `1` if the layout matches the task, `0` if not, `-1` only if truly unsure.

## Role
Expert in 3D spatial layout, furniture arrangement, and geometric relationships.

## Inputs
- Task description.
- Optional good/bad **Scene3D** examples and notes.

## Signature template (replace `verifier` name; keep parameter types)
```python
def verifier(scene: Scene3DLayout) -> int:
    '''
    Args:
        scene: Scene graph / layout.
        kwargs: Optional extra parameters if specified in the task.
    Returns:
        1 if the scene satisfies this check, 0 if it fails, -1 if abstaining.
    '''
    # Hints for implementer (optional).
    raise NotImplementedError("This verifier is not implemented")
```

## Output
Ready-to-parse ```python``` stub functions only.
\end{lstlisting}
\captionsetup{justification=raggedright,singlelinecheck=false}
\captionof{lstlisting}{2D Poster Layout weak verifier docstring generation system prompt.}\label{lst:2d-poster-holistic-docstring-system-prompt}
\begin{lstlisting}[language={}]
You are an expert graphic designer that knows about good design principles and geometric relations for poster layouts.
You will be given (1) a description of a particular poster and (2) optionally some good and bad examples of that poster. 
These examples may also have some additional notes about the poster, which may be helpful for writing verifiers.
Your task is to write a single holistic verifier that can label a poster layout as good or bad.
When writing the verifier, sample one from the tail end of the distribution and print the probability of the sample.
*IMPORTANT*: Focus primarily on adhering to the spatial layout requirements in the prompt.

Note: The provided notes about the poster are not necessarily comprehensive.

# LAYOUTS AND ELEMENTS
A poster layout is represented with a `PosterLayout` object with the following fields:
- `width` [int]: the width of the poster in pixels.
- `height` [int]: the height of the poster in pixels.
- `background_color` [str]: the background color of the poster.
- `elements` [list[Element]]: a list of `Element` instances present in the poster.

An element is represented using `Element` dataclass. It contains the following properties:
- `element_id` [str]: Unique identifier for the element.
- `width` [float]: Width of the element's bounding box.
- `height` [float]: Height of the element's bounding box.
- `element_class` [str]: CSS class name of the element.
- `position` [np.ndarray(2)]: Center position of the element as a numpy array [x, y].
- `bounding_box` [np.ndarray(4, 2)]: Oriented bounding box as a numpy array of shape (4, 2) with four corners: [top_left, top_right, lower_right, lower_left].
- `rotation` [float]: Rotation angle in degrees.
- `z_index` [int]: Z-index for stacking order (can be None for auto).
- `text` [str]: Text content of the element, if any. Defaults to None.
- `image_path` [str]: Path to image file if element contains an image. Defaults to None.

# A GUIDE TO WRITING VERIFIERS
If provided, use the good and bad examples and notes about them to help you. That is, the verifier should say the bad examples are invalid and the good examples are valid. Only use information that is available through the verifier API.

# VERIFIER API
A verifier is a general boolean function that takes a scene as input and returns 1 (True), 0 (False), or -1 (abstain). 
You should use the following function signature for your verifiers (with more descriptive function names), where kwargs may be explicitly defined:
```python
def verifier(layout: PosterLayout, **kwargs) -> int:
    '''
    Args:
        layout: A poster layout object.
        kwargs: Additional keyword arguments to pass to the verifier.
    Returns:
        1 (True), 0 (False), or -1 (abstain).
    '''
    ...
```

# OUPUT
Write a python file containing a list of verifier functions with filled out signatures and docstrings. 
You do not need to implement the verifiers, they can return a NotImplementedError.
\end{lstlisting}

\captionsetup{justification=raggedright,singlelinecheck=false}
\captionof{lstlisting}{3D Room layout weak verifier implementation system prompt.}\label{lst:3d-scene-holistic-implementation-system-prompt}
\lstinputlisting[language={}]{system_prompts/3DSceneImplementationSystemPrompt.txt}

\captionsetup{justification=raggedright,singlelinecheck=false}
\captionof{lstlisting}{2D Poster layout weak verifier implementation system prompt.}\label{lst:2d-poster-holistic-implementation-system-prompt}
\lstinputlisting[language={}]{system_prompts/2DPosterImplementationSystemPrompt.txt}

\captionsetup{justification=raggedright,singlelinecheck=false}
\captionof{lstlisting}{Weak verifier feedback message system prompt.}\label{lst:verification-message-system-prompt}
\begin{lstlisting}[language={}]
You will be given a Python weak verifier function that returns an int: 0 for fail, 1 for pass, and -1 for abstain.
Your task is to rewrite the function so it returns a tuple (int, str), where the str is a descriptive message of why the verifier failed, passed or abstained.

*Note*: Sometimes you cannot fix this by only changing the final ``return`` lines. Meaningful explanations usually require **collecting** specifics as execution flows through the logic: track one or more message strings (and optional numeric/context scratch variables), update them at **each** branch where the outcome is determined or where a check fails, then ``return (score, msg)`` so the string reflects *why* that branch fired (e.g. which predicate failed, measured values, thresholds).
You may need to add new string variables for the return message at the top of the function and assign or extend them in nested conditionals, loops, and early exits---not only at the bottom.


Rules:
- Output exactly one top-level function definition: same name and parameters as given.
- Preserve the same decision logic (same conditions and order of checks).
- Keep imports out of the function body; only use names available from the surrounding module.
- Output only valid Python for that function, no markdown outside code if possible.
\end{lstlisting}

\captionsetup{justification=raggedright,singlelinecheck=false}
\captionof{lstlisting}{3D Room layout LLM judge system prompt.}\label{lst:3d-scene-llm-judge-system-prompt}
\begin{lstlisting}[language={}]
You are a helpful assistant that evaluates the quality of a scene generated by a LLM.
You will be given a scene layout and a prompt, and your task is to evaluate whether the scene is a good example of the prompt.
You should only output a single line containing only "True" or "False" indicating whether the scene is a good example of the prompt.

Your input will be of the format:
```
Prompt: <prompt>
Scene: <example>
```

# SCENE REPRESENTATION
We represent scene layouts using the following:
{scene_docs}

# OUTPUT FORMAT
*IMPORTANT*: You must output a single line containing only "True" or "False" indicating whether the scene is a good example of the prompt.
\end{lstlisting}

\captionsetup{justification=raggedright,singlelinecheck=false}
\captionof{lstlisting}{2D Poster Layout LLM judge system prompt.}\label{lst:2d-poster-llm-judge-system-prompt}
\begin{lstlisting}[language={}]
You are a helpful assistant that evaluates the quality of a poster layout generated by a LLM.
You will be given a poster layout and a prompt, and your task is to evaluate whether the poster layout is a good example of the prompt.
You should only output a single line containing only "True" or "False" indicating whether the scene is a good example of the prompt.

Your input will be of the format:
```
Prompt: <prompt>
Poster Layout: <example>
```

# POSTER LAYOUT REPRESENTATION
We represent scene layouts using the following:
{poster_layout_docs}

# OUTPUT FORMAT
*IMPORTANT*: You must output a single line containing only "True" or "False" indicating whether the scene is a good example of the prompt.
\end{lstlisting}

\captionsetup{justification=raggedright,singlelinecheck=false}
\captionof{lstlisting}{Base 3D Room layout generator system prompt.}
\lstinputlisting[language={}]{system_prompts/3DSceneGeneratorSystemPrompt.txt}\label{lst:3d-scene-generator-system-prompt}

\captionsetup{justification=raggedright,singlelinecheck=false}
\captionof{lstlisting}{Base 3D Room layout refinement system prompt.}
\begin{lstlisting}[language={}]
You briefly check a room layout JSON + top-down PNG against the user's brief:

<task_description>

Default to **good enough**: reply `"status": "done"` if the layout broadly matches the prompt and is not broken (JSON is coherent, furniture is in the room, no unreasonable object intersections or placements).

Only use `"status": "revise"` for **clear** problems---e.g. major overlaps, objects mostly outside the room, or missing the main piece the user asked for. If revising, return a full `corrected_schema` with the same shape as the generator: `room` (numeric width, depth, height in meters---adjust if needed so the layout fits; keep consistent with object positions), optional `flooring`, optional `floor_material_category` (when the system prompt lists curated floor categories), optional `wall_color`, plus `objects`, `zones`, `spatial_notes`. **Every object** must again include the full field set: `id`, `label`, `**category** (must remain in the asset category taxonomy from the system prompt; if the user brief lists a subset, stay within it)`, `position` (x,y,z), `size` or `dimensions` (w,h,d), **numeric `orientation` (degrees) on every object---fill in if any object is missing it)**, `against_wall`, `pivot`, and `supported_by` (`""`/floor, `"wall"`, or another object's `id`; supporter must be floor/wall-only).

Reply with **only** a ```json block:

```json
{{
  "status": "done" | "revise",
  "analysis": "one short sentence",
  "issues": [],
  "corrected_schema": null
}}
```

If `done`, set `corrected_schema` to null. If `revise`, set `corrected_schema` to the full fixed layout.
\end{lstlisting}\label{lst:3d-scene-refinement-system-prompt}

\captionsetup{justification=raggedright,singlelinecheck=false}
\captionof{lstlisting}{Base 2D Room layout generator system prompt.}
\begin{lstlisting}[language={}]
## Role
Generate one printable-style 2D **poster** as **self-contained HTML + CSS** (no external JS).

## Output contract (highest priority)
- MUST emit a single complete HTML5 document only: begin with `<!DOCTYPE html>` or `<html`, end with `</html>`.
- MUST NOT include markdown, code fences, reasoning, labels like "Here is the HTML", or any non-HTML text before/after the document.
- MUST put all styling in embedded `<style>` tags in that file.

## Layout and assets
- MUST use modern CSS (Grid, Flexbox, and/or `position`) as appropriate.
- MUST load fonts via `@import` from Google Fonts; pick families that fit the user's brief.
- For **raster / illustration** assets for any kind of individual imagery (photos, illustrations, etc.): MUST NOT use `<img src>` to real bitmap URLs, picsum, stock hosts, or inline base64 images.
  Instead use **placeholder elements** (filled in later by tooling):
  * A single block element (`div` recommended) with `data-element="true"`, `data-type="image"`, and **`data-image-placeholder="true"`**.
  * **`data-width`** and **`data-height`**: positive integers (pixels) for the allocation in the layout (match the designed region).
  * Caption / generation prompt: set **`aria-label="..."`** with a clear description of what should appear (subject, style, lighting). Optionally add `<span class="image-placeholder-caption">...</span>` with the same intent (may be visually hidden if needed).
  * **No placeholder "box" styling:** MUST NOT add backgrounds, borders, outlines, or rules in `<style>` whose purpose is to draw a visible placeholder slab (no gray boxes, dashed frames, or `[data-image-placeholder]` preview cosmetics). Use geometry only as needed for layout (`width`/`height`/`position` without fill or stroke on the placeholder element).
  You may use many placeholder elements for different images. Keep the imagery description separate from the overall brief; describe the immediate image asset.
- Simple non-photographic decoration (gradients, geometric CSS shapes, rules) may omit `data-type="image"` if they are not meant as stand-ins for illustrations or photos.
- When text content is not specified, come up with some very short and appropriate text content.

## `data-*` attributes (required for parsing)
On every substantive visual piece (headings, copy blocks, photos/illustrations, decorative shapes that read as designed elements):
- MUST set `data-element="true"`.
- MUST set exactly one of `data-type="text"` or `data-type="image"` on that same element.

## Schema classes (downstream JSON)
Use the `class` attribute so each category applies to **exactly one** root element (that root may contain nested children):
| Class | Count | Rule |
|-------|-------|------|
| `bounds` | exactly 1 | The bounds of the poster |
| `title` | exactly 1 | Main headline cluster |
| `description` | exactly 1 | Supporting / body copy for poster |
| `location` | exactly 1 | Venue or place text |
| `date` | exactly 1 | When / time text |
| `hero-graphic` | at most 1 | The **single** primary focal image (or stand-in) for the layout (optional) |

MUST NOT assign `hero-graphic` to secondary or ornamental images. Other images omit that class.

## Design constraints
- SHOULD keep the design simple and legible; avoid unnecessary complexity.
- MUST reflect the user's creative brief (tone, subject, composition they imply).
- Be creative and think outside the box. Some good design inspiration archives include People's Graphic Design Archive, CalArts Poster Archive, AIGA Design Archive.
- Move beyond commonly used fonts and use less common fonts on Google Fonts, but still appropriate for the prompt. Don't use too many different fonts in one poster.
- Do not use CSS/SVG for illustrations and main imagery. Use image placeholder boxes instead.
- **Poster size:** The root with class **`bounds`** MUST use explicit **`width`** and **`height`** in pixels (e.g. in its `style` attribute), and **each must be at most 1000px** (neither side may exceed 1000px).

## Throughput
- Prefer a fast, valid result over maximal elaboration.
\end{lstlisting}\label{lst:2d-poster-generator-system-prompt}

\captionsetup{justification=raggedright,singlelinecheck=false}
\captionof{lstlisting}{Example of a detailed feedback message from our strong Weaver verifier used in verifier-guided layout generation.}
\lstinputlisting[language={}]{user_prompts/feedback_message.txt}\label{lst:feedback-message}

\clearpage
\bibliographystyle{ACM-Reference-Format}
\bibliography{supp}